\documentclass[superscriptaddress,preprintnumbers,amssymb,twocolumn]{revtex4} 
\pdfoutput=1
\usepackage{graphicx}
\usepackage{epsfig}
\usepackage{amsmath}
\usepackage{amsfonts}
\usepackage{amssymb}

\usepackage{braket}
\usepackage{cancel}
\usepackage{pstricks}
\usepackage{color}

\usepackage{ulem}

\def\lsim{\:\raisebox{-0.5ex}{$\stackrel{\textstyle<}{\sim}$}\:}
\def\gsim{\:\raisebox{-0.5ex}{$\stackrel{\textstyle>}{\sim}$}\:}

\usepackage{epstopdf}

\begin{document}
{\tiny }\global\long\def\bra#1{\Bra{#1}}
{\tiny }\global\long\def\ket#1{\Ket{#1}}
{\tiny }\global\long\def\set#1{\Set{#1}}
{\tiny }\global\long\def\braket#1{\Braket{#1}}
{\tiny }\global\long\def\norm#1{\left\Vert #1\right\Vert }
{\tiny }\global\long\def\rmto#1#2{\cancelto{#2}{#1}}
{\tiny }\global\long\def\rmpart#1{\cancel{#1}}
{\tiny \par}

\title {Light-quarks Yukawa couplings and new physics in exclusive high-$p_T$ Higgs + jet and Higgs + $b$-jet events}
\author{Jonathan Cohen}
\email{jcohen@tx.technion.ac.il}
\affiliation{Physics Department, Technion-Institute of Technology, Haifa 32000, Israel}
\author{Shaouly Bar-Shalom}
\email{shaouly@physics.technion.ac.il}
\affiliation{Physics Department, Technion-Institute of Technology, Haifa 32000, Israel}
\author{Gad Eilam}
\email{eilam@physics.technion.ac.il}
\affiliation{Physics Department, Technion-Institute of Technology, Haifa 32000, Israel}
\author{Amarjit Soni}
\email{adlersoni@gmail.com}
\affiliation{Physics Department, Brookhaven National Laboratory, Upton, NY 11973, USA}

\date{\today}

\begin{abstract}
We suggest that the exclusive Higgs + light (or b)-jet production at the LHC,
$pp \to h+j(j_b)$, is a rather sensitive probe
of the light-quarks Yukawa couplings and of other forms of new physics (NP) in the
Higgs-gluon $hgg$ and quark-gluon $qqg$ interactions.
We study the Higgs $p_T$-distribution in $pp \to h+j(j_b) \to \gamma \gamma + j(j_b)$,
i.e., in $h+j(j_b)$ production followed by the Higgs
decay $h \to \gamma \gamma$, employing
the ($p_T$-dependent) signal strength formalism
to probe various types of NP which are relevant to these processes and
which we parameterize
either as scaled Standard Model (SM) couplings (the kappa-framework) and/or
 through new higher dimensional effective operators (the SMEFT framework).
We find that the exclusive $h+j(j_b)$ production
at the 13 TeV LHC is sensitive to various NP scenarios,
with typical scales ranging from a few TeV to ${\cal O}(10)$ TeV, depending
on the flavor, chirality and Lorentz structure of the underlying physics.
\end{abstract}


\maketitle

\section{Introduction \label{sec1}}

The next runs of the LHC will be dedicated to two primary tasks:
the search for new physics (NP) and the detailed scrutiny of the Higgs properties,
which might shed light on NP specifically related to the origin of
mass and flavor and to the observed hierarchy between the two disparate Planck
and ElectroWeak (EW) scales.
Indeed, the study of Higgs systems is in particular challenging,
since it requires precision examination of some of its weakest couplings
(within the SM) and measurements of highly non-trivial
processes involving high jet multiplicities, large backgrounds and
low detection efficiencies.

The s-channel Higgs production and its subsequent decays, $pp \to h \to ff$,
which led to its discovery, are relatively inefficient for NP searches. In particular,
if the NP scale, $\Lambda$, is of ${\cal O}({\rm TeV})$ and larger, then
its effect in
these processes is expected to be suppressed by at least $\sim m_h^2/\Lambda^2$,
since most of these events come from the dominant gluon fusion s-channel production mechanism
and are, therefore, clustered around $\sqrt{s} \simeq m_h$.
However, in some fraction of the events, the Higgs recoils
against one or more hard jets and, thus, carries a large $p_T$,
which may play a key role in the hunt for NP and/or for background
rejection in Higgs studies. Indeed, a key observable for Higgs boson events is
the number of jets produced in the
event. For that reason, and since the Higgs $p_T$ distribution is sensitive to the production
mechanism, there has recently been a growing interest,
both experimentally \cite{ATLAS1,ATLAS2,ATLAS3,ATLAS4,CMS1,CMS2} and theoretically
\cite{hjth1,schulz2,schulz3,hjth2,hjth3,hjth4,hjth5,highptrecent,1703.03886}, in
the behavior of the Higgs $p_T$ distribution in inclusive and exclusive Higgs production,
where the Higgs carries a substantial fraction of transverse momentum (for earlier work
see \cite{veryearly1,veryearly2,early1,early2}). In particular, the Higgs $p_T$ distribution in the
exclusive Higgs + jets production, $pp \to h + nj$, was one of the prime
targets of the measurements performed recently by ATLAS and CMS
\cite{ATLAS1,ATLAS2,ATLAS3,ATLAS4,CMS1,CMS2}.

In this paper we will thus focus on the exclusive Higgs + 1-jet production, $pp \to h + j$,
where $j$ stands for either a ``light-jet" defined as any
non-flavor tagged jet originating
from a gluon or light-quarks $j = g,u,d,c,s$ (i.e., assuming them to be
indistinguishable from the observational point of view) or a $b$-quark jet ($j_b$).
It is interesting to note that there has been some hints in the LHC 8 TeV data for an excess
in the $h+j$ channel \cite{ATLAS3,schulz3}, although the statistics are still limited
and the theoretical uncertainties are relatively large.
Indeed, a significant effort has been dedicated in recent years, from the theory side, towards
understanding and reducing the uncertainties pertaining to the Higgs+jet production cross-section
at the LHC \cite{hjth1,schulz1,schulz2,hjth2,hjth3,hjth4,hjth5,hbjet1,1704.06620}, with
special attention given to higher transverse momentum of the Higgs, where NP effects
are expected to become more apparent. In particular,
the high-$p_T$ Higgs spectrum in $pp \to h+j(j_b)$ can be sensitive to various well motivated
NP scenarios, such as supersymmetry \cite{susyhj,0309204,dawson1,dawson2},
heavy top-partners \cite{toppartner},
higher dimensional effective operators
\cite{1312.3317_SMEFT,1411.2029,1409.6299_dawson,1308.2225_dim7,hbjet2_bMM} and NP
in Higgs-top quark and Higgs-gluon interactions in the so-called ``kappa-framework",
where one assumes that the $hgg$ and $htt$ interactions
are scaled by some factor with respect to the SM \cite{1309.5273,1405.4295,1405.7651,1612.00283}.

\begin{figure}[htb]
\begin{center}
\includegraphics[scale=0.2]{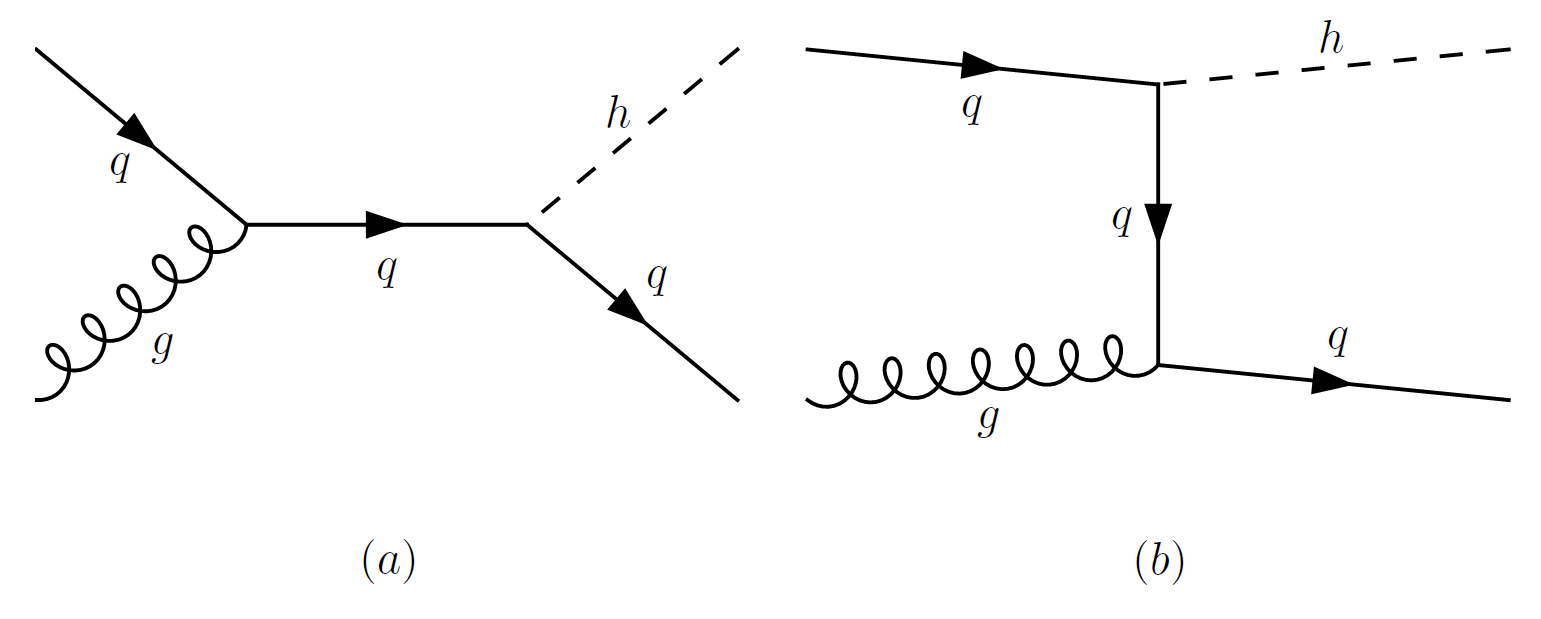}
\end{center}
\vspace{-0.8cm}
\caption{The tree-level SM diagrams for $gq \to q h$, where $q = u,d,c,s,b$.
The diagrams corresponding to $g \bar q \to h \bar q $ and $q \bar q \to g h$ can be
obtained by crossing symmetry, see also text.}
\label{SMTL}
\end{figure}

In general, there is
a tree-level contribution to $pp \to h+j(j_b)$ in the SM from the hard processes
$gq \to q h, ~g \bar q \to \bar q h$ and $q \bar q \to g h$ ($q = u,d,c,s,b$).
The corresponding SM tree-level diagrams, which are depicted in Fig.~\ref{SMTL}, are
proportional
to the light-quarks Yukawa couplings, $y_q$, so that the SM tree-level
contribution to the overall $pp \to h+j(j_b)$ cross-section is small
(e.g., in the case of $pp \to h+c$, it is at the percent level). In particular,
the squared matrix elements, summed and averaged over spins and colors, for these tree-level
hard processes are:
\begin{eqnarray}
\sum \overline{ \left| {\cal M}_{SM}^{q \bar q \to gh} \right|^2}  &=&
\frac{2 g_s^2 y_q^2}{{\cal C}_{qq}}
\frac{m_h^4+\hat s^2}{\hat t \hat u} ~, \label{dsigb1} \\
\sum \overline{ \left| {\cal M}_{SM}^{q g \to q h} \right|^2} &=&
- \frac{{\cal C}_{qq}}{{\cal C}_{qg}} \sum \overline{ \left| {\cal M}_{SM}^{q \bar q \to gh} \right|^2} (\hat s \leftrightarrow \hat t)~,  \label{dsigb2} \\
\sum \overline{ \left| {\cal M}_{SM}^{\bar q g \to \bar q h} \right|^2} &=&
- \frac{{\cal C}_{qq}}{{\cal C}_{qg}} \sum \overline{ \left| {\cal M}_{SM}^{q \bar q \to gh} \right|^2} (\hat s \leftrightarrow \hat u)~,  \label{dsigb3}
\end{eqnarray}
where $\hat s=(p_1+p_2)^2,~\hat t=(p_1+p_3)^2$ and $\hat u=(p_2+p_3)^2$,
defined for the process $q(-p_1)+ \bar q(-p_2) \to h + g(p_3)$. Also,
$g_s$ is the strong coupling constant and
${\cal C}_{qq}=N^2$, ${\cal C}_{qg}=NV$ are the color average factors, where
$V=N^2-1=8$ corresponds to the number of gluons in the adjoint representation
of the SU(N) color group.

Thus, in the limit $y_q \to 0$, the dominant
and leading order (LO) SM contribution to the Higgs + light-jet cross-section, $\sigma(pp \to h+j)$,
arises from the
1-loop process $gg \to gh$, which is generated by 1-loop top-quark exchanges
(and the subdominant
b-quark loops \cite{bquark-loops}), and can be parameterized by an effective Higgs-gluon
$ggh$ interaction Lagrangian:
\begin{eqnarray}
{\cal L}_{eff}^{ggh} = C_g h G_{\mu \nu}^a G^{\mu \nu, a} ~, \label{Leffggh}
\end{eqnarray}
where $C_g$ is the Higgs-gluon point-like effective coupling,
which at lowest order in the SM is \cite{veryearly1,veryearly2}:
$C_g=\alpha_s/(12\pi v)$, where $v=246$ GeV is the Higgs vacuum expectation value (VEV).
In what follows we will use the point-like $ggh$ effective coupling of Eq.~\ref{Leffggh}
with $C_g$ given as an asymptotic expansion in $1/m_t$ up to $m_t^{-6}$,
as implemented in MADGRAPH5 for the
Higgs effective field theory (HEFT) model \cite{HEFTggh}.
We will neglect throughout this work the 1-loop effects of the b-quark and of the
lighter quarks with enhanced Yukawa couplings (i.e., as large as the b-quark Yukawa),
which are expected to yield a correction at the level of a few percent compared
to the dominant top-quark loops, when the
Higgs transverse momentum is larger than $\sim m_h/2$ \cite{bquark-loops,1606.09253}.

This prescription for the Higgs-gluon coupling
is a good approximation for a Higgs produced with a $p_T(h) \lsim 200$ GeV, see e.g.,
\cite{highptrecent,overestimate}, whereas, as will be shown in this work, the harder $p_T(h) \gsim 200$ GeV
regime is important for probing NP in Higgs +jet production.
However, since the exact form of the loop induced $ggh$ interaction
(i.e., using a finite top-quark mass) is currently unknown beyond LO (1-loop),
we choose to work with the effective $ggh$ point-like interaction (as described above)
in order to simplify the calculation and the presentation of our analysis.
Given the exploratory nature of this work and the type of study
presented, this approximation is not expected to have an effect on our results
at a level which changes the main outcome and conclusions of this work.
In particular, in order to give an estimate
of the sensitivity of our results to the
calculation scheme, we will also study and analyse some samples of our
results using the exact LO calculation of the 1-loop diagrams (mass dependent top-quark exchanges)
which involve the $ggh$ interaction vertex. Indeed, since this LO 1-loop calculation
is the only currently available exact (mass dependent)
calculational setup for $pp \to h+j(j_b)$, a comparison
between the NP effects calculated with the point-like $ggh$ approximation and with the
mass dependent 1-loop diagrams can serve as a yardstick for
the uncertainty and sensitivity of our results
to the calculational setup.

The subprocesses $gq \to q h$, $g \bar q \to \bar q h$ and $q \bar q \to g h$
(which, as can be seen from Eqs.~\ref{dsigb1}-\ref{dsigb3}, are proportional to $y_q^2$ at tree-level)
also receive a 1-loop contribution from the above $ggh$ effective vertex (i.e., from the
top-quark loops), which is, however,
small compared to the $gg \to gh$ \cite{veryearly1,veryearly2,early1,early2}.
In particular, the $gg \to gh$ contribution to $\sigma(pp \to h+j)$ at the LHC is
about an order of magnitude larger than the one from $gq \to q h$ and more than two orders of magnitude
larger than the two other channels $g \bar q \to \bar q h$ and $q \bar q \to g h$.
\begin{figure}[htb]
\begin{center}
\includegraphics[scale=0.20]{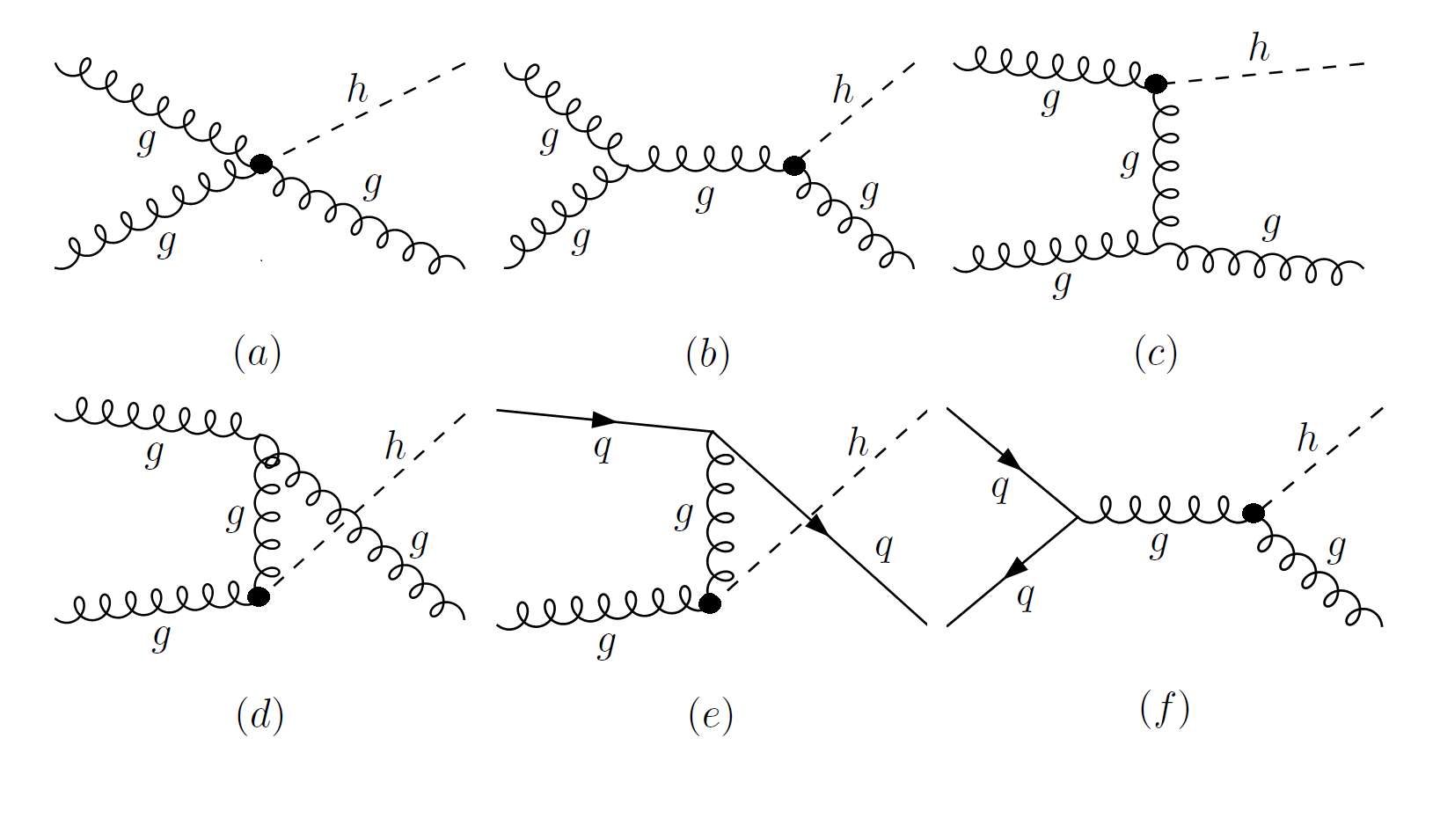}
\end{center}
\vspace{-1.0cm}
\caption{The 1-loop SM diagrams for
$gg \to gh, ~gq \to q h$ and $q \bar q \to g h$,
in the effective Higgs-gluon description,
where the loop-induced $ggh$ and $gggh$ vertices are represented by
a heavy dot. See also text.}
\label{SM1L}
\end{figure}

The 1-loop (and LO for $y_q =0$) SM differential hard cross-sections for
$gg \to gh, ~gq \to q h, g \bar q \to \bar q h$ and
$q \bar q \to g h$
(the corresponding SM diagrams for all channels are shown in Fig.~\ref{SM1L}), expressed
in terms of the above effective $ggh$ interaction and neglecting the light-quark masses,
are given by \cite{veryearly1,veryearly2}:
\begin{eqnarray}
\sum \overline{ \left| {\cal M}_{SM}^{gg \to gh} \right|^2} &\simeq& \frac{96 g_s^2 C_g^2}{{\cal C}_{gg}}
\frac{m_h^8+\hat s^4+\hat t^2+\hat u^2}{\hat s \hat t \hat u} ~, \label{dsig1} \\
\sum \overline{ \left| {\cal M}_{SM}^{q \bar q \to gh} \right|^2} &\simeq&
\frac{16 g_s^2 C_g^2}{{\cal C}_{qq}}
\frac{\hat t^2+\hat u^2}{\hat s} ~, \label{dsig2} \\
\sum \overline{ \left| {\cal M}_{SM}^{q g \to q h} \right|^2} &\simeq&
- \frac{{\cal C}_{qq}}{{\cal C}_{qg}} \sum \overline{ \left| {\cal M}_{SM}^{q \bar q \to gh} \right|^2} (\hat s \leftrightarrow \hat t)~,  \label{dsig3} \\
\sum \overline{ \left| {\cal M}_{SM}^{\bar q g \to \bar q h} \right|^2} &\simeq&
- \frac{{\cal C}_{qq}}{{\cal C}_{qg}} \sum \overline{ \left| {\cal M}_{SM}^{q \bar q \to gh} \right|^2} (\hat s \leftrightarrow \hat u)~,  \label{dsig4}
\end{eqnarray}
where ${\cal C}_{gg}=V^2=64$ and
$\hat s=(p_1+p_2)^2,~\hat t=(p_1+p_3)^2,~\hat u=(p_2+p_3)^2$, with momenta
defined via $h \to g(p_1)+g(p_2)+g(p_3)$ for $gg \to gh$ and via
$q(-p_1)+ \bar q(-p_2) \to h + g(p_3)$ for $q \bar q \to gh$.

Turning now to the possible manifestation of NP in Higgs + jet production at the LHC,
there are, in principle, two ways in which
$pp \to h+j(j_b)$ can be modified:
\begin{itemize}
\item when the NP generates new interactions that are absent
in the SM and that can potentially change the SM kinematic distributions
in this process.
\item when the NP comes in the form of scaled SM couplings, corresponding to the
previously mentioned kappa-framework.
\end{itemize}
We will explore both types of NP effects in $pp \to h+j$ and $pp \to h+j_b$
 and, in particular,
focus on NP that modifies the light and b-quarks Yukawa couplings and/or the light and b-quarks
interactions with the gluon, as well as the Higgs-gluon effective vertex in
Eq.~\ref{Leffggh}.
Indeed, the Higgs mechanism of the SM implies that the fermion's Yukawa couplings
are proportional to the ratio between their masses and the EW VEV, i.e.,
$y_f \propto m_f/v$. Thus, at least for the light fermions of the
1st and 2nd generations [where $m_f/v \sim {\cal O}(10^{-5})$ and $m_f/v \sim {\cal O}(10^{-4}-10^{-3})$, respectively],
any signal which can be associated with their Yukawa couplings would stand out
as clear evidence for NP beyond the SM.
The current experimental bounds on the Yukawa couplings of light-quark's
of the 1st and 2nd generations, $y_u,y_d,y_s,y_c$, coming from fits
to the measured Higgs data,
allow them to be as large as the b-quark Yukawa $y_b$ \cite{bounds1}.
From the phenomenological point of view, it is, therefore, important
to explore the possibility that the light-quark Yukawa couplings and/or
their interactions with the gauge boson's are significantly enhanced or modified with
respect to the SM.
Indeed, there has recently been a growing interest in the study
of light-quark's Yukawa couplings,
see e.g., \cite{perez1,yotam1,1606.09253,1608.04376,felix,han,1608.01746,1405.0990,han1}.
For example, in \cite{yotam1,1606.09253}, the Higgs
$p_T$ distributions in inclusive Higgs production, $pp \to h+X$, was used
to study the sensitivity to $y_q$,
where it was shown that the measurements from the 8 TeV LHC run constrain the
Yukawa couplings of the 1st generation quarks and the c-quark to
be $y_u,y_d \lsim 0.5 y_b$ \cite{yotam1}
and $y_c \lsim 5 y_b$ \cite{1606.09253}, respectively.
Slightly improved bounds are expected in the inclusive
channel at the future LHC Runs:
$y_u,y_d \lsim 0.3 y_b$ \cite{yotam1,1608.04376} and
$y_c \lsim y_b$ \cite{1606.09253}.
As we will see below, a $p_T$-dependent ratio between the NP and SM cross-sections
(the signal strength)
for the exclusive Higgs + jet production
cross-section, $\sigma(pp \to h +j)$, followed by the Higgs decays to e.g.,
$\gamma \gamma$ and $WW^\star$, may be used to put
comparable and, in some cases, stronger constraints on $y_q$. In particular,
we will show that,
if the $ggh$ effective coupling also
deviates from its SM value, then
significantly stronger bounds on $y_q$ are expected.

We also explore exclusive Higgs + jet production in the SMEFT, defined as the expansion of the SM
Lagrangian with an infinite series of higher dimensional effective operators.
We find that the exclusive $pp \to h+j(j_b)$ signal can probe
the NP scenarios portrayed by the SMEFT with typical scales ranging from a few to ${\cal O}(10)$ TeV,
depending on the details of underlying physics.

The paper is organized as follows: in section \ref{sec2}
we outline our notation and define our observables for the study of NP
in $pp \to h+j$ and $pp \to h+j_b$.
In sections \ref{sec3} and \ref{sec4} we discuss the NP effects in
$pp \to h+j(j_b)$ within the kappa and the SMEFT frameworks, respectively, and
in section \ref{sec5} we summarize.

\section{Notation and observables \label{sec2}}

We define the signal strength for $pp \to h+j$ (and similarly for $pp \to h+j_b$),
followed
by the Higgs decay $h \to ff$, where
$f$ can be any of the SM Higgs decay
products (e.g.,  $f=b,~\tau,~\gamma,~W,~Z$), as the ratio of the number of
$pp \to h+j \to ff + j$ events
in some NP scenario relative
to the corresponding number of Higgs events in the SM:
%
\begin{eqnarray}
\mu_{hj}^f = \frac{{\cal N}(pp \to h+j \to ff+j)} {{\cal N}_{SM}(pp \to h+j \to ff+j)}
 ~. \label{sig-strength0}
\end{eqnarray}
%

In particular, ${\cal N}$ is the event yield ${\cal N}={\cal L} \sigma {\cal A} \epsilon$,
where ${\cal L}$ is the luminosity, ${\cal A}$ is the acceptance in the signal analysis (i.e.,
the fraction of events that ''survive" the cuts) and $\epsilon$ is the efficiency which
represents the probability that the fraction of events that pass the set of cuts are
correctly identified.
Clearly, the luminosity and efficiency factors, ${\cal L}$ and $\epsilon$, cancel by definition
in $\mu_{hj}^f$ of Eq.~\ref{sig-strength0}, whereas the acceptance factors,
${\cal A}$ and ${\cal A}_{SM}$,
do not in general, unless the NP in the numerator
of $\mu_{hj}^f$ does not change the kinematics of the events.
Given the exploratory nature of this work, we will assume, for simplicity, that
${\cal A} \simeq {\cal A}_{SM}$ in Eq.~\ref{sig-strength0}, in which case
one obtains:$^{[1]}$\footnotetext[1]{The effect of ${\cal A} \neq {\cal A}_{SM}$ can be estimated by simulating
the detector acceptance in the actual analysis, and scaling our results below
(for the signal strength $\mu_{hj}^f$) by the factor ${\cal A}/{\cal A}_{SM}$.}
%
\begin{eqnarray}
\mu_{hj}^f \simeq
\frac{\sigma(pp \to h+j)} {\sigma_{SM}(pp \to h+j)} \cdot \frac{BR(h \to ff)} {BR_{SM}(h \to ff)}
 ~. \label{sig-strength}
\end{eqnarray}
%

We further assume that there is no NP in the Higgs decay $h \to ff$
and, for definiteness, we will occasionally consider the decay channel $h \to \gamma \gamma$ (i.e., with a SM rate),
at the LHC with a luminosity
of 300 $fb^{-1}$ and/or 3000 $fb^{-1}$ (corresponding to the high-luminosity LHC, HL-LHC), representing the lower and higher statistics cases for the Higgs + jet signal $pp \to h + j \to \gamma \gamma +j$.

We will henceforward use the $p_T$-dependent ``cumulative cross-section", satisfying
a given lower Higgs $p_T$ cut, as follows:
\begin{eqnarray}
\sigma(p_T^{cut}) \equiv \sigma \left( p_T(h) > p_T^{cut} \right) =
\int_{p_T(h) \geq p_T^{cut}} dp_T \frac{d\sigma}{dp_T} ~, \label{dsigpt}
\end{eqnarray}
which turns out to be useful for minimizing the
ratio between the higher-order and LO $pp \to h+j$ cross-sections
(i.e., the K-factor)
for values of $p_T^{cut} \gsim 150$ GeV \cite{schulz2,hjth3}.
Furthermore,
as was mentioned earlier and will be shown below, the $p_T$-distribution of
the Higgs may be sensitive to the specific type of the underlying NP,
so that the cumulative cross-section
of Eq.~\ref{dsigpt} gives an extra
handle for extracting the NP effects in $pp \to h+j$, without having to
analyze fully differential quantities associated with $pp \to h+j$.

All cross-sections are calculated using MadGraph5 \cite{madgraph5}
at LO parton-level,
where a dedicated universal FeynRules output (UFO) model
was produced for the MadGraph5 sessions
using FeynRules \cite{FRpaper}, for both the kappa and SMEFT frameworks.
The analytical results were cross-checked with Formcalc \cite{FormCalc},
while intermediate steps were validated using FeynCalc \cite{FeynCalc}.
We use the LO MSTW 2008 PDF set \cite{mstw2008}, in the
4 flavor and 5 flavor schemes ${\rm MSTW2008lo68cl}$\_nf4 and ${\rm MSTW2008lo68cl}$,
respectively, with a dynamical scale choice for the central value of the factorization ($\mu_F$)
and renormalization ($\mu_R$) scales, corresponding to the sum of
the transverse mass in the hard-process level:
$\mu_F = \mu_R = \mu_T \equiv \sum_i \sqrt{m_i^2+p_{T}^{2}(i)} = \sqrt{m_h^2+p_{T}^{2}(h)} + p_T(j)$.
The uncertainty in
$\mu_F$  and $\mu_R$ is evaluated by varying them in the range
$\frac{1}{2} \mu_T \leq \mu_F, \mu_R \leq 2 \mu_T$.
As mentioned above, all cross-sections were calculated with a lower
$p_{T}(h)$ cut and, in some instances, an overall invariant
mass cut was imposed using Mad-Analysis5 \cite{MGA}.

To study the sensitivity of $\mu_{hj}^f$ to NP we
define our NP signal to be (recall that $\mu_{hj}^f(SM) = 1$):
\begin{eqnarray}
\Delta\mu_{hj}^f \equiv \mid \mu_{hj}^f - 1\mid ~, \label{deltamu}
\end{eqnarray}
and assume that $\mu_{hj}^f$ will be measured to a given accuracy $\delta\mu_{hj,exp}^f(1\sigma)$,
with a central value $\hat\mu_{hj,exp}^f$:
\begin{eqnarray}
\mu_{hj,exp}^f=\hat\mu_{hj,exp}^f \pm \delta\mu_{hj,exp}^f(1\sigma) ~. \label{muexp}
\end{eqnarray}

Thus, taking $\hat\mu_{hj,exp}^f = \mu_{hj}^f$
($\mu_{hj}^f$ being our prediction for the measured
value $\hat\mu_{hj,exp}^f$),
the statistical significance of the NP signal is:
\begin{eqnarray}
N_{SD} = \frac{\Delta\mu_{hj}^f}{\delta\mu_{hj}^f} ~, \label{NSD}
\end{eqnarray}
which we will use in the following analysis,
where $\delta\mu_{hj}^f$ represents the combined experimental and theoretical $1\sigma$ error, e.g.,
$\delta\mu_{hj}^f =
\sqrt{\left(\delta\mu_{hj,theory}^f\right)^2 +
\left(\delta\mu_{hj,exp}^f\right)^2}$.
In particular, in the spirit of the ultimate goal of the Higgs physics program, which is to reach a percent level
accuracy in the measurements and calculations of Higgs production and decay modes \cite{Higgsplan},
we will assume throughout this work that
the signal strength
defined above, for Higgs+jet production followed by the Higgs decay, will be measured and known
to a 5\%($1\sigma$) accuracy. That is, that the combined
experimental and theoretical uncertainties will be pushed down to $\delta\mu_{hj}^f = 0.05(1\sigma)$.
Indeed, achieving such an accuracy is both a theoretical and experimental challenge,
which, however, seems to be feasible in the LHC era with the large statistics expected in the future
runs and in light of the recent progress made in higher-order calculations.

\begin{figure}[htb]
\begin{center}
\includegraphics[scale=0.55]{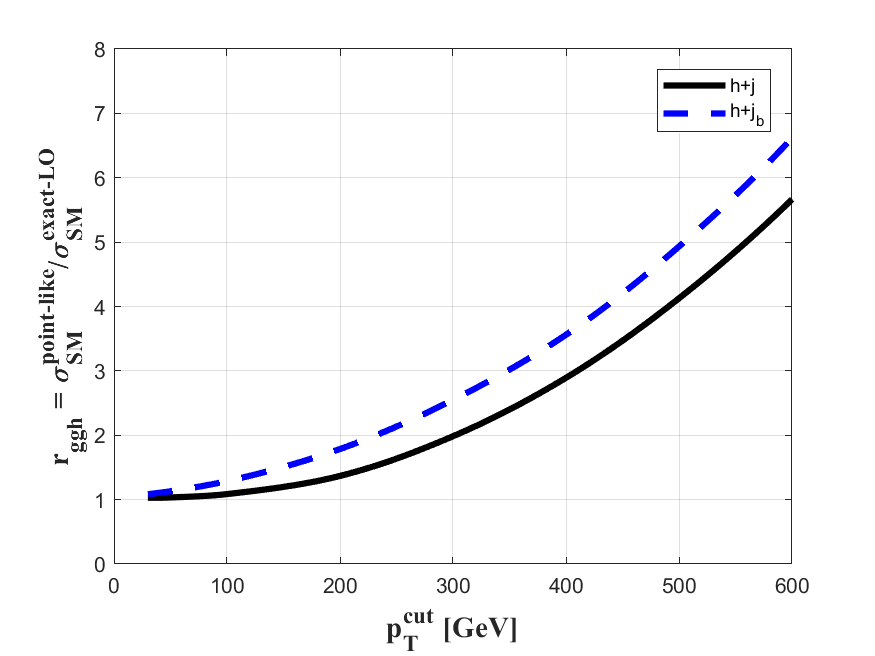}
\end{center}
\caption{The ratio $r_{ggh}$ defined in Eq.~\ref{rggh}, as a function
of $p_T^{cut}$:
$r_{ggh} = \sigma_{SM}^{point-like}(p_T^{cut})/\sigma_{SM}^{exact-LO}(p_T^{cut})$,
where $\sigma_{SM}^{point-like}(p_T^{cut})$ and $\sigma_{SM}^{exact-LO}(p_T^{cut})$
are the cumulative SM cross-sections which are
calculated for a given $p_T^{cut}$, using the
point-like $ggh$ approximation and the full LO 1-loop set of diagrams,
respectively.
See also text.}
\label{compsigma}
\end{figure}

Finally, we wish to briefly address
the uncertainty associated with
the effective point-like $ggh$ approximation which we use for the calculation of
all the SM-like diagrams for $pp \to h+j(j_b)$ that
involve the $ggh$ interaction
(i.e., all diagrams in Fig.~\ref{SM1L} in the $pp \to h+j$ case and
diagram (e) in Fig.~\ref{SM1L} for the $pp \to h+j_b$ case).
As mentioned earlier, for the differential $p_T(h)$ distribution,
$d \sigma/dp_T(h)$, this approximation is accurate
up to $p_T(h) \lsim 200$ GeV. As a result,
the $p_T$-dependent cumulative cross-section
defined in Eq.~\ref{dsigpt} accrues an error which depends on
the $p_T^{cut}$ used.
To estimate the corresponding uncertainty in $\sigma_{SM}(p_T^{cut})$,
we plot in Fig.~\ref{compsigma} the ratio:
\begin{eqnarray}
r_{ggh} \equiv \frac{\sigma_{SM}^{point-like}(p_T^{cut})}{\sigma_{SM}^{exact-LO}(p_T^{cut})} \label{rggh} ~,
\end{eqnarray}
as a function
of $p_T^{cut}$ for both $pp \to h+j$ and $pp \to h+j_b$,
where $\sigma_{SM}^{point-like}(p_T^{cut})$ and
$\sigma_{SM}^{exact-LO}(p_T^{cut})$ are
the cumulative cross-sections which are
calculated for a given $p_T^{cut}$, using the
point-like $ggh$ approximation and the full LO 1-loop set of diagrams
(i.e., top-quark loops with a finite top-quark mass), respectively.
The loop-induced SM cross-sections were calculated
using the loopSM model of MadGraph5.

We see that the point-like $ggh$ approximation
overestimates the cumulative cross-sections
for exclusive Higgs + jet production, in particular
at large $p_T(h)$, and that the effect is more pronounced in the
Higgs + b-jet case. In particular,
for $p_T^{cut} = 100,~200,~400$ GeV, we find
$r_{ggh} \sim 1,~1.4,~2.9$ for $pp \to h+j$ and
$r_{ggh} \sim 1.3,~1.8,~3.6$ for $pp \to h+j_b$.
Thus, by using the effective point-like $ggh$ vertex we are
overestimating the Higgs + jet cross-sections (which
are dominated by the SM diagrams involving the $ggh$ interaction)
and, therefore, the
corresponding expected number of Higgs + jet events, roughly by a factor
of $r_{ggh}$. On the other hand, as will be shown later,
the statistical significance of the signals
($N_{SD}$ defined in Eq.~\ref{NSD} above) only mildly
depend on the calculation scheme (i.e., on $r_{ggh}$).
We will address these issues in a more quantitative manner below.

\section{Higgs + jet production in the kappa-framework \label{sec3}}

The kappa-framework is defined by multiplying the SM couplings $g_i$ by a
scaling factor $\kappa_i$,
which parameterizes the effects of NP when it has the same Lorentz structure
as the corresponding SM interactions \cite{kappa_framework,1612.00269}. In the case of
$pp \to h+j(j_b)$, the
relevant scaling factors apply to
the effective (1-loop) Higgs-gluon interaction of Eq.~\ref{Leffggh}
and to the light and/or b-quark Yukawa couplings.
In particular, the effective interaction Lagrangian for $pp \to h+j(j_b)$ in the kappa-framework,
takes the form:
\begin{eqnarray}
{\cal L}_{eff}^{h+j} = - \sum_{q=u,d,s,c,b} \kappa_q \frac{m_b}{v} h \bar q q + \kappa_g C_g h G_{\mu \nu}^a
G^{\mu \nu, a} ~, \label{Lkappa}
\end{eqnarray}
where
we have scaled the light-quark Yukawa coupling, $y_q$, with
the SM b-quark Yukawa:
\begin{eqnarray}
\kappa_q \equiv \frac{y_q}{y_b^{SM}} ~,  \label{kappaq}
\end{eqnarray}
and $y_b^{SM}= \sqrt{2}m_b/v$. In particular, $\kappa_{g}=1$,
$\kappa_{b}=1$, $\kappa_{c} \sim 0.3$, $\kappa_{s} \sim {\cal O}(10^{-2})$ and $\kappa_{u,d} \sim {\cal O}(10^{-3})$
are the SM strengths for the corresponding couplings. In what follows,
we will refer to the SM case by $\kappa_{u,d,c,s} = 0$, since the effect of
the small SM values for $\kappa_{u,d,c,s}$ in $pp \to h+j$ are negligible.

\subsection{The case of Higgs + light-jet production \label{subsec31}}

As mentioned earlier,
in the case of $pp \to h+j$, where $j=g,u,d,s,c$ is
a non-flavor tagged light-jet originating from a gluon or
any quark of the 1st and 2nd generations,
the SM tree-level diagrams involving
the light-quarks Yukawa couplings are vanishingly small
(see Eqs.~\ref{dsigb1}-\ref{dsigb3}).
Therefore,
the dominant SM contribution to $\sigma(pp \to h+j)$
arises at 1-loop via the sub-processes
$gg \to gh$, $gq \to q h$, $g \bar q \to \bar q h$ and $q \bar q \to g h$
(the corresponding diagrams are depicted in Fig.~\ref{SM1L}, where the loops are
represented by an effective $ggh$ vertex). In particular,
using the Higgs-gluon effective Lagrangian of Eq.~\ref{Leffggh}, the
corresponding total SM cross-section for $pp \to h+j$ can be written as:
\begin{eqnarray}
\sigma_{SM}^{hj} = C_g^2
\left(\sigma^{gg}_{SM} + \sigma^{gq}_{SM} + \sigma^{g \bar q}_{SM} +
\sigma^{q \bar q}_{SM} \right) ~, \label{sigSM0}
\end{eqnarray}
where $\sigma^{ij}_{SM}$, for $ij=gg,gq,g \bar q,q \bar q$,
can be obtained from the corresponding squared amplitudes given in Eqs.~\ref{dsig1}-\ref{dsig4}.
For example, $\sigma^{gg}_{SM}$ is part of the SM cross-section
coming from $gg \to gh$, which is the dominant sub-process in the SM.

On the other hand, turning on the light-quark $qqh$ Yukawa couplings
and allowing for deviations
also in the Higgs-gluon $ggh$ interaction, within the kappa-framework of
Eq.~\ref{Lkappa}, we obtain the total NP cross-section for $pp \to h+j$:

\begin{eqnarray}
\sigma^{hj} = \kappa_g^2 \sigma_{SM}^{hj} +
\kappa_q^2 \sigma_{qqh}^{hj} ~, \label{sigNP}
\end{eqnarray}
where $\sigma_{SM}^{hj} \simeq \sigma^{hj}(\kappa_g=1,\kappa_q=0)$ is given in Eq.~\ref{sigSM0}
and $\sigma_{qqh}^{hj}=\sigma^{hj}(\kappa_g=0,\kappa_q=1)$ arises from the
the s-channel and t-channel tree-level $gq \to qh$ diagrams, depicted
in Fig.~\ref{SMTL}, where
only the (scaled) light-quarks $qqh$ Yukawa couplings contribute.
The interference term between the diagrams involving the $ggh$ and $qqh$ couplings is proportional
to the light-quark mass and is, therefore, neglected in Eq.~\ref{sigNP}. In particular,
$\sigma^{hj}$ is practically insensitive to the signs of $\kappa_g$ and $\kappa_q$.

Furthermore,
in the $hgg - h \bar q q $ kappa-framework of Eq.~\ref{Lkappa},
the ratio of branching ratios in Eq.~\ref{sig-strength} is given by:
\begin{eqnarray}
 \mu_{h \to ff} &\equiv& \frac{BR(h \to ff)} {BR_{SM}(h \to ff)} \nonumber \\
&=& \frac{1}{1 + \left(\kappa_g^2 -1 \right) BR_{SM}^{gg} +
\kappa_q^2 BR_{SM}^{bb}} ~,
\label{muBR}
\end{eqnarray}
where $BR_{SM}^{gg,bb} = BR_{SM}(h \to gg,bb)$ and
we will assume no NP in the Higgs decay $h \to ff$.
In particular, as mentioned above, we assume that the Higgs decays via
$h \to \gamma \gamma$ with a SM decay rate.

Collecting the expressions from Eqs.~\ref{sig-strength}, \ref{sigNP} and
\ref{muBR}, we obtain the signal strength in the kappa-framework:
\begin{eqnarray}
\mu_{hj}^f = \left( \kappa_g^2 + \kappa_q^2 R^{hj} \right) \cdot \mu_{h \to ff} ~, \label{mufinal}
\end{eqnarray}
where
\begin{eqnarray}
R^{hj} \equiv  \frac{\sigma_{qqh}^{hj}}{\sigma_{SM}^{hj}} ~, \label{RNP}
\end{eqnarray}
is the NP contribution scaled with the SM cross-section and
calculated using cumulative cross-sections, as defined
in Eq.~\ref{dsigpt}, i.e., for a given $p_T^{cut}$
in both numerator and denominator:
$R^{hj} = R^{hj}(p_T^{cut}) =
\sigma_{qqh}^{hj}(p_T^{cut})/\sigma_{SM}^{hj}(p_T^{cut})$.
The ratio $R^{hj}$ contains
all the dependence of $\mu_{hj}^f$
on the Higgs $p_T$
and, as will be further discussed below, is where all the uncertainties
reside, i.e.,
the higher order corrections (K-factor), the theoretical uncertainty of the PDF
due to variations of the renormalization and factorization scales and the
acceptance factors.
\begin{figure}[htb]
\begin{center}
\includegraphics[scale=0.45]{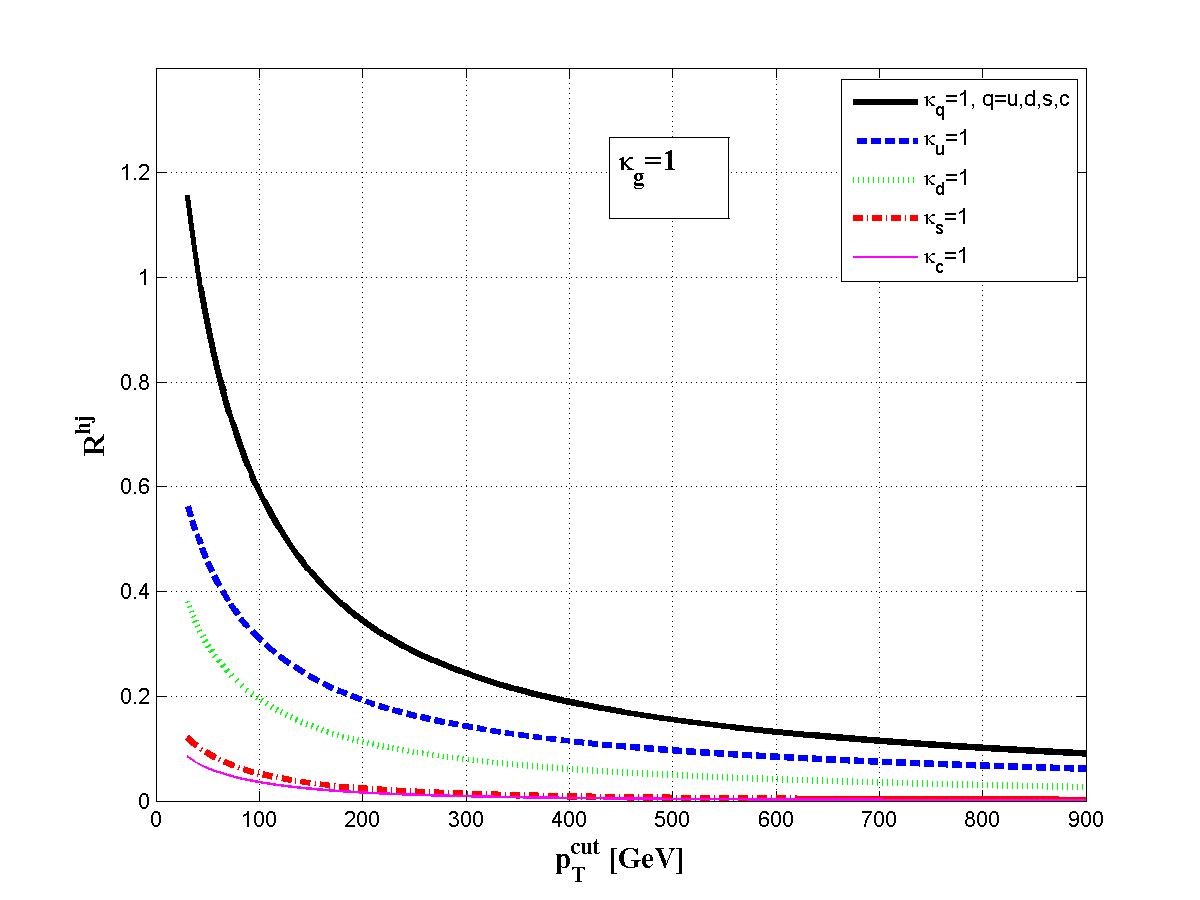}
\includegraphics[scale=0.45]{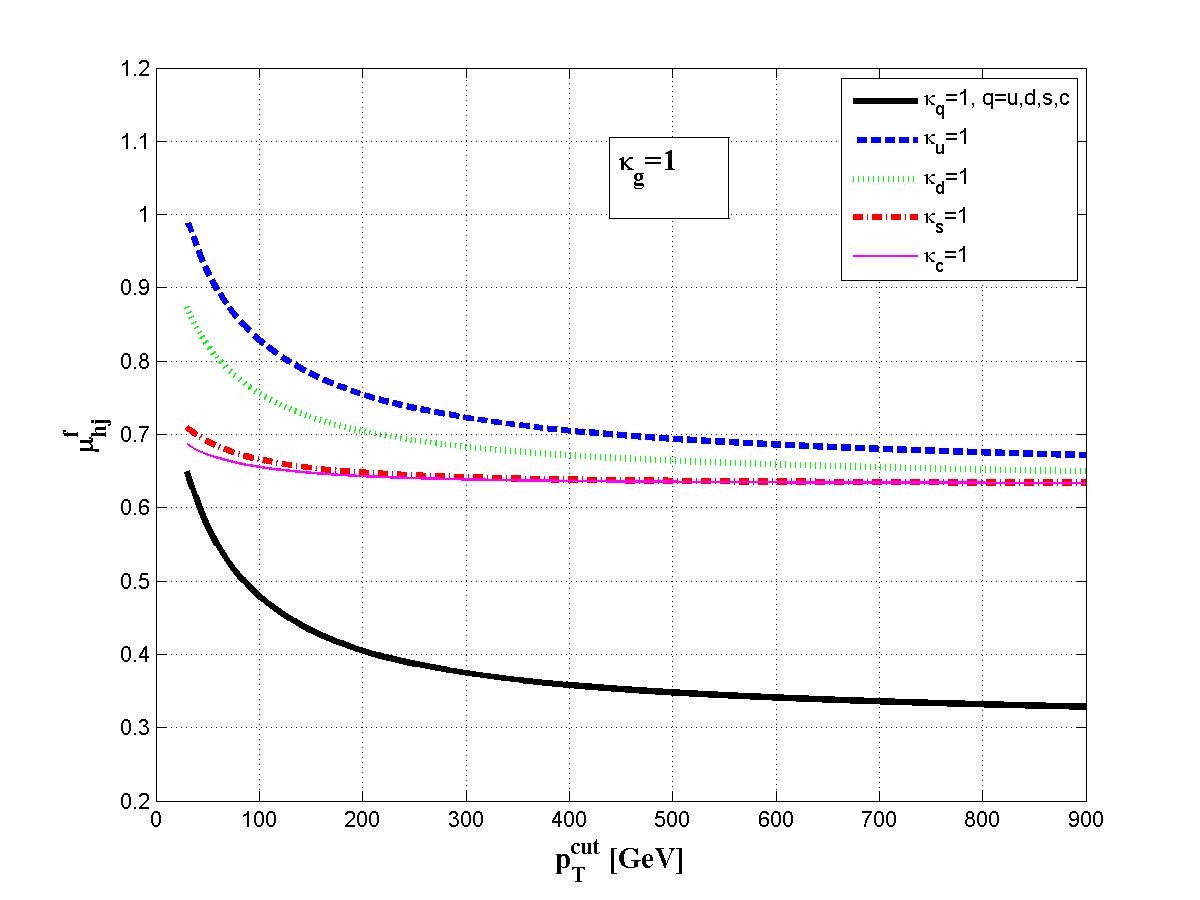}
\end{center}
\caption{The $p_T^{cut}$ dependence of $R^{hj}$ (top) and
$\mu_{hj}^f$ (bottom),
for $\kappa_g=1$ (i.e., assuming no NP in the $ggh$ interaction)
and the cases $\kappa_q=1$ for all $q=u,d,s,c$ (solid line),
$\kappa_u=1$ (dashed line), $\kappa_d=1$ (dotted line),
$\kappa_s=1$ (dot-dashed line) and $\kappa_c=1$ (thin solid line).
}\label{fig2}
\end{figure}

In Fig.~\ref{fig2} we show the dependence of $R^{hj}$ and the signal strength,
$\mu_{hj}^f$, on $p_T^{cut}$, assuming no NP in the $hgg$ interaction ($\kappa_g=1$) and
for the cases in which either a single or all light-quark Yukawa couplings are modified,
i.e., $\kappa_q=1$ for any one of the light-quarks $q=u,d,s,c$ or $\kappa_q=1$ for all $q=u,d,s,c$.
We find that the effect of $\kappa_q \neq 0$ is to change
the softer $p_T(h)$ spectrum, so that $R^{hj}$ drops when
$p_T^{cut}$ is increased.
As a result, the contribution
of $\kappa_q$ to $pp \to h+j$ sharply drops
in the harder $p_T(h)$ region, $p_T(h) \gsim 300$ GeV, where
$R^{hj} \lsim {\cal O}(0.1)$, see Fig.~\ref{fig2}.

Note, however, that
the signal strength approaches an asymptotic value as $p_T^{cut}$ is further
increased, which corresponds
to the region where the $\kappa_q$ dependence of $\mu_{hj}^f$ is dominated
by the decay factor $\mu_{h \to ff}$ in Eq.~\ref{muBR}.
In particular, $\mu_{hj}^f \to 0.6-0.7$ in the single $\kappa_q =1$ case and
$\mu_{hj}^f \to 0.3$ when $\kappa_q =1$ for all light-quarks.
Thus, in the high Higgs $p_T$ regime, the difference between the effects of
a single $\kappa_q \neq 0$
is small, i.e., for either of the quark flavors $q=u,d,c,s$.
The advantage of monitoring the high $p_T(h)$ spectrum, where
$R^{hj}$ is suppressed is, therefore,
reducing the theoretical and experimental uncertainties which, as mentioned above,
reside only in $R^{hj}$.
Indeed, this will be illustrated in Table \ref{tab1} below, where we
show the sensitivity of the signal to the theoretical
uncertainty obtained by scale variations.
\begin{figure*}[htb]
\begin{center}
\includegraphics[scale=0.4]{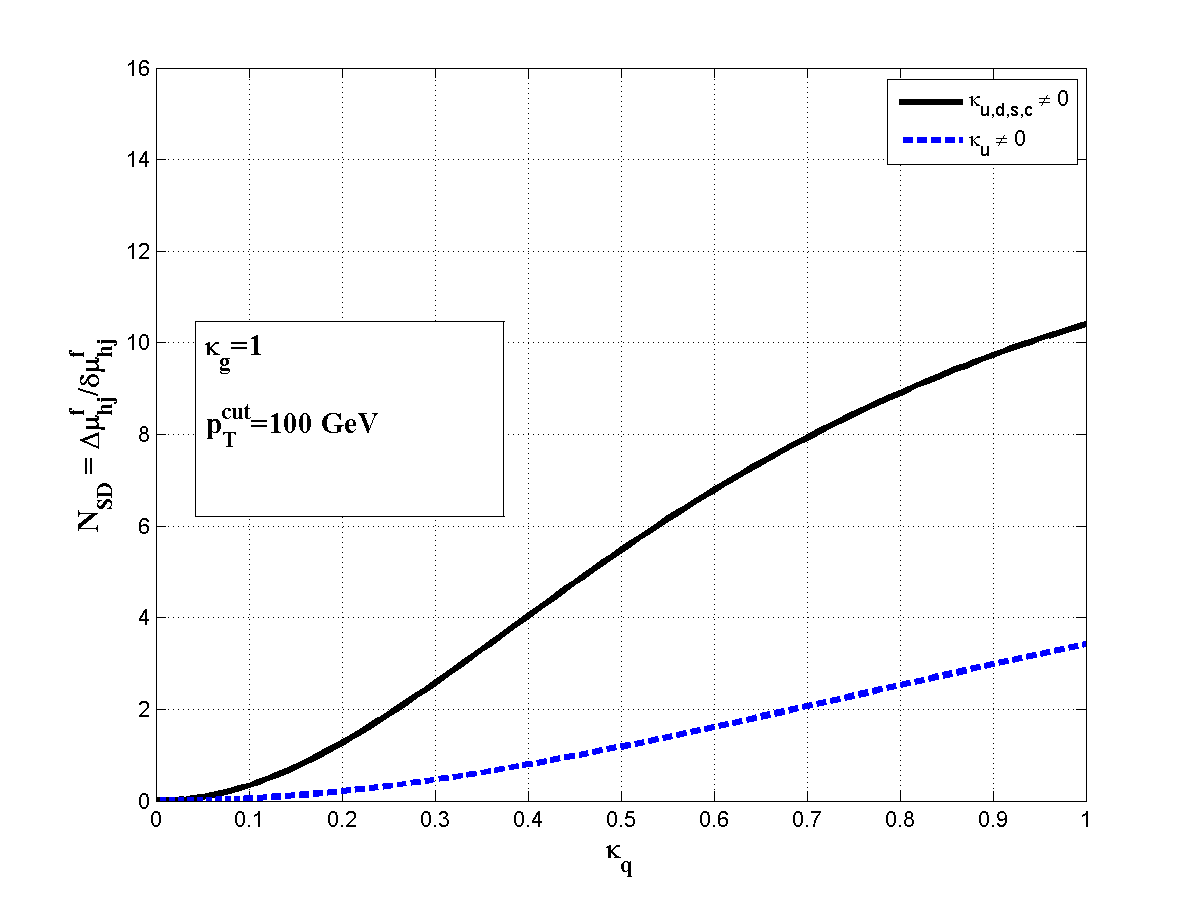}
\includegraphics[scale=0.4]{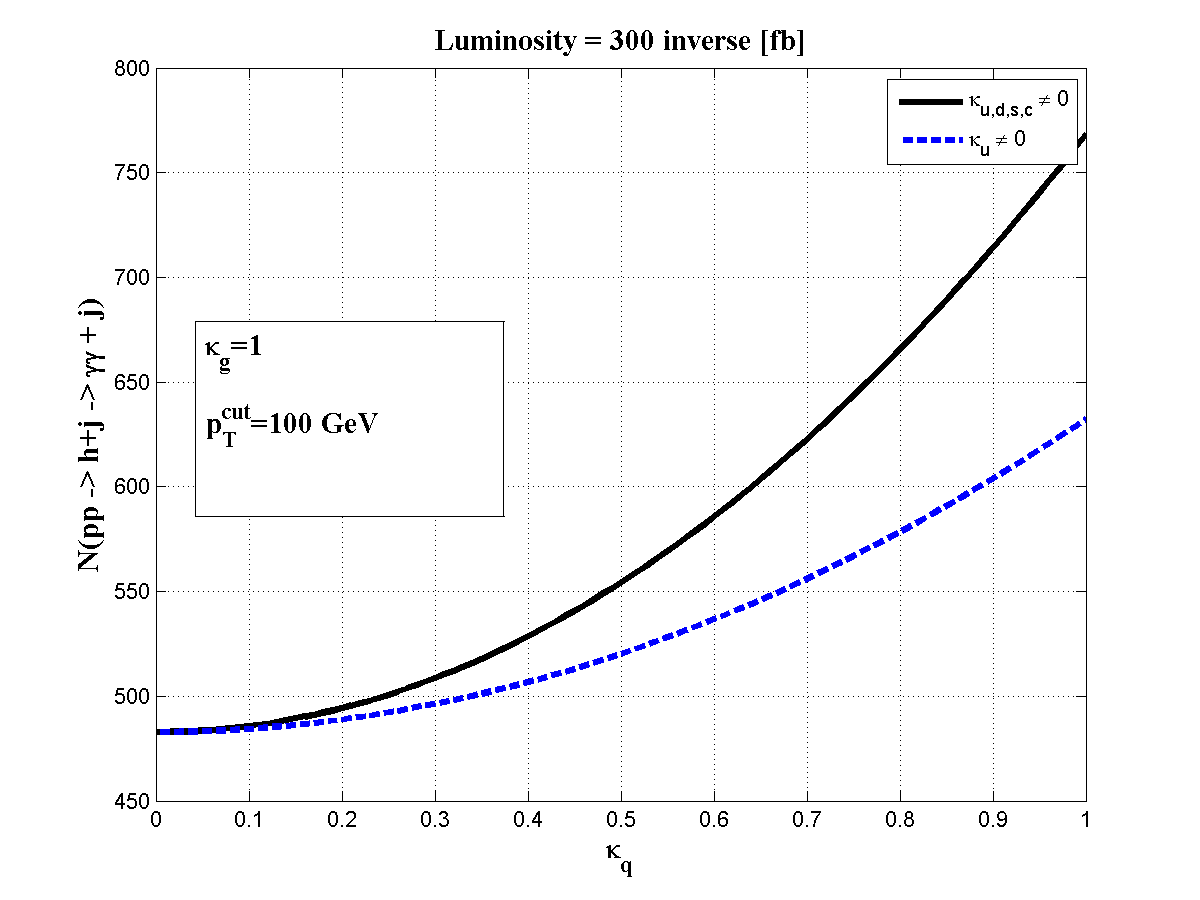}
\includegraphics[scale=0.4]{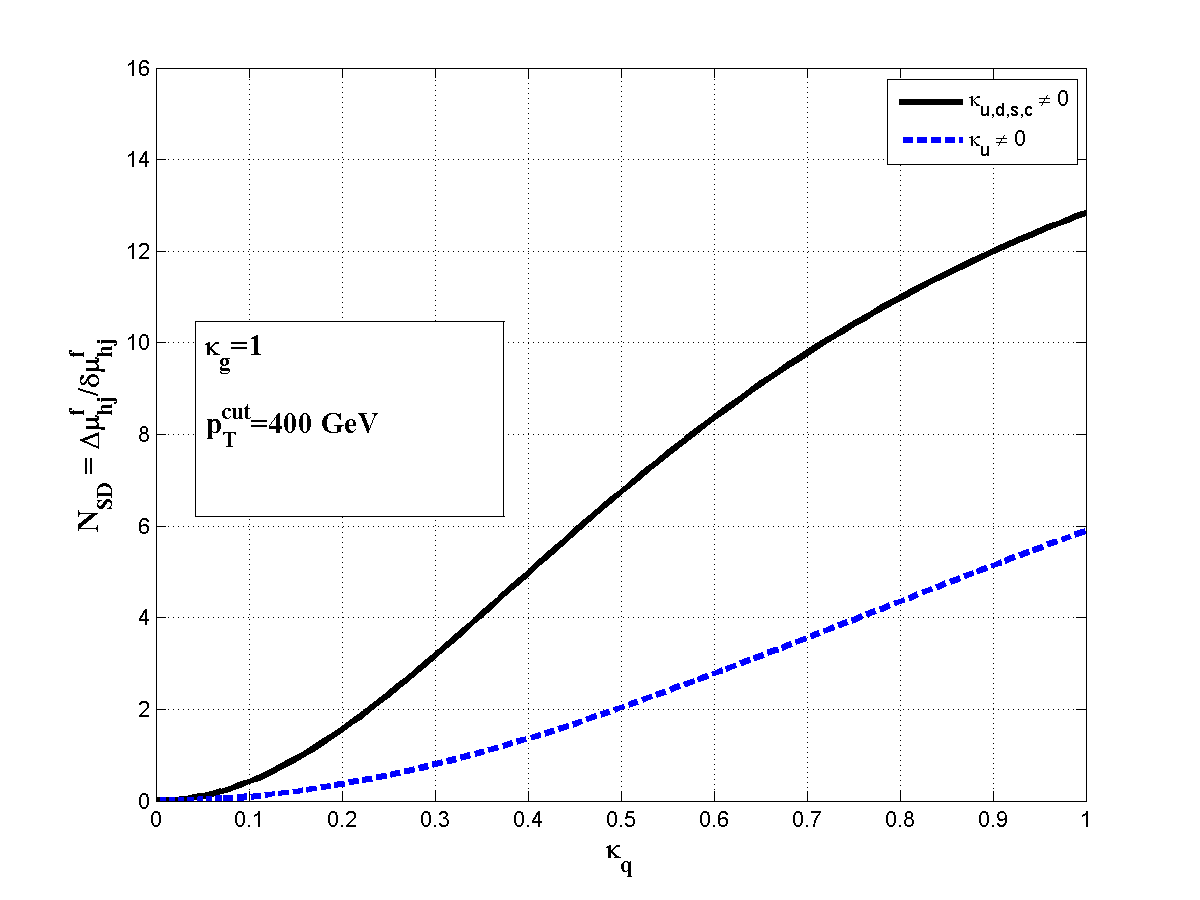}
\includegraphics[scale=0.4]{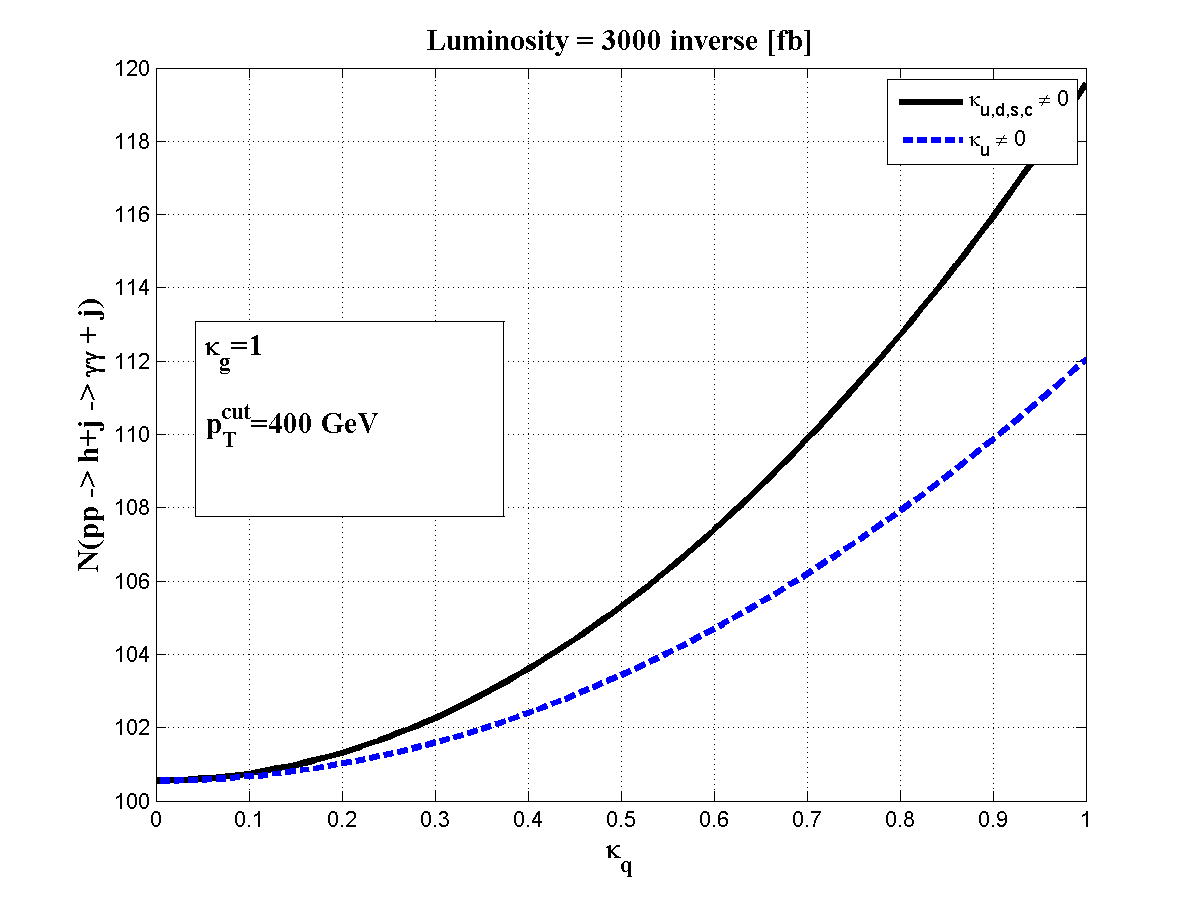}
\end{center}
\caption{The expected statistical significance, $N_{SD}=\Delta \mu_{hj}^f/\delta \mu_{hj}^f$, and the number of
$pp \to h + j \to \gamma \gamma +j$ events,
as a function of
$\kappa_q$, for $\kappa_g=1$ (i.e., assuming no NP in the $hgg$ interaction)
and for $p_T^{cut}=100$ GeV (upper plots) and $p_T^{cut}=400$ GeV (lower plots).
The two cases of a single $\kappa_u \neq 0$ (dashed line) and
$\kappa_q \neq 0$ for all $q=u,d,s,c$ (solid line), are considered.
We assume a 5\% relative error ($\delta\mu_{hj}^f=0.05$) and an acceptance of 50\%
in the event yield, with a luminosity of ${\cal L}=300$ fb$^{-1}$
for the $p_T^{cut}=100$ GeV case and
${\cal L}=3000$ fb$^{-1}$
in the $p_T^{cut}=400$ GeV case.}
\label{fig4}
\end{figure*}
\begin{figure}[htb]
\begin{center}
\includegraphics[scale=0.45]{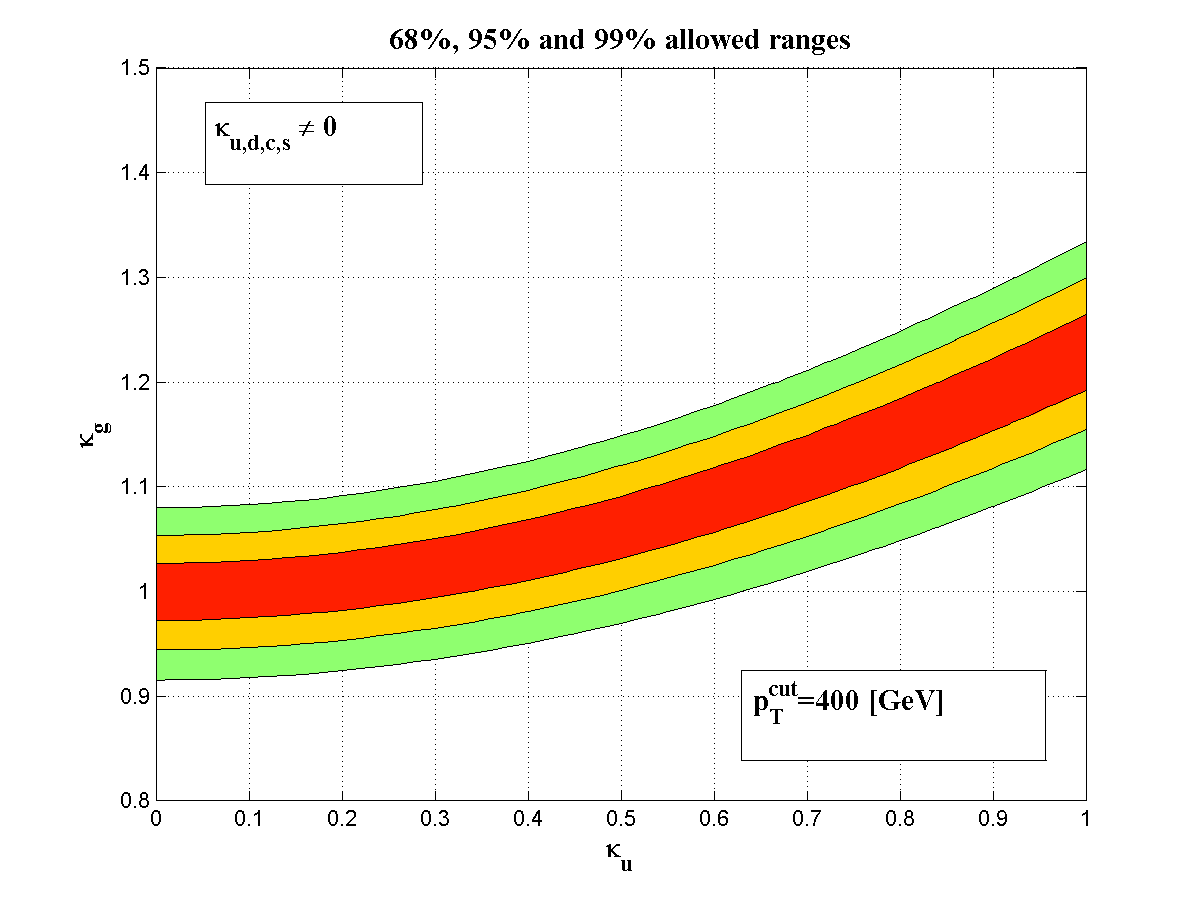}
\includegraphics[scale=0.45]{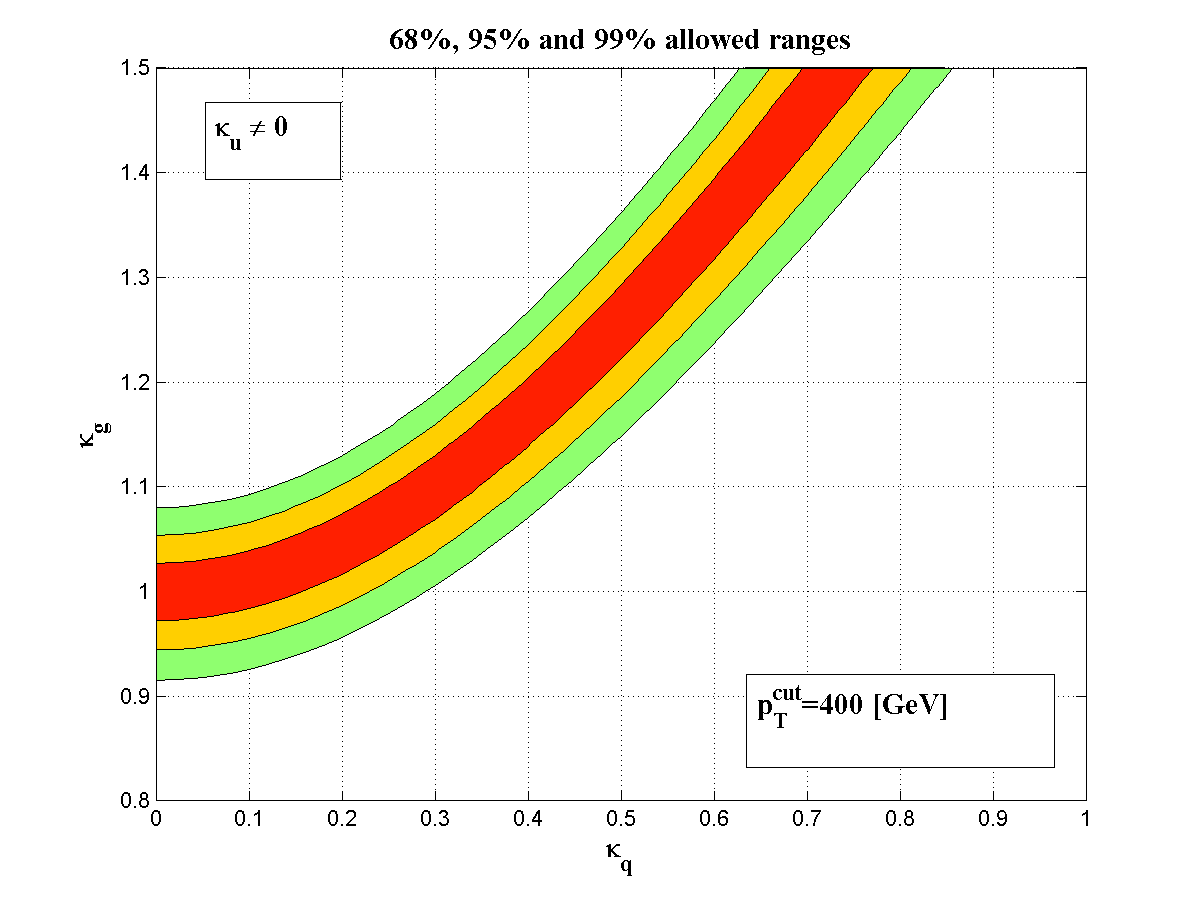}
\end{center}
\caption{The 68\%(red), 95\%(orange) and 99\%(green) CL sensitivity ranges,
corresponding to $\Delta\mu_{hj}^f \equiv \left|\mu_{hj}^f -1 \right| \leq 0.05,~0.1$ and $0.15$, respectively,
with $p_T^{cut}=400$ GeV,
in the $\kappa_u-\kappa_g$ plane for $\kappa_d=\kappa_s=\kappa_c=0$ (top) and
in the
$\kappa_q-\kappa_g$ plane for $\kappa_q=\kappa_u=\kappa_d=\kappa_s=\kappa_c$ (bottom).
Recall that $\kappa_g=1$ and $\kappa_{u,d,c,s} \to 0$ represent the SM
case, see also text.}
\label{fig5}
\end{figure}
    	
In Fig.~\ref{fig4} we plot the expected statistical significance,
$N_{SD}$ defined in Eq.~\ref{NSD}, assuming a 5\% relative error ($\delta\mu_{hj}^f=0.05$),
as a function of $\kappa_q$ for two cases:
(i) $\kappa_q \neq 0$ for all $q=u,d,s,c$ and
(ii) only $\kappa_u \neq 0$.
In both cases we
assume no NP
in the Higgs-gluon coupling ($\kappa_g=1$) and we use
two different $p_T^{cut}$ values $p_T^{cut}=100,~400$ GeV.
We see that, in the single $\kappa_u \neq 0$ case,
there is a $3\sigma$ sensitivity to values of $\kappa_u \gsim 0.6$,
for $\kappa_g =1$ and using $p_T^{cut}=400$ GeV.
In the case where the NP modifies $\kappa_q$ for all $q=u,d,c,s$,
one can expect a deviation of more than $3 \sigma$ for values of $\kappa_q \gsim 0.3$.
We also show in Fig.~\ref{fig4} the corresponding
expected number of $pp \to h + j \to \gamma \gamma +j$ events,
as a function of $\kappa_q$ for cases (i) and (ii) considered above,
with $p_T^{cut}=100$ and 400 GeV and an integrated luminosity of 300 and 3000 fb$^{-1}$,
respectively, assuming a signal acceptance of 50\%. We can see
that around $1000$($100$) $pp \to h + j \to \gamma \gamma +j$ events
with $p_T(h) > 100(400)$ GeV are expected
at the LHC(HL-LHC), i.e.,
with ${\cal L} =300(3000)$ fb$^{-1}$. Thus,
in both cases it should be possible
to probe the NP effects when the Higgs decays via $ h \to \gamma \gamma$.

The signal strength $\mu_{hj}^f$ is more sensitive to NP in
the Higgs-gluon coupling, i.e., to $\kappa_g$.
We find, for example, that
if $\mu_{hj}^f$ is known to
a $5\%(1\sigma)$ accuracy, then a deviation of more than $3\sigma$ is expected
for $\kappa_g \lsim 0.9$
for any value of $\kappa_q$ and for any $p_T^{cut} \lsim 500$ GeV.
This is illustrated in Fig.~\ref{fig5} where we plot
the 68\%, 95\% and 99\% confidence level (CL) allowed ranges in the
$\kappa_q-\kappa_g$ plane, for $p_T^{cut}=400$ GeV and
assuming that the signal strength has been measured to be
$\mu_{hj}^f \sim 1 \pm 0.05(1\sigma)$, i.e., with
a SM central value and to an accuracy of $\delta\mu_{hj}^f=5\%(1\sigma)$.
Here also, we consider both the single $\kappa_u$ case where $\kappa_u \neq 0$ and
$\kappa_d=\kappa_s=\kappa_c=0$ and the case where $\kappa_q \neq 0$ for
all $q=u,d,s,c$. In particular, values of $\{\kappa_q,\kappa_g\}$ outside the
shaded 99\% contour will be excluded at more than
$3\sigma$, if the
signal strength will be measured to lie within $ 0.85 < \mu_{hj}^f < 1.15$.
\begin{table}[htb]
\begin{center}
\begin{tabular}{|c||c|c|c|}
  \multicolumn{4}{|c|}{Statistical significance $N_{SD}=\frac{\Delta\mu_{hj}^f}{\delta\mu_{hj}^f}$} \\
\hline \hline
  \multicolumn{4}{c}{$\kappa_u \neq 0$, $\kappa_d=\kappa_s=\kappa_c=0$} \\
\hline
 & $\kappa_u=0$ & $\kappa_u=0.25$ & $\kappa_u=0.5$  \\
\hline \hline
 $\kappa_g=0.8 $ & $6.79$  & $7.12^{-0.03}_{+0.03}$ & $8.0^{-0.11}_{+0.10}$ \\
\hline
$\kappa_g=0.9 $  & $3.53$ & $3.97^{-0.03}_{+0.03}$ & $5.14^{-0.11}_{+0.10}$ \\
\hline
$\kappa_g=1.0 $ & $0$ & $0.56^{-0.03}_{+0.03}$ & $2.03^{-0.11}_{+0.10}$ \\
\hline
$\kappa_g=1.1 $ & $3.78$ & $3.09^{+0.03}_{-0.03}$ & $1.30^{+0.11}_{-0.10}$ \\
\hline
$\kappa_g=1.2 $  & $7.75$ & $6.95^{+0.03}_{-0.03}$ & $4.84^{+0.11}_{-0.10}$ \\
\hline
\end{tabular}
\medskip
\begin{tabular}{|c||c|c|c|}
  \multicolumn{4}{c}{$\kappa_q \neq 0$ for all $q=u,d,s,c$} \\
\hline
 & $\kappa_q=0$ & $\kappa_q=0.25$ & $\kappa_q=0.5$  \\
\hline \hline
 $\kappa_g=0.8 $ & $6.79$  & $8.30^{-0.05}_{+0.04}$ & $11.13^{-0.13}_{+0.12}$ \\
\hline
$\kappa_g=0.9 $  & $3.53$ & $5.43^{-0.05}_{+0.04}$ & $9.03^{-0.13}_{+0.12}$ \\
\hline
$\kappa_g=1.0 $ & $0$ & $2.32^{-0.05}_{+0.04}$ & $6.74^{-0.13}_{+0.12}$ \\
\hline
$\kappa_g=1.1 $ & $3.78$ & $1.01^{+0.05}_{-0.04}$ & $4.26^{-0.13}_{+0.12}$ \\
\hline
$\kappa_g=1.2 $  & $7.75$ & $4.55^{+0.04}_{-0.04}$ & $1.61^{-0.13}_{+0.11}$ \\
\hline
\end{tabular}
\caption{The statistical significance of the NP signal for $pp \to h+j$,
$N_{SD}=\Delta\mu_{hj}^f/\delta\mu_{hj}^f$,
assuming a 5\% error ($\delta\mu_{hj}^f=0.05(1\sigma)$),
for $p_T^{cut}=400$ GeV and for values of the scaled
couplings $\kappa_q=0,0.25,0.5$ and $\kappa_g=0.8,0.9,1,1.1,1.2$,
in the single $\kappa_u$ case assuming $\kappa_d=\kappa_s=\kappa_c=0$ (top table)
and in the case where $\kappa_q \neq 0$ for all $q=u,d,s,c$ (bottom table).
The errors indicate the theoretical dependence on the PDF scale, where the superscript(subscript) corresponds
to twice(half) the nominal scale $\mu_F = \mu_R = \mu_T \equiv \sum_i \sqrt{m_i^2+p_{T}^{2}(i)}$,
see also text.}
\label{tab1}
\end{center}
\end{table}

In Table \ref{tab1} we list the statistical significance of the NP signal, $N_{SD}=\Delta\mu_{h_b}^f/\delta\mu_{h_b}^f$,
as defined in Eq.~\ref{NSD}, again
assuming 5\% error ($\delta\mu_{hj}^f=0.05(1\sigma)$),
for $p_T^{cut}=400$ GeV and some discrete values of the scaled
couplings: $\kappa_q=0,0.25,0.5$ and $\kappa_g=0.8,0.9,1,1.1,1.2$.
Here also, results are given
in the single $\kappa_u$ case and in the case where $\kappa_q \neq 0$ for all $q=u,d,s,c$. We
include the theoretical
uncertainty obtained by scale variations and (although of little use)
write $N_{SD}$ up to the 2nd digit
to illustrate the small uncertainty due the scale variation.
Note that for $\kappa_q=0$ there is no dependence on the scale of the PDF since,
in this case, it is cancelled in the ratio of cross-sections as defined
in the signal strength $\mu_{hj}^f$.
We see that indeed the effect of the variation of scale with which the PDF is evaluated
is negligible
due to the smallness of $R^{hj}$ in the harder $p_T$ spectrum, in particular
for $p_T^{cut}=400$ GeV used in the Table \ref{tab1} (see also discussion above).

All the results presented in this section were obtained using the
effective point-like $ggh$ approximation, which as was shown in section \ref{sec2}
(see Fig.~\ref{compsigma}), overestimates the contribution of the SM-like diagrams
involving the 1-loop $ggh$ vertex when compared to the 1-loop induced (top-mass dependent)
terms.
In particular, this approximation effects
the denominator of the scaled NP ratio $R^{hj}$ in Eq.~\ref{RNP},
i.e., the SM cumulative cross-section $\sigma_{SM}^{hj}(p_T^{cut})$.
To give a feeling for the sensitivity of our results
to the underlying calculation setup
at the high $p_T(h)$ regime, where the point-like $ggh$ approximation
shows ${\cal O}(1)$ deviations, we recalculate the statistical significance
$N_{SD}$ in Table \ref{tab1} using the top-mass dependent 1-loop result for
$\sigma_{SM}^{hj}(p_T^{cut})$ in Eq.~\ref{RNP}. In this case,
the scaled NP ratio $R^{hj}$ changes to:
\begin{eqnarray}
R^{hj} \to \tilde R^{hj} = r_{ggh} R^{hj} ~, \label{RNPtilda}
\end{eqnarray}
where $r_{ggh}$, which is defined in Eq.~\ref{rggh}, is the ratio
between the point-like and the LO loop-induced (mass dependent) SM cross-sections.
Thus, replacing $R^{hj} \to \tilde R^{hj}$ in the expression of Eq.~\ref{mufinal} for the
signal strength and using the definition for $N_{SD}$ in Eq.~\ref{NSD},
we obtained the statistical significance in the exact 1-loop case:
\begin{eqnarray}
\tilde N_{SD} = r_{ggh} N_{SD} - \frac{\left(r_{ggh}-1 \right)
\left( \tilde\kappa_g^2 \mu_{h \to ff} - 1 \right) }{\delta\mu_{hj}^f} ~, \label{NSDtilda}
\end{eqnarray}
where $\mu_{h \to ff}$ is the scaled Higgs decay branching ratio defined in
Eq.~\ref{muBR}
and $\delta\mu_{hj}^f$ is the assumed $1 \sigma$ error (see
Eq.~\ref{NSD}).
Note that in Eq.~\ref{NSDtilda} above we have denoted the the modified $ggh$ interaction by $\tilde\kappa_g$ (rather than $\kappa_g$),
since caution has to be taken when interpreting
the NP associated with the $ggh$ vertex in the exact
top-quark 1-loop case.
In particular,
in the calculation of $\sigma^{hj} = \sigma(pp \to h+j)$ using the effective point-like $ggh$ interaction,
$\kappa_g$ simply corresponds to the scaling of the effective $ggh$ SM vertex (see Eq.~\ref{Lkappa})
and, therefore, to the ratio
$\kappa_g = \sqrt{\sigma^{hj}/\sigma_{SM}^{hj}}$ (see Eq.~\ref{sigNP} for $\kappa_q=0$).
On the other hand, in the exact LO (1-loop) calculation,
the diagrams in Fig.~\ref{SM1L} involving
NP in the effective $ggh$ interaction should be added at the amplitude
level to the SM 1-loop diagrams (i.e., with the top-quark loops). Thus, in this case,
generic NP effects associated with the $ggh$ vertex in $\sigma^{hj}$ can be parameterized
as follows \cite{1312.3317_SMEFT,1405.7651}:
\begin{eqnarray}
\sigma^{hj}(\kappa_q=0) = \left( \kappa_t^2 + A \kappa_t \kappa_g + B \kappa_g^2 \right) \sigma_{SM}^{hj}
\equiv \tilde\kappa_g^2 \sigma_{SM}^{hj} ~,\label{tkappag}
\end{eqnarray}
where $\kappa_t \equiv y_t/y_t^{SM}$ is the $tth$ coupling modifier
(which parameterizes potential NP in the SM top-quark loop diagrams) and $A,B$ are phase-space
coefficients which depend on the lower Higgs $p_T$ cut ($p_T^{cut}$), see \cite{1312.3317_SMEFT}.
Thus, when considering NP in $pp \to h+j$ within the exact 1-loop calculation,
the $ggh$ coupling modifier $\tilde\kappa_g$ (defined in
Eq.~\ref{tkappag}), which appears
in Eq.~\ref{NSDtilda} and in Table \ref{tab1comp} should be interpreted as the overall NP effect in the $ggh$ interaction,
where $\tilde\kappa_g = \kappa_t$ corresponds to NP which modifies
only the $tth$ Yukawa coupling while
$\tilde\kappa_g = \sqrt{1 + A \kappa_g + B\kappa_g^2 }$ applies to the
case where $\kappa_t=1$ and the NP arises from some other
underlying heavy physics which is integrated out and generates the $ggh$ effective interaction of Eq.~\ref{Lkappa}.
This interpretation of $\tilde\kappa_g$ applies to all instances below where we discuss our results for the NP effect
in $pp \to h+j(j_b)$ within the exact LO 1-loop case.

In Table \ref{tab1comp} we list the statistical significance $\tilde N_{SD}$
calculated according to Eq.~\ref{NSDtilda},
again taking a 5\% error $\delta\mu_{hj}^f=0.05(1\sigma)$,
$p_T^{cut}=400$ GeV and the same values of the scaled
couplings as in Table \ref{tab1comp},
where here only the single $\kappa_u \neq 0$ case is considered.
We also list in Table \ref{tab1comp} the values of $N_{SD}$
of Table \ref{tab1} (i.e., corresponding to the case
where the diagrams involving the $ggh$ interaction are calculated
with the point-like $ggh$ interaction).
We see that the expected significance of the NP signal in $pp \to h+j$
is mildly sensitive to the calculation scheme. In particular,
variations at the level of $0.1\sigma -1 \sigma$ are observed in $N_{SD}$
depending on the values of the scaled NP couplings $\kappa_q$ and $\kappa_g$
(note that $\tilde N_{SD} = N_{SD}$ for $\kappa_u=0$),
so that the point-like $ggh$ approximation is indeed useful for estimating
the NP effect in $pp \to h+j$ even for events with $p_T(h) > 400$ GeV.

\begin{table}[htb]
\begin{center}
\begin{tabular}{|c||c|c|c|}
  \multicolumn{4}{|c|}{$\tilde N_{SD}$ $\left( N_{SD} \right) $} \\
\hline \hline
  \multicolumn{4}{c}{$\kappa_u \neq 0$, $\kappa_d=\kappa_s=\kappa_c=0$} \\
\hline
 & $\kappa_u=0$ & $\kappa_u=0.25$ & $\kappa_u=0.5$  \\
\hline \hline
 $\tilde\kappa_g=0.8 $ & $6.8(6.8)$  & $6.8(7.1)$ & $7.0(8.0)$ \\
\hline
$\tilde\kappa_g=0.9 $  & $3.5(3.5)$ & $3.7(4.0)$ & $4.1(5.1)$ \\
\hline
$\tilde\kappa_g=1.0 $ & $0(0)$ & $0.3(0.6)$ & $1.0(2.0)$ \\
\hline
$\tilde\kappa_g=1.1 $ & $3.8(3.8)$ & $3.4(3.1)$ & $2.3(1.3)$ \\
\hline
$\tilde\kappa_g=1.2 $  & $7.8(7.8)$ & $7.2(7.0)$ & $5.8(4.8)$ \\
\hline
\end{tabular}
\caption{The statistical significance of the NP signal for $pp \to h+j$,
$\tilde N_{SD}$, corresponding to the case where
the SM cross-section is calculated exactly (mass dependent) at 1-loop (LO) and given in Eq.~\ref{NSDtilda}.
As in Table \ref{tab1}, results are shown for 5\% error ($\delta\mu_{hj}^f=0.05(1\sigma)$),
$p_T^{cut}=400$ GeV and for values of the scaled
couplings $\kappa_u=0,0.25,0.5$ and $\tilde\kappa_g=0.8,0.9,1,1.1,1.2$,
in the single $\kappa_u \neq 0$ case assuming $\kappa_d=\kappa_s=\kappa_c=0$.
We also list in parenthesis the corresponding values of the statistical significance
$N_{SD}$ for the case where the SM cross-section is calculated with the point-like
$ggh$ approximation.
See also text.}
\label{tab1comp}
\end{center}
\end{table}

\subsection{The case of Higgs + b-jet production \label{subsec32}}

We next turn to Higgs + b-jet production,
which can be described in the five flavor scheme (5FS),
where one treats the b-quark as a massless parton while keeping its
Yukawa coupling finite \cite{5FS_hbjet}, see also
\cite{1010.2977_hbjet4LO,1007.5411_hbjet3_NLO}.
In particular, the
LO contribution to $pp \to h + j_b$ arises at tree-level by the same diagrams that
drive the subprocess $qg \to hq$ (and the charged conjugate one $g \bar b \to \bar b h$),
shown in Fig.~\ref{SMTL} with $q=b$. The cross-section
for these diagrams is proportional to the $bbh$ Yukawa coupling (squared)
and can be obtained
from the corresponding squared amplitudes
which are given in Eqs.~\ref{dsigb1}-\ref{dsigb3}.
The 1-loop contribution to $gb \to b h$, which, in the infinite
top-quark mass limit, can be described by the effective $ggh$ vertex
(see Fig.~\ref{SM1L}), is given in Eqs.~\ref{dsig2}-\ref{dsig4}.
It is comparable to the LO tree-level one at low $p_T(h) \lsim 100$ GeV, while it
dominates at the higher $p_T(h)$ spectrum (see below).$^{[2]}$\footnotetext[2]{Note that
the Higgs + light-jet
processes (in particular, the dominant gluon-fusion
process $gg \to hg$) may ''contaminate" the Higgs + b-jet signal, when the
light jet is mistagged as a b-jet. The probability for that is, however, expected
to be at the sub-percent level for a b-tagging efficiency
of $\epsilon_b \sim 60-70\%$
and is, therefore, neglected.}

Let us denote the corresponding tree-level and 1-loop cumulative cross-sections
(following Eq.~\ref{dsigpt}) for $pp \to h + j_b$ as
$\sigma_{bbh}^{hj_b} \equiv \sigma_{bbh}^{hj_b}(p_T^{cut})$
and $\sigma_{ggh}^{hj_b} \equiv \sigma_{ggh}^{hj_b}(p_T^{cut})$, respectively.
Thus,
in the kappa-framework where $\kappa_b$ and $\kappa_g$ are the only NP scaled
couplings,
the total Higgs + b-jet cross-section is (again
there is negligible interference between the diagrams
involving the $bbh$ and $ggh$ interactions):
\begin{eqnarray}
\sigma^{hj_b} = \kappa_g^2 \sigma_{ggh}^{hj_b} + \kappa_b^2 \sigma_{bbh}^{hj_b} ~, \label{sighb}
\end{eqnarray}
so that the SM cross-section is obtained for $\kappa_g=\kappa_b=1$, i.e.,
$\sigma_{SM}^{hj_b} = \sigma_{ggh}^{hj_b} + \sigma_{bbh}^{hj_b}$.

The signal strength for $pp\to h + j_b \to ff +j_b$ is then given by:
%
\begin{eqnarray}
\mu_{hj_b}^f &=& \frac{{\cal N}(pp \to h+j_b \to ff+j_b)} {{\cal N}_{SM}(pp \to h+j_b \to ff+j_b)}
\nonumber \\
&\simeq&
\left( \frac{\kappa_g^2}{1+R^{hj_b}} + \frac{\kappa_b^2}{1+\left(R^{hj_b}\right)^{-1}} \right)
\cdot \mu_{h \to ff}^b ~, \label{sig-strength-b}
\end{eqnarray}
%
where
\begin{eqnarray}
R^{hj_b} \equiv \frac{\sigma_{bbh}^{hj_b}}{\sigma_{ggh}^{hj_b}} ~, \label{Rbg}
\end{eqnarray}
and
\begin{eqnarray}
 \mu_{h \to ff}^b &\equiv& \frac{BR(h \to ff)} {BR_{SM}(h \to ff)}  \nonumber \\
&= & \frac{1}{1 + \left(\kappa_g^2 -1 \right) BR_{SM}^{gg} +
\left(\kappa_b^2 -1 \right) BR_{SM}^{bb}} ~.
\label{muBRb}
\end{eqnarray}

Once again, all the uncertainties associated with the measurement of $\mu_{hj_b}^f$
reside in the ratio of cross-sections $R^{hj_b}$ and
in the limit $R^{hj_b} \ll 1$, we get
an expression for $\mu_{h j_b}^f$ which is similar to
the one obtained for the Higgs + light-jet case in
Eq.~\ref{mufinal}, with the replacement
$\kappa_q \to \kappa_b$:
\begin{eqnarray}
\mu_{hj_b}^f (R^{hj_b} \ll 1) \simeq \left( \kappa_g^2 + \kappa_b^2 R^{hj_b} \right)
\cdot \mu_{h \to ff}^b ~. \label{mufinalb}
\end{eqnarray}

In particular, we find that, as in the Higgs + light-jet case, the $\kappa_b$ term is
important for softer $p_T(h)$  for which $R^{hj_b} \sim {\cal O}(1)$, while
the $\kappa_g$ contribution
is dominant at the harder $p_T(h)$ regime, where $R^{hj_b} \ll 1$.
For example, we obtain $R^{hj_b} \sim 2$ for
$p_T^{cut} \sim 35$ GeV, dropping to $R^{hj_b} \sim 1$ at
$p_T^{cut} \sim 90$ GeV (i.e., the point where $\sigma_{bbh}^{hj_b}$ is comparable to
$\sigma_{ggh}^{hj_b}$), then
to $R^{hj_b} \sim 0.4$ for $p_T^{cut} \sim 200$ GeV and further to
$R^{hj_b} \sim 0.15$ at $p_T^{cut} \sim 400$ GeV.
Thus, here also, the effects of higher-order corrections and variation of scales,
as well as the acceptance factors,
become insignificant when the signal strength is evaluated for a
high $p_T^{cut} \sim 400$ GeV, for which $R^{hj_b} \sim {\cal O}(0.1)$.

\begin{figure}[htb]
\begin{center}
\includegraphics[scale=0.45]{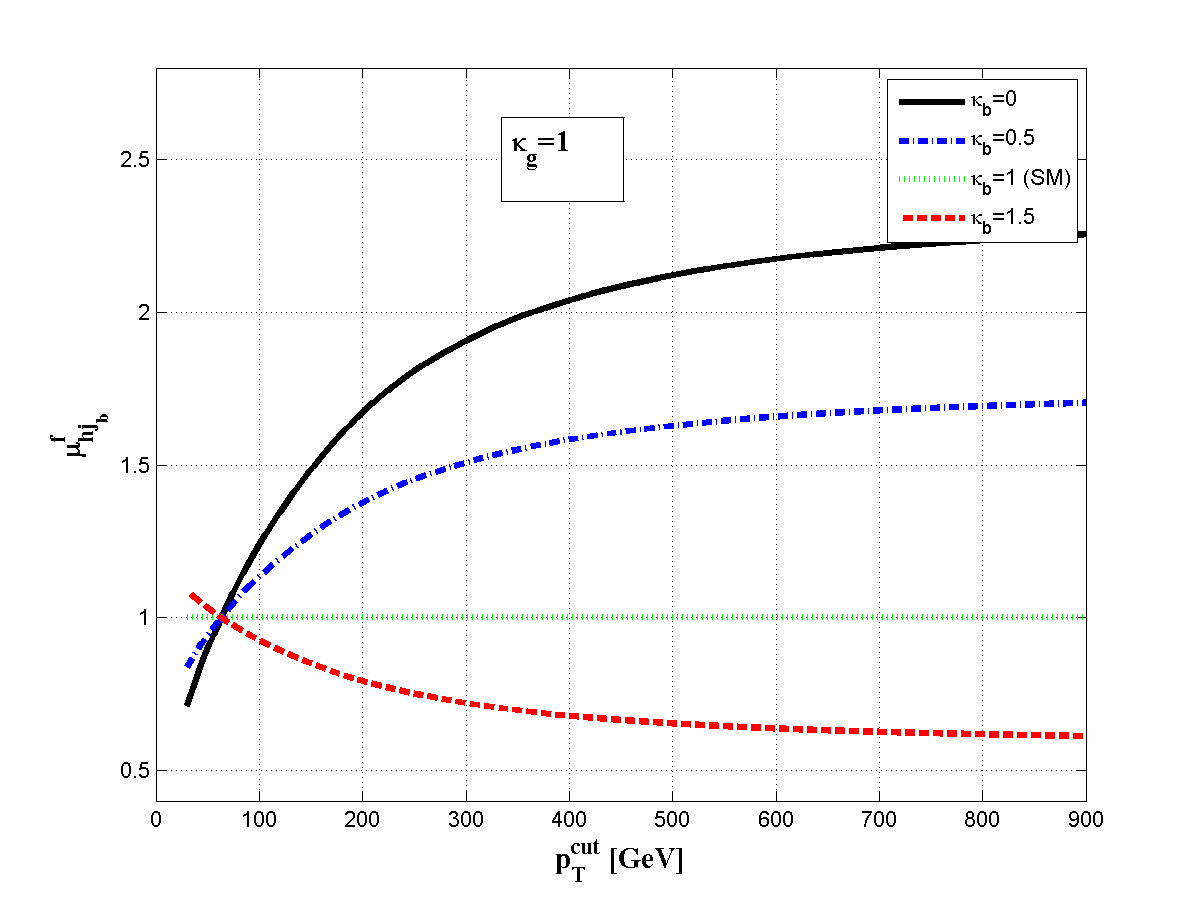}
\includegraphics[scale=0.45]{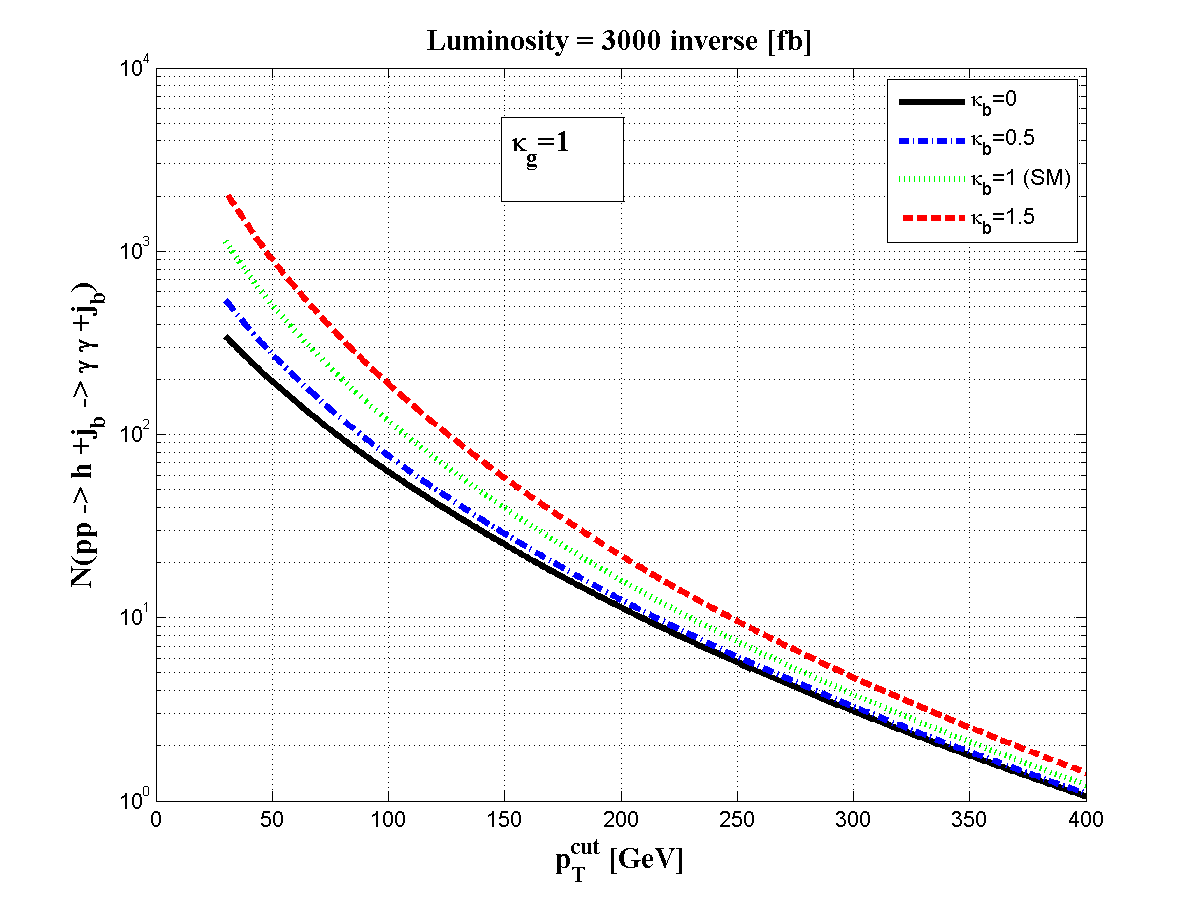}
\end{center}
\caption{The $p_T^{cut}$ dependence of
$\mu_{hj_b}^f$ (top) and of the expected number of events
$N(pp \to h + j_b \to \gamma \gamma +j_b) =
{\cal L} \cdot \sigma(pp \to h + j_b ) \cdot BR(h \to \gamma \gamma) \cdot {\cal A} \cdot \epsilon_b$ (bottom)
at the HL-LHC with ${\cal L}=3000~{\rm fb}^{-1}$, an acceptance of ${\cal A} = 0.5$ and a
b-jet tagging efficiency of $\epsilon_b=0.7$.
The curves are
for $\kappa_g=1$ (i.e., assuming no NP in the $ggh$ interaction)
and for $\kappa_b=0,0.5,1,1.5$ ($\kappa_b=1$ corresponds to the SM case where $\mu_{hj_b}^f=1$).}
\label{fighjb1}
\end{figure}

In Fig.~\ref{fighjb1} we show the dependence of the signal strength
$\mu_{hj_b}^f$ on $p_T^{cut}$, assuming no NP in the Higgs-gluon $ggh$ interaction ($\kappa_g=1$)
and for values of $\kappa_b$ within $0 < \kappa_b < 1.5$,
which are consistent with the current measurements of the
125 GeV Higgs production and decay processes \cite{1606.02266}.
We see that, once again, the signal strength approaches an asymptotic value
(for a given $\kappa_b$ value) as $p_T^{cut}$ is
increased, which is
where the $\kappa_g$ term dominates and the
$\kappa_b$ dependence arises mostly from the decay factor
$\mu_{h \to ff}^b$ in Eq.~\ref{muBRb}.
\begin{figure}[htb]
\begin{center}
\includegraphics[scale=0.45]{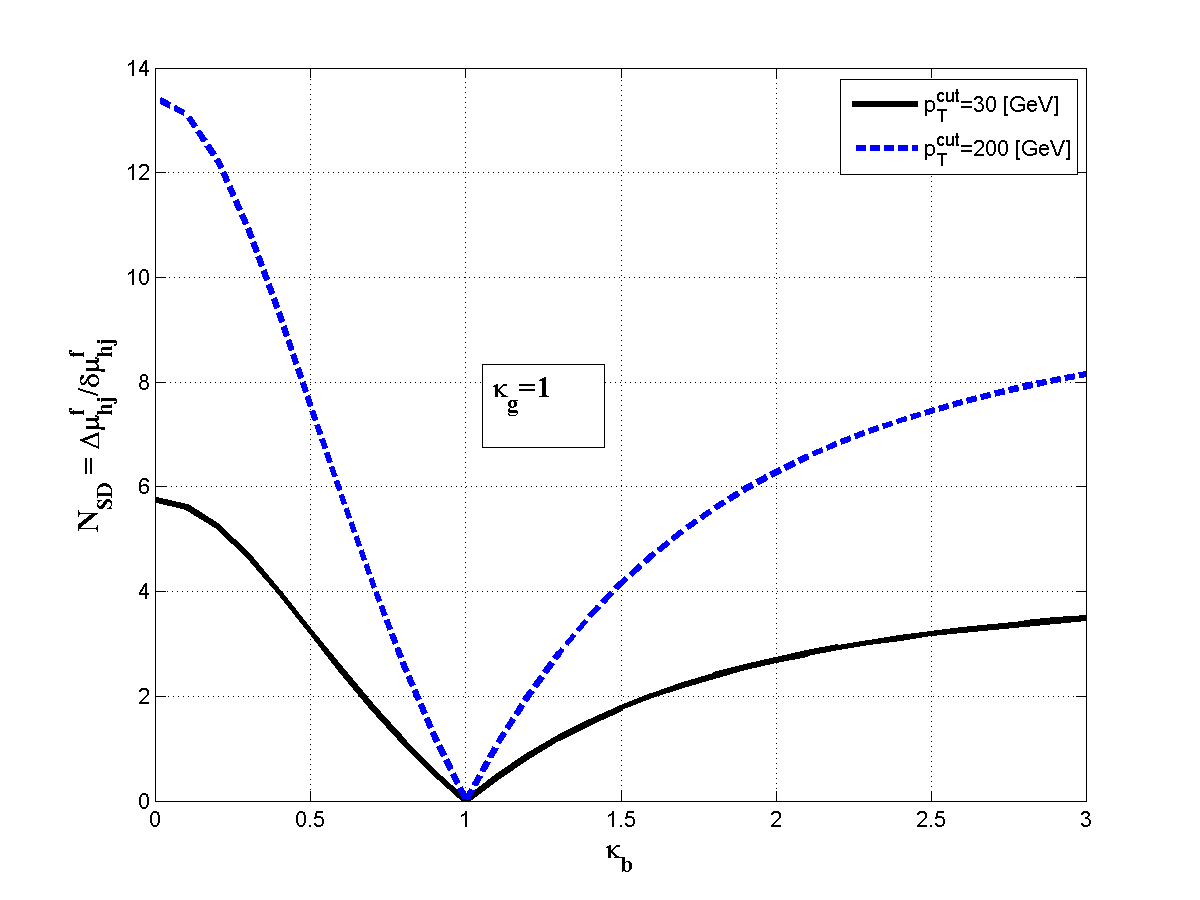}
\end{center}
\caption{The statistical significance, $N_{SD}=\Delta\mu_{hj_b}^f/\delta\mu_{hj_b}^f$,
as a function of $\kappa_b$ for $p_T^{cut}=30$ GeV (solid line) and
 $p_T^{cut}=200$ GeV (dashed line),
assuming $\kappa_g=1$ and a
$5\%(1 \sigma)$ error $\delta\mu_{hj_b}^f = 0.05$. See also text.}
\label{fighjb2}
\end{figure}

We also show in Fig.~\ref{fighjb1}
the expected number of $pp \to h +j_b \to \gamma \gamma + j_b$ events, $N(pp \to h + j_b \to \gamma \gamma + j_b) =
{\cal L} \cdot \sigma(pp \to h + j_b ) \cdot BR(h \to \gamma \gamma) \cdot {\cal A} \cdot \epsilon_b$,
as a function of $p_T^{cut}$ at the HL-LHC with
${\cal L}=3000$ fb$^{-1}$, an acceptance of
${\cal A} = 0.5$ and a
b-jet tagging efficiency of 70\%, i.e., $\epsilon_b=0.7$.
We see that, under these conditions
and for the values of $\kappa_g$ and $\kappa_b$ considered,
a $p_T^{cut} \lsim 100$ GeV is required
to ensure ${\cal O}(100)$
$pp \to h +j_b \to \gamma \gamma + j_b$ events. In particular,
${\cal N}(pp \to h +j_b \to \gamma \gamma + j_b) \sim {\cal O}(1000)$
for $p_T^{cut} \sim 30$ GeV  and
${\cal N}(pp \to h +j_b \to \gamma \gamma + j_b) \sim {\cal O}(10)$
for $p_T^{cut} \sim 200$ GeV, respectively.
\begin{figure}[htb]
\begin{center}
\includegraphics[scale=0.45]{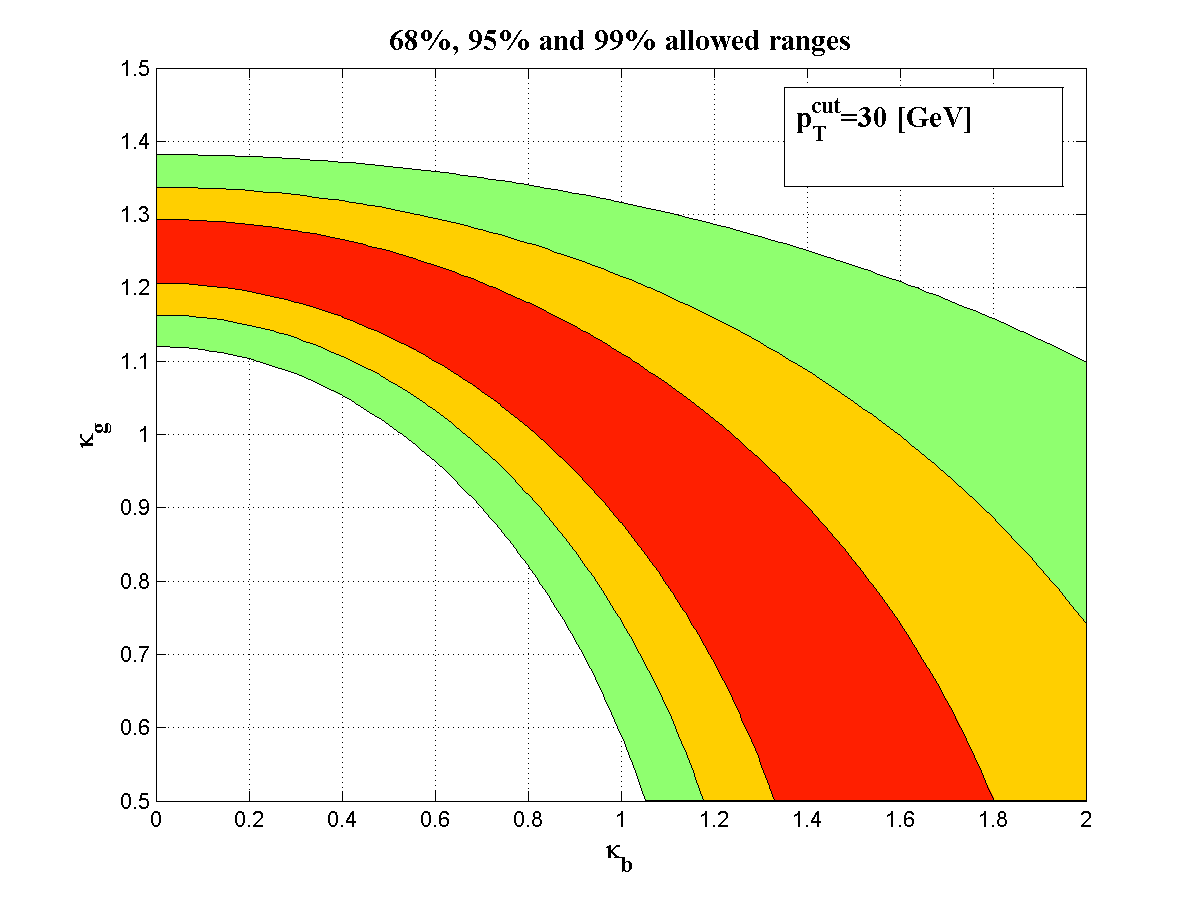}
\includegraphics[scale=0.45]{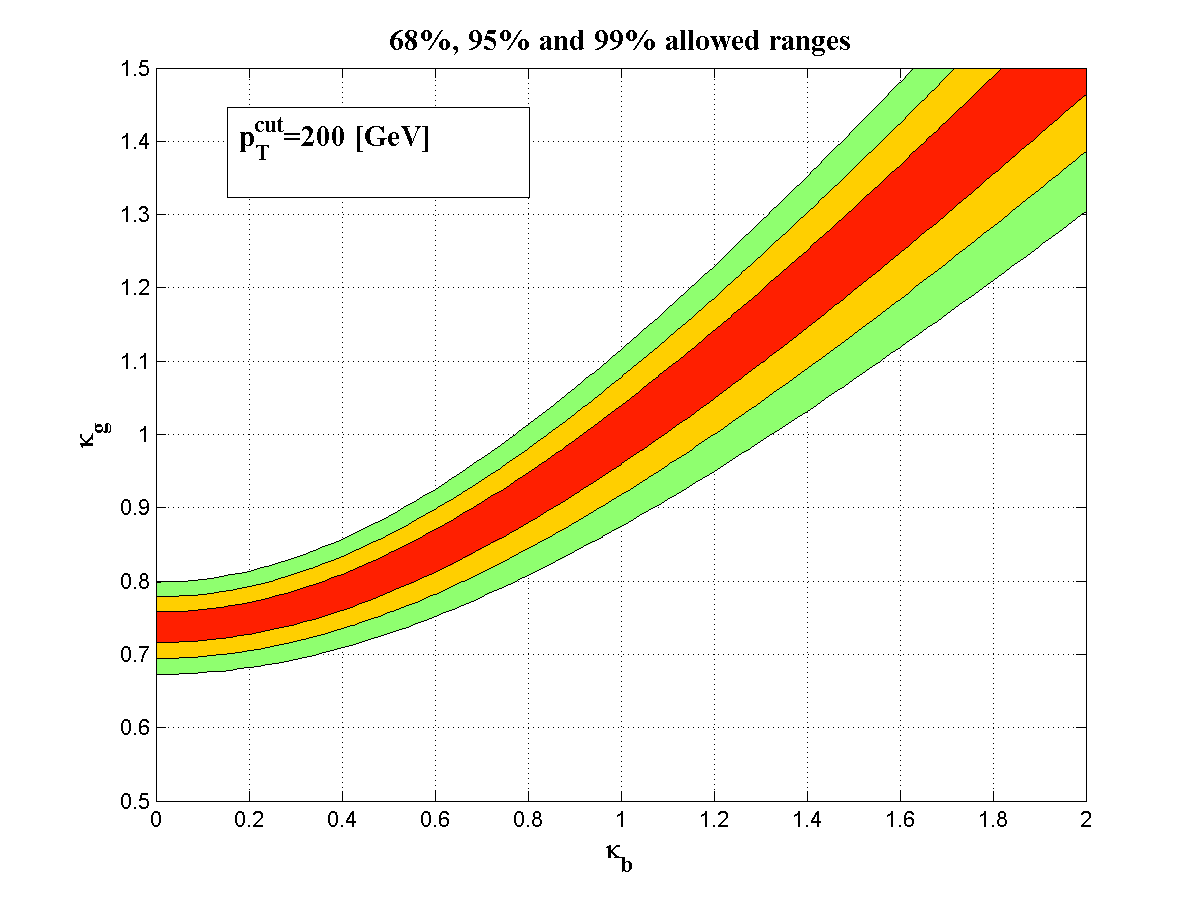}
\end{center}
\caption{The 68\%(red), 95\%(orange) and 99\%(green) CL
allowed ranges in the
$\kappa_b-\kappa_g$ plane,
corresponding to $\Delta\mu_{hj}^f \equiv \left| \mu_{hj}^f -1 \right| \leq 0.05,~0.1$ and $0.15$, respectively,
for $pp \to h + j_b$ events
with $p_T(h) > 30$ GeV (top) and $p_T(h) > 200$ GeV (bottom).}
\label{fighjb3}
\end{figure}
\begin{table}[htb]
\begin{center}
\begin{tabular}{|c||c|c|c|c|c|}
  \multicolumn{6}{|c|}{Statistical significance $N_{SD}=\frac{\Delta\mu_{hj_b}^f}{\delta\mu_{hj_b}^f}$} \\
\hline \hline
 & $\kappa_b=0.5$ & $\kappa_b=0.75$ & $\kappa_b=1$ & $\kappa_b=1.25$ & $\kappa_b=1.5$ \\
\hline \hline
 $\kappa_g=0.8 $ & $0.4^{+0.6}_{-0.3}$  & $2.8^{+0.08}_{-0.08}$ & $4.6^{-0.3}_{+0.3}$  &
 $6.0^{-0.6}_{+0.6}$ & $6.9^{-0.7}_{+0.7}$ \\
\hline
$\kappa_g=0.9 $  & $3.5^{-0.8}_{+0.8}$ & $0.2^{-0.08}_{+0.3}$ & $2.4^{-0.2}_{+0.2}$  & $4.3^{-0.4}_{+0.4}$ & $5.6^{-0.7}_{+0.7}$ \\
\hline
$\kappa_g=1.0 $ & $7.5^{-1.0}_{+1.0}$ & $3.3^{-0.4}_{+0.5}$ & $0$  & $2.4^{-0.3}_{+0.3}$ & $4.2^{-0.6}_{+0.6}$ \\
\hline
$\kappa_g=1.1 $ & $11.8^{-1.3}_{+1.3}$ & $6.7^{-0.7}_{+0.7}$ & $2.6^{-0.2}_{+0.2}$  & $0.4^{-0.2}_{+0.2}$ & $2.6^{-0.5}_{+0.5}$ \\
\hline
$\kappa_g=1.2 $  & $16.1^{-1.5}_{+1.5}$ & $10.2^{-0.9}_{+0.9}$ & $5.3^{-0.3}_{+0.3}$  & $1.7^{-0.07}_{+0.07}$ & $0.9^{-0.4}_{+0.4}$ \\
\hline
\end{tabular}
\caption{The statistical significance of the NP signal
for $pp \to h+j_b$, $N_{SD}=\Delta\mu_{hj_b}^f/\delta\mu_{hj_b}^f$,
assuming $\delta\mu_{hj_b}^f=0.05(1\sigma)$,
for $p_T^{cut}=200$ GeV and for values of the scaled
couplings $\kappa_b=0.5,0.75,1,1.25,1.5$ and $\kappa_g=0.8,0.9,1,1.1,1.2$.
The errors indicate the theoretical
uncertainty obtained by scale variations, where the superscript(subscript) corresponds
to twice(half) the nominal scale $\mu_F = \mu_R = \mu_T \equiv \sum_i \sqrt{m_i^2+p_{T}^{2}(i)}$, see also text.}
\label{tab2}
\end{center}
\end{table}

In the following, we will therefore use $p_T^{cut} = 30$ GeV and
200 GeV as two representative extreme
cases, where the former can be detected in the
$pp \to h +j_b \to \gamma \gamma + j_b$ channel, while the latter
is more suited for a higher statistics channel, such as
$pp \to h +j_b \to WW^\star + j_b$ followed by
the leptonic W-decays $WW^\star \to 2\ell 2\nu$, which
has a rate about five times larger than
$pp \to h +j_b \to \gamma \gamma + j_b$.
In Fig.~\ref{fighjb2} we plot
the statistical significance of the signals,
$N_{SD}=\Delta\mu_{hj_b}^f/\delta\mu_{hj_b}^f$,
for $p_T^{cut} = 30$ and 200 GeV,
as a function of $\kappa_b$, assuming $\kappa_g=1$ and a
$5\%(1\sigma)$ error $\delta\mu_{hj_b}^f = 0.05$.
We see that, for $p_T^{cut} = 200$ GeV
a $3 \sigma$ effect is expected
if $\kappa_b \lsim 0.8$ and/or $\kappa_b \gsim 1.3$, while
for $p_T^{cut} = 30$ GeV a larger deviation from the SM is required,
i.e., $\kappa_b \lsim 0.5$ and/or $\kappa_b \gsim 2.2$,
for a statistically significant signal of NP
in $pp \to h +j_b \to \gamma \gamma + j_b$.

In Fig.~\ref{fighjb3} we plot the 68\%, 95\% and 99\% CL
sensitivity ranges of NP in the
$\kappa_b-\kappa_g$ plane,
for $pp \to h+ j_b$ with
$p_T^{cut}=30$ GeV and $p_T^{cut}=200$ GeV,
assuming again that
$\mu_{hj}^f \sim 1 \pm 0.05(1\sigma)$, i.e., around the SM value
with a $5\%(1\sigma)$ accuracy.
We see that the two $p_T^{cut}$ cases probe different regimes in the
$\kappa_g - \kappa_b$ plane and are, therefore, complementary.

Finally, in Table \ref{tab2} we list the statistical significance of NP in
$pp \to h +j_b$, for $\delta\mu_{hj_b}^f=0.05(1\sigma)$,
$p_T^{cut}=200$ GeV and for several discrete values of the scaled
couplings: $\kappa_b=0.5,0.75,1,1.25,1.5$ and $\kappa_g=0.8,0.9,1,1.1,1.2$.
We include again the theoretical
uncertainty obtained by scale variations, which we find to be somewhat higher
than in the case of $pp \to h +j$.

Here also we can estimate the sensitivity of the signal to the calculational
setup, using the prescription described in the previous section. In particular,
we find that calculating $R^{hj_b}$ in Eq.~\ref{Rbg} with the exact
1-loop finite top-quark mass effect in
$\sigma_{ggh}^{hj_b}$,
the statistical significance values quoted in Table \ref{tab2} can vary
by up to a few standard deviations depending on the values of the scaled couplings
$\tilde\kappa_g$ and $\kappa_b$. For example, for
$(\kappa_b,\tilde\kappa_g)=(0.5,0.8),(0.5,1.0),(0.75,1.1),(1.0,1.2),(1.25,0.8)$ (see the
definition of $\tilde\kappa_g$ in Eq.~\ref{tkappag} and discussion therein), the
expected statistical significance changes from $N_{SD}=0.4,7.5,6.7,5.3,6.0$ in the
point-like $ggh$ approximation to $\tilde N_{SD}=2.3,4.0,4.4,4.1,4.0$
in the loop-induced (top-quark mass dependent) case.

\section{Higgs + jet production in the SMEFT \label{sec4}}

The SMEFT is defined by expanding the SM Lagrangian with an infinite series
of higher dimensional operators, ${\cal O}_i^{(n)}$ (using only the SM fields),
as \cite{EFTpapers,warsha}:
\begin{eqnarray}
{\cal L}_{SMEFT} ={\cal L}_{SM} + \sum_{n=5}^\infty \frac1{\Lambda^{(n-4)}}  {\sum_i
f_i^{(n)}  {\cal O}_i^{(n)}}\label{eff1} ~,
\end{eqnarray}
where $\Lambda$ is the scale of the NP that underlies the SM,
$n$ denotes the dimension and $i$ all other distinguishing labels.

Considering the expansion up to operators of dimension 6 (for a complete list
of dimension 6 operators in the SMEFT, see e.g. \cite{warsha}),
we will study here the following subset of operators
that can potentially modify the Higgs + jet production
processes:
\begin{eqnarray}
{\cal O}_{u \phi} &=& \left( \phi^{\dagger} \phi \right) \left(\bar Q_L \tilde\phi u_R \right) + h.c. ~, \\
{\cal O}_{d \phi} &=& \left( \phi^{\dagger} \phi \right)
\left(\bar Q_L \phi d_R \right) + h.c. ~, \\
{\cal O}_{u g} &=& \left(\bar Q_L \sigma^{\mu \nu} T^a u_R \right)
\tilde\phi G_{\mu \nu}^a + h.c. ~, \\
{\cal O}_{d g} &=& \left(\bar Q_L \sigma^{\mu \nu} T^a d_R \right)
\phi G_{\mu \nu}^a +h.c. ~, \\
{\cal O}_{\phi g} &=& \left(\phi^{\dagger} \phi \right)
G_{\mu \nu}^a G^{a, \mu \nu} ~,
\end{eqnarray}
where $\phi$ is the SM Higgs doublet
(with $\tilde\phi \equiv i \sigma_2 \phi^\star$), $G^{a, \mu \nu}$ denotes the QCD gauge-field strength and $Q_L$ and $u_R(d_R)$ are the ${\rm SU(2)}_L$ quark doublet and charge 2/3(-1/3) singlets, respectively.

In particular, we assume that the physics which underlies Higgs+jet production
is contained within (dropping the dimension index $n=6$):
\begin{eqnarray}
{\cal L}_{SMEFT} ={\cal L}_{SM} + \sum_{i=u \phi,d \phi,u g,dg,\phi g} \frac{f_i}{\Lambda_i^2}
{\cal O}_i \label{effours} ~,
\end{eqnarray}
and, to be as general as possible, we allow different scales of the NP which underly
the different operators. For example, $\Lambda_{u \phi}$ corresponds to the typical scale
of ${\cal O}_{u \phi}$, where by ``typical scale" we mean that
the corresponding Wilson coefficient is $f_{u \phi} \sim {\cal O}(1)$.

The effects of the operators ${\cal O}_{u \phi}, ~ {\cal O}_{d \phi}$ and
${\cal O}_{\phi g}$ can be ``mapped" into the kappa-framework, satisfying:
\begin{eqnarray}
\kappa_q \simeq \frac{y_q^{SM}}{y_b^{SM}} - \frac{f_{q \phi}}{y_b^{SM}} \frac{v^2}{\Lambda_{q \phi}^2} ~~,~~
\kappa_g = 1+ \frac{12 \pi f_{\phi g}}{\alpha_s} \frac{v^2}{\Lambda_{\phi g}^2} ~,
\label{SMEFTkappa}
\end{eqnarray}
where $y_q^{SM}/y_b^{SM} \to 0$ for e.g., $q=u~{\rm or}~ d$, while $y_q^{SM}/y_b^{SM} =1$ for the b-quark.
Thus,
the sensitivity of the signal strength $\mu_{hj}^f$ for $pp \to h+j$ (defined in Eqs.~\ref{sig-strength0} and \ref{sig-strength}) to the effective Lagrangian containing the operators ${\cal O}_{u \phi}, ~ {\cal O}_{d \phi}$ and
${\cal O}_{\phi g}$ can be obtained from the analysis that has been
performed for the kappa-framework in the previous section.
For example, it follows from Eq.~\ref{SMEFTkappa} that,
for $f_{u \phi},~f_{\phi g} \sim {\cal O}(1)$, one expects
$|\kappa_u| \lsim 0.5$ and $\Delta \kappa_g = | \kappa_g -1| \gsim 0.1$,
if the corresponding scales of NP are $\Lambda_{u \phi} \gsim 3$ TeV and $\Lambda_{\phi g} \lsim 15$ TeV, respectively.

On the other hand, the (flavor diagonal) operators
${\cal O}_{u g}$ and ${\cal O}_{d g}$ induce new chromo-magnetic dipole moment (CMDM) type,
$qqg$ and contact $qqgh$ interactions, which have a new Lorentz structure
and, therefore,
cannot be described by scaling the SM couplings.
In particular, these new CMDM-like operators give rise to
different Higgs + jet kinematics with respect to the SM.
The effects of the light-quarks and b-quark CMDM-like effective operators,
${\cal O}_{q g}$ ($q=u,d,c,s,b$),
in Higgs production at the LHC was studied in \cite{1411.0035,hbjet2_bMM},
where it was found that the inclusive Higgs production, $pp \to h+X$,
and Higgs + b-jets events
can be used to probe the CMDM-like interactions if its typical scale is
$\Lambda_{qg} \sim {\rm few}$ TeV. Here we will show that a better
sensitivity to the scale of the effective quark CMDM-like
operators, $\Lambda_{qg}$, can be achieved by analysing the exclusive
$pp \to h + j(j_b) \to \gamma \gamma + j(j_b)$ Higgs production
and decay channels and using the signal strength formalism with the
cumulative cross-sections for a high $p_T^{cut} \sim 200-300$ GeV.

Note that, in the general case where the Wilson coefficients $f_{u \phi}$, $f_{d \phi}$,
$f_{u g}$ and $f_{d g}$ are arbitrary $3 \times 3$ matrices in flavor space,
the operators ${\cal O}_{u \phi}$, ${\cal O}_{d \phi}$,
${\cal O}_{u g}$ and ${\cal O}_{d g}$ will generate tree-level flavor-violating
$u_i \to u_j$ and $d_i \to d_j$ transitions ($i,j =1-3$ are flavor indices).
One way to avoid that is to assume proportionality of these Wilson coefficients to
the corresponding $3 \times 3$ Yukawa coupling matrices ($Y_u$ and $Y_d$),
in which case the field redefinitions which diagonalize
the quark matrices also diagonalize these operators and the effective theory is
automatically minimally-flavor-violating (MFV). That is,
\begin{eqnarray}
\frac{f}{\Lambda^2} {\cal O}_{u g} \to  Y_q \cdot \frac{f_{MFV}}{\Lambda^2_{MFV}} {\cal O}_{u g} ~, \label{MFV1}
\end{eqnarray}
so that the relation
between generic NP parameters $(f,\Lambda)$ and the corresponding
parameters in the MFV effective theory is (for a single flavor $q$):
\begin{eqnarray}
\frac{\Lambda_{MFV}^2}{\Lambda^2} = y_q \frac{f_{MFV}}{f}~.
\end{eqnarray}

Thus, if $f_{MFV} \sim f$, then $\Lambda_{MFV} \sim \sqrt{y_q} \cdot \Lambda$,
in which case $\Lambda_{MFV} \ll \Lambda$ for $q \neq t$. On the other hand,
for $\sqrt{y_q f_{MFV}/f}  \sim {\cal O}(1)$ we have
$\Lambda_{MFV} \sim \Lambda$.
In what follows
we would like to keep our discussion as general as possible,
not restricting to any assumption
about the possible flavor structure of the Wilson coefficients.
In particular, we will focus below on a single flavor (diagonal element)
of these operators and assume that flavor violation
is controlled by some underlying mechanism in the high-energy theory (not necessarily
MFV), thereby suppressing the non-diagonal elements of these operators to an acceptable level.
\begin{figure}[htb]
\begin{center}
\includegraphics[scale=0.45]{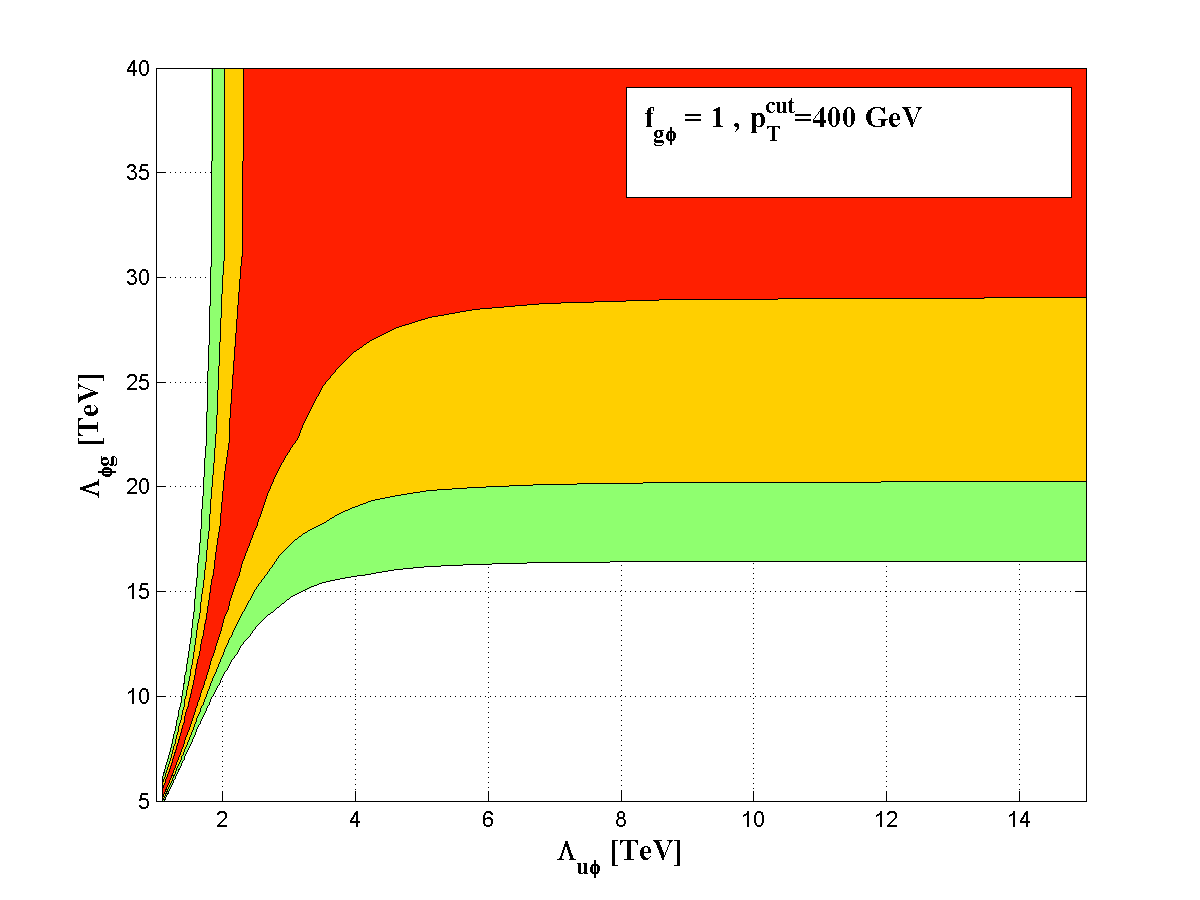}
\includegraphics[scale=0.45]{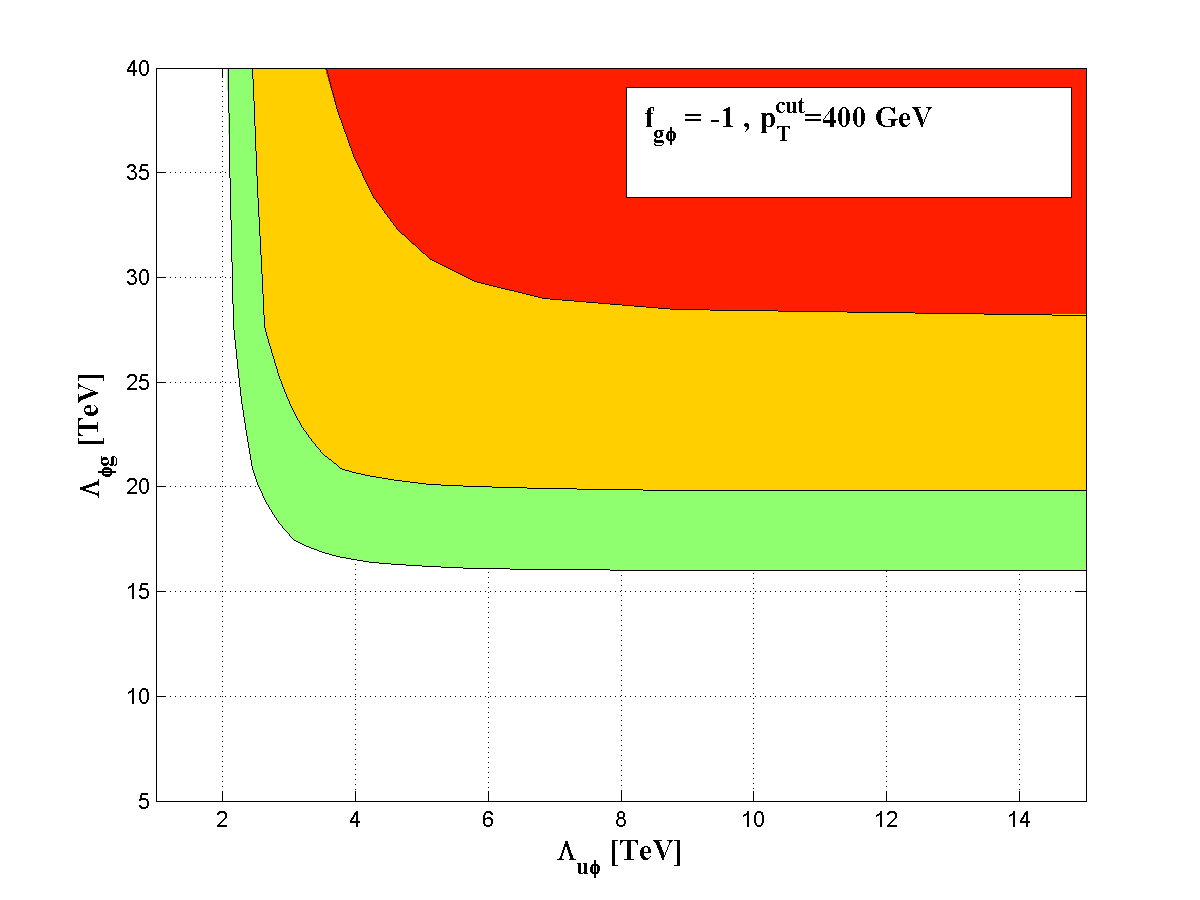}
\end{center}
\caption{The 68\%(red), 95\%(orange) and 99\%(green) CL ranges in the
$\Lambda_{q \phi}-\Lambda_{\phi g}$ plane,
corresponding to $\Delta\mu_{hj}^f \equiv \left| \mu_{hj}^f -1 \right| \leq 0.05,~0.1$ and $0.15$, respectively,
with $p_T^{cut}=400$ GeV and for
$f_{\phi g}=1$ (upper plot)
and $f_{\phi g}=-1$ (lower plot). In both cases
$|f_{u \phi}|=1$, see text.}
\label{fig7}
\end{figure}
\begin{figure}[htb]
\begin{center}
\includegraphics[scale=0.2]{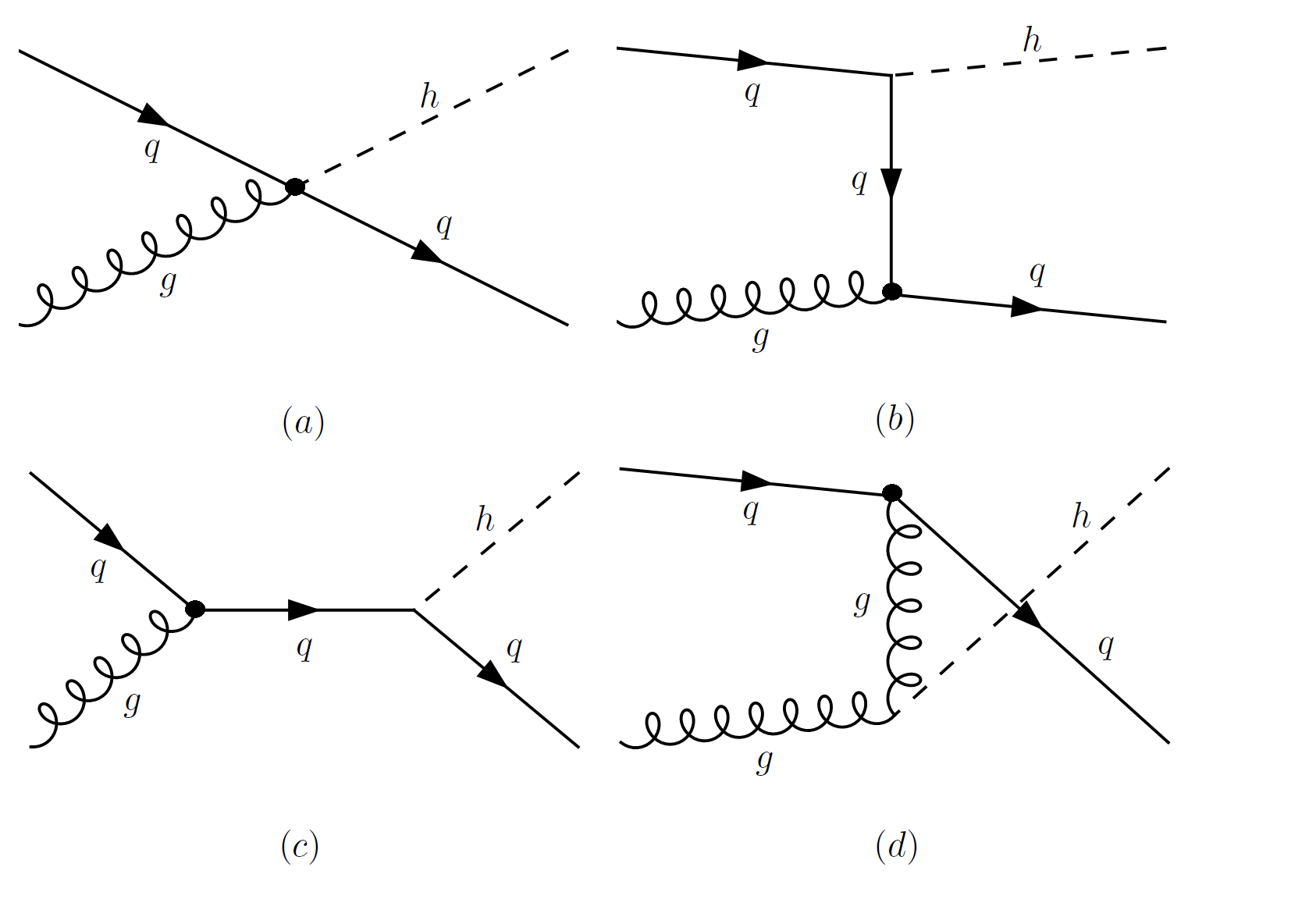}
\end{center}
\vspace{-0.8cm}
\caption{Sample of tree-level diagrams for $gq \to hq$, $q = u,d,c,s,b$
generated by the CMDM-like effective operator ${\cal O}_{q g}$, where
the heavy dot represents the CMDM-like vertices.
There are additional diagrams for the subprocess
$q \bar q\to hg$ and $g \bar q \to h \bar q $
that can also be obtained by crossing symmetry.
In the case of a Higgs + light jet production, $pp \to h+j$,
diagrams (b) and (c) are essentially absent (i.e., $y_q \to 0$).}
\label{figCMDMu}
\end{figure}

\subsection{The case of Higgs + light-jet production \label{subsec41}}

Let us consider first the operators
${\cal O}_{u \phi}$ and ${\cal O}_{\phi g}$, which, as seen from Eq.~\ref{SMEFTkappa}, modify the SM
$uuh$ and $ggh$ couplings in a way that is equivalent to the kappa-framework
(we will focus below only on the case of the 1st generation u-quark operator ${\cal O}_{u \phi}$).$^{[3]}$\footnotetext[3]{The
effects of ${\cal O}_{\phi g}$ and the top and bottom quarks operators
${\cal O}_{t \phi}$ and ${\cal O}_{b \phi}$ on the subprocess $gg \to hg$
were considered in \cite{1411.2029}, in the context of
Higgs-$p_T$ distribution in Higgs + jet production at the LHC.}
In particular, using Eq.~\ref{SMEFTkappa} and the
analysis performed in the previous section for NP in the kappa-framework, we plot
in Fig.~\ref{fig7} the 68\%, 95\% and 99\% CL sensitivity ranges in the
$\Lambda_{u \phi}-\Lambda_{\phi g}$ plane, for $p_T^{cut}=400$ GeV, assuming that
$\mu_{hj}^f \sim 1 \pm 0.05(1\sigma)$.
The sensitivity ranges are shown for the two cases
$f_{\phi g}= \pm 1$, where in both cases we set
$|f_{u \phi}|=1$, since the cross-section is
$\propto \kappa_q^2$ (see Eq.~\ref{sigNP}) so that
there is no dependence on the sign
of $f_{u \phi}$ for $y_u^{SM}/y_b^{SM} \to 0$ (see
Eq.~\ref{SMEFTkappa}).

We see that a measured value of $\mu_{hj}^f$ which is consistent with the SM at $3\sigma$ (i.e,
with $0.85 \leq \mu_{hj}^f \leq 1.15$) will exclude NP
with typical scales of $\Lambda_{\phi g} \lsim 15$ TeV (equivalent to $\kappa_u \gsim 0.6$) and
$\Lambda_{u \phi} \lsim 2$ TeV (equivalent to $\kappa_g \gsim 1.1$), for $f_{\phi g}=-1$.
In the case of $f_{\phi g}=1$,
there is an allowed narrow band in the
$\Lambda_{u \phi}-\Lambda_{\phi g}$ plane, stretching
down to NP scales of $\Lambda_{\phi g} \sim 5$ TeV and
$\Lambda_{u \phi} \sim 1$ TeV, which are consistent
with $0.85 \leq \mu_{hj}^f \leq 1.15$.
We note that, as in the kappa-framework analysis, these sensitivity ranges
in the $\Lambda_{u \phi}-\Lambda_{\phi g}$ plane mildly depend on
the calculation scheme of the SM-like diagrams involving the $ggh$
interaction, i.e., on the difference between
the point-like $ggh$ approximation and the exact 1-loop results.

\begin{figure}[htb]
\begin{center}
\includegraphics[scale=0.45]{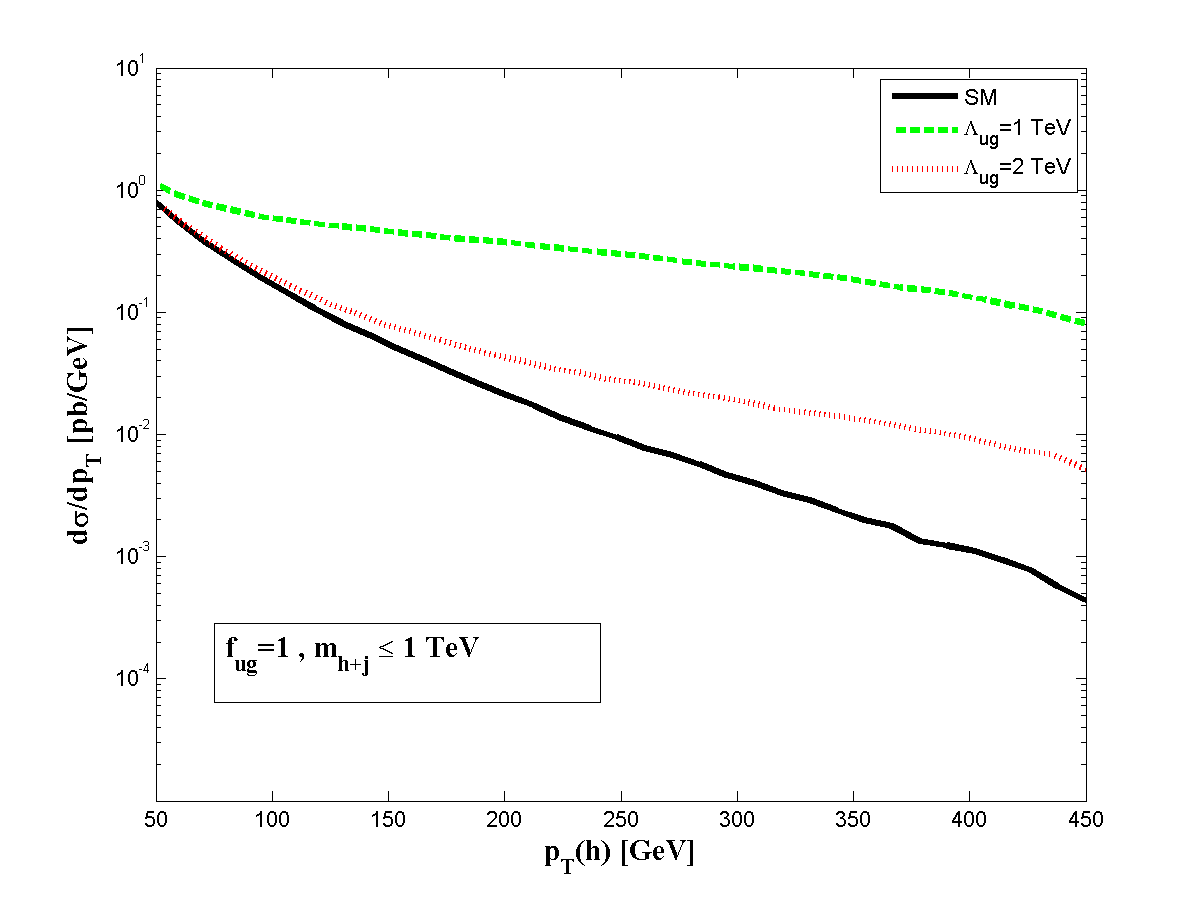}
\end{center}
\caption{The differential $p_T(h)$ distribution, $d\sigma(pp \to h+j)/dp_T(h)$,
 in the SM and with NP in the form
of ${\cal O}_{ug}$, for $\Lambda_{ug}=1$ and 2 TeV with $f_{ug}=1$ and
with an invariant mass cut of $m_{h+j} \leq 1$ TeV.
The SM curve was obtained using the point-like $ggh$ approximation which,
as mentioned earlier, overestimates the SM cross-section for $p_T(h) \gsim 200$ GeV.}
\label{CMDMu_pt}
\end{figure}
\begin{figure}[htb]
\begin{center}
\includegraphics[scale=0.45]{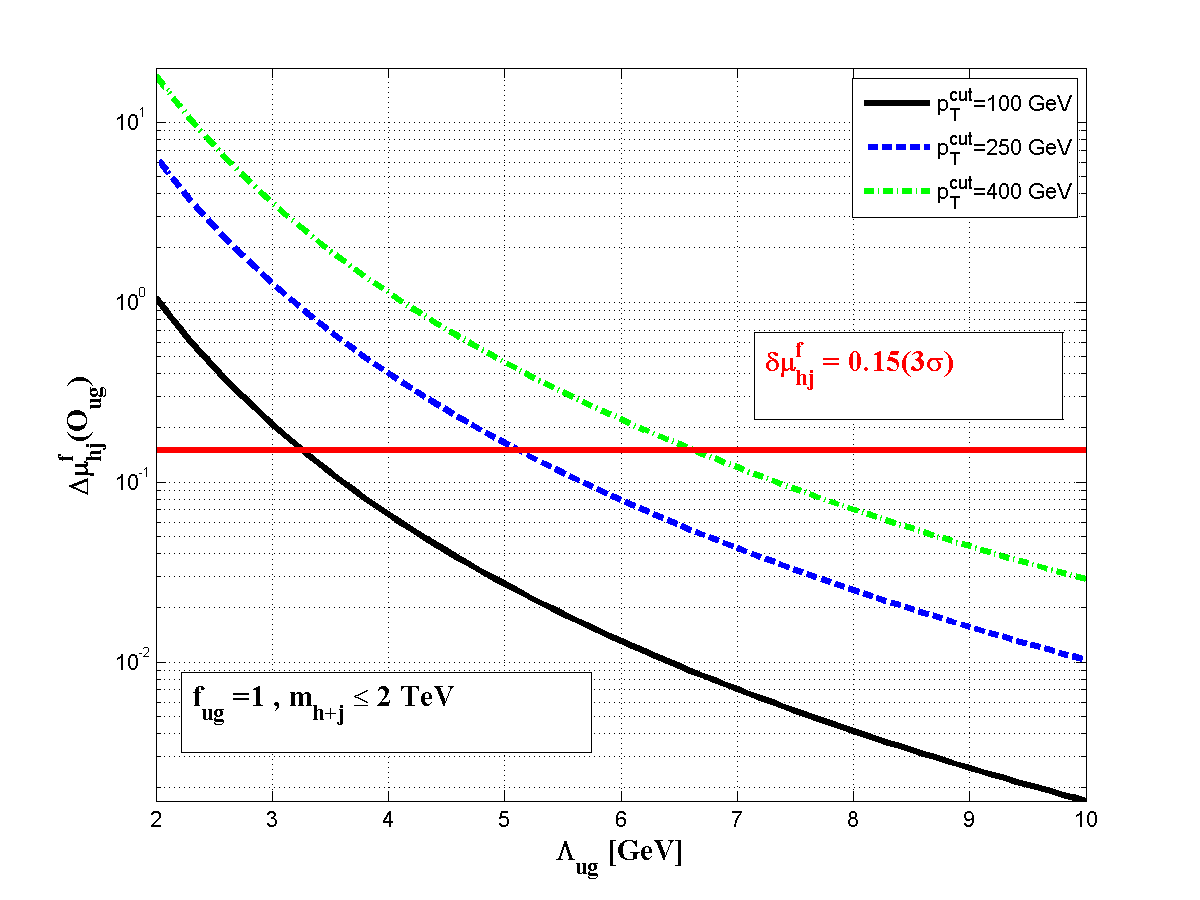}
\end{center}
\caption{The NP signal $\Delta\mu_{hj}^f ({\cal O}_{u g})$, as a function of $\Lambda_{ug}$
for $f_{ug}=1$, $p_T^{cut}=100,~250,~400$ GeV and an invariant mass cut of $m_{h+j} \leq 2$ TeV.
The horizontal red line
indicates the $5\%$ accuracy level. See also text.}
\label{CMDMu_dmu}
\end{figure}
\begin{figure}[htb]
\begin{center}
\includegraphics[scale=0.45]{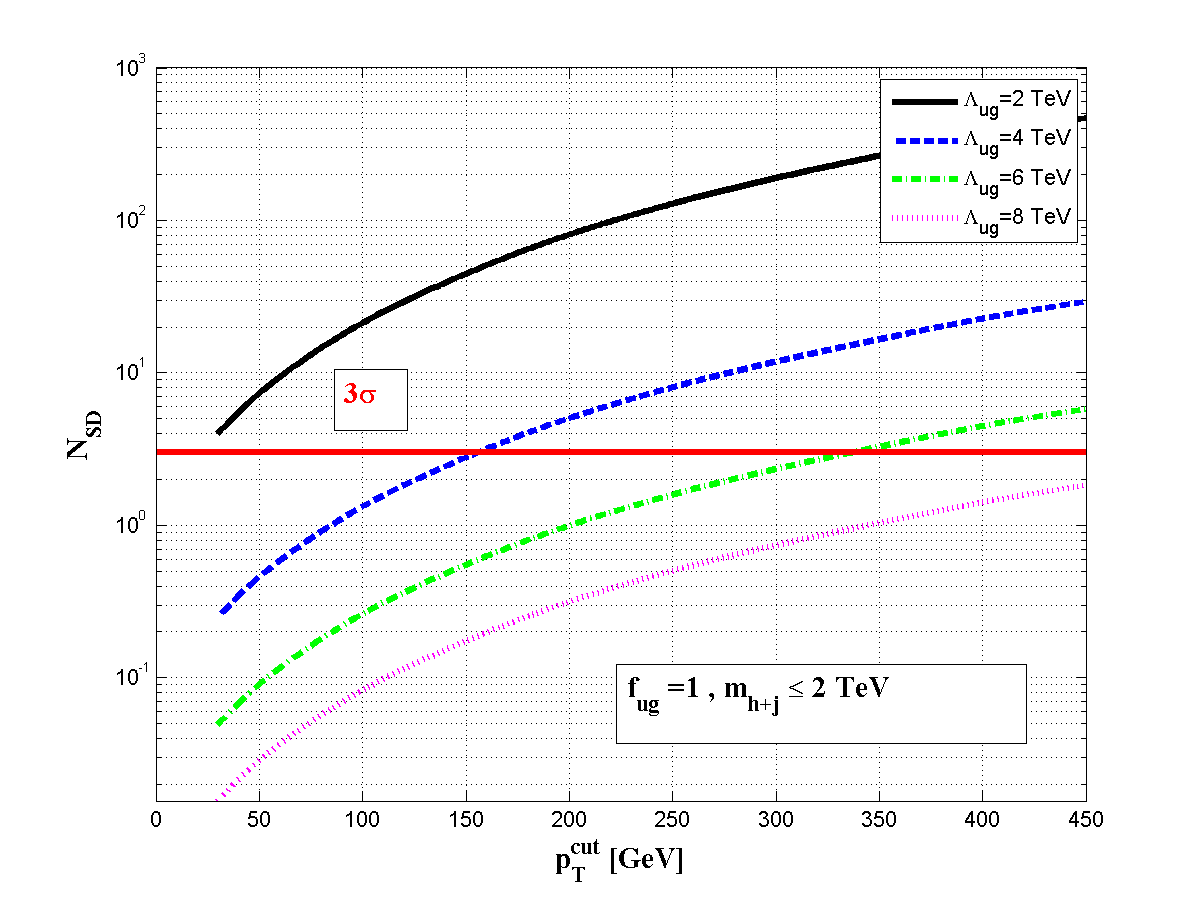}
\includegraphics[scale=0.45]{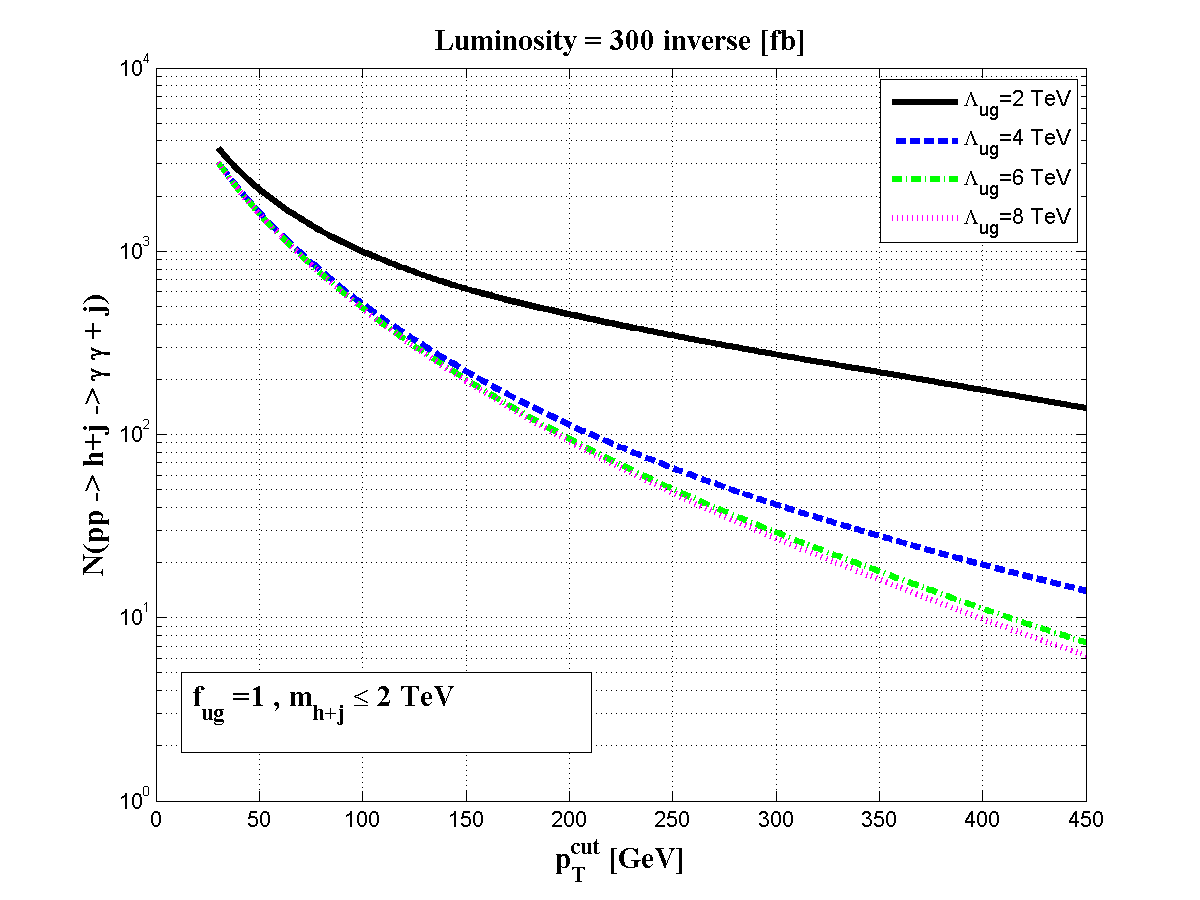}
\end{center}
\caption{The statistical significance of the signal (upper plot),
$N_{SD}=\mu_{hj}^f/\delta\mu_{hj}^f$,
for $\delta\mu_{hj}^f = 0.05(1\sigma)$,
and the expected number of $pp \to h +j \to \gamma \gamma +j$ events (lower plot),
as a function of $p_T^{cut}$,
for $\Lambda_{ug}=2,~4,~6$ and 8 TeV with $f_{ug}=1$ and with
${\cal L}=300$ fb$^{-1}$, a signal acceptance of 50\% and an invariant
mass cut of $m_{h+j} \leq 2$ TeV.
See also text.}
\label{CMDMu_nsd}
\end{figure}

We study next the effect of the CMDM-like operator ${\cal O}_{u g}$
on $pp \to h+j$ (again focusing only on the u-quark operator). The tree-level diagrams corresponding to
the contribution of ${\cal O}_{u g}$ to $pp \to h+j$ are depicted
in Fig.~\ref{figCMDMu}. They contain the momentum dependent CMDM-like
$uug$ vertex and $uugh$ contact interaction, which do not
interfere with the SM diagrams in the limit
of $m_u \to 0$. In particular,
in the presence of ${\cal O}_{u g}$, the total $pp \to h+j$ cross-section
can be written as:
\begin{eqnarray}
\sigma^{hj} = \sigma_{SM}^{hj} + \left(\frac{f_{u g}}{\Lambda_{u g}^2}\right)^2 \sigma_{ug}^{hj} ~, \label{sigCMDMu}
\end{eqnarray}
where the squared amplitudes for $\sigma_{SM}^{hj}$ are given in
Eqs.~\ref{dsig2}-\ref{dsig4} (see also Eq.~\ref{sigSM0})
and $\sigma_{ug}^{hj}$ is the NP cross-section corresponding to the square of the
CMDM-like amplitude, which is generated by the tree-level diagrams for
$q \bar q \to gh,~q g \to q h$ and $\bar q g \to \bar q h$ shown
in Fig.~\ref{figCMDMu}, with an insertion of the effective CMDM-like
$uug$ and $uugh$ vertices.
In particular, $\sigma_{ug}^{hj}$ is composed of
$\sigma_{ug}^{hj} = \sigma_{ug}^{hj} \left(q \bar q \to gh \right) +
\sigma_{ug}^{hj} \left(q g \to q h \right) +
\sigma_{ug}^{hj} \left(\bar q g \to \bar q h \right)$,
where the corresponding amplitude squared (summed and averaged over spins and colors)
are given by:
\begin{eqnarray}
\sum \overline{ \left| {\cal M}_{ug}^{q \bar q \to gh} \right|^2} &=& \frac{8}{{\cal C}_{qq}} \hat u \hat t
\left[ 1- 4 v C_g
+ 8 v^2 C_g^2 \right] ~, \label{dsigug1} \\
\sum \overline{ \left| {\cal M}_{ug}^{q g \to q h} \right|^2} &=&
- \frac{{\cal C}_{qq}}{{\cal C}_{qg}} \sum \overline{ \left| {\cal M}_{ug}^{q \bar q \to gh} \right|^2} (\hat s \leftrightarrow \hat t)~,  \label{dsigug2} \\
\sum \overline{ \left| {\cal M}_{ug}^{\bar q g \to \bar q h} \right|^2} &=&
- \frac{{\cal C}_{qq}}{{\cal C}_{qg}} \sum \overline{ \left| {\cal M}_{ug}^{q \bar q \to gh} \right|^2} (\hat s \leftrightarrow \hat u)~,  \label{dsigug3}
\end{eqnarray}
with
$\hat s=(p_1+p_2)^2,~\hat t=(p_1+p_3)^2$ and $\hat u=(p_2+p_3)^2$, defined for
$q(-p_1)+ \bar q(-p_2) \to h + g(p_3)$.
\begin{figure*}[htb]
\begin{center}
\includegraphics[scale=0.4]{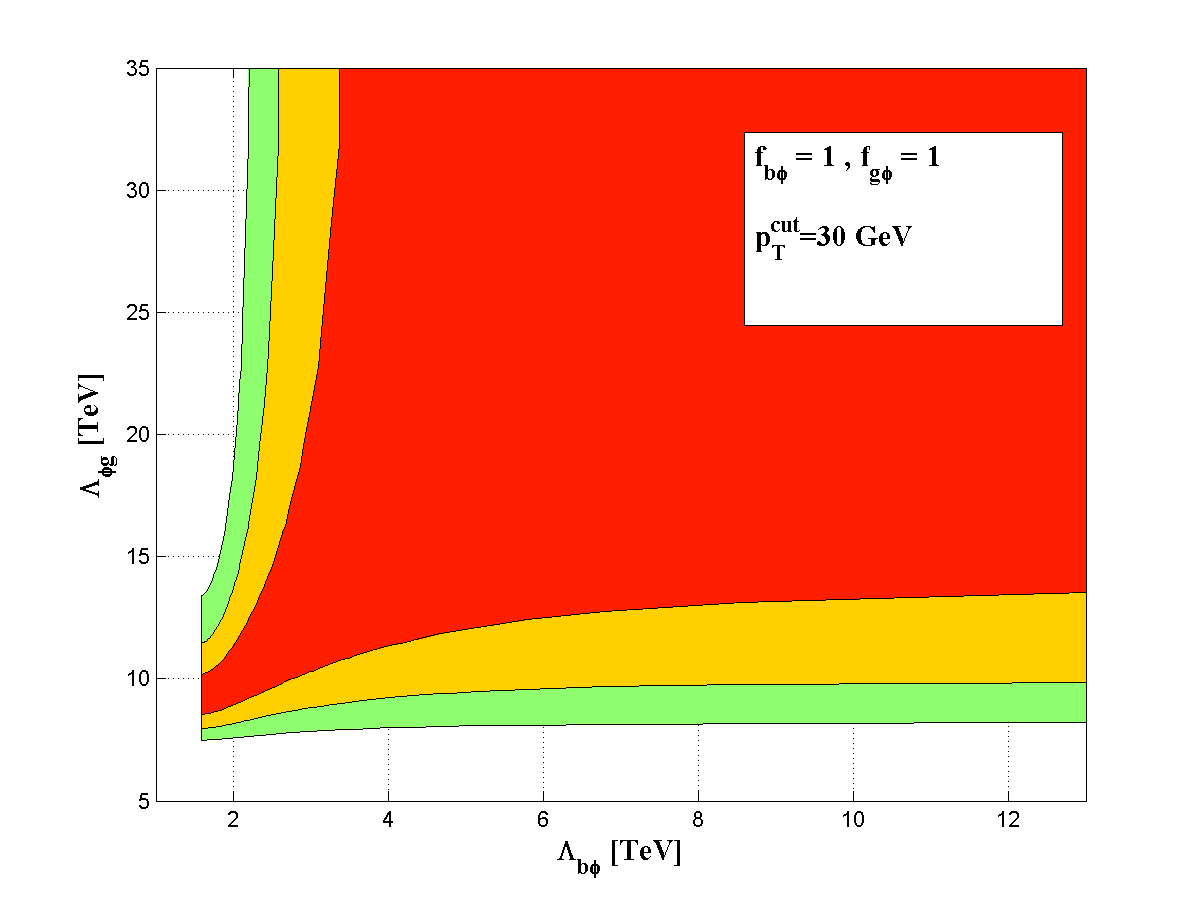}
\includegraphics[scale=0.4]{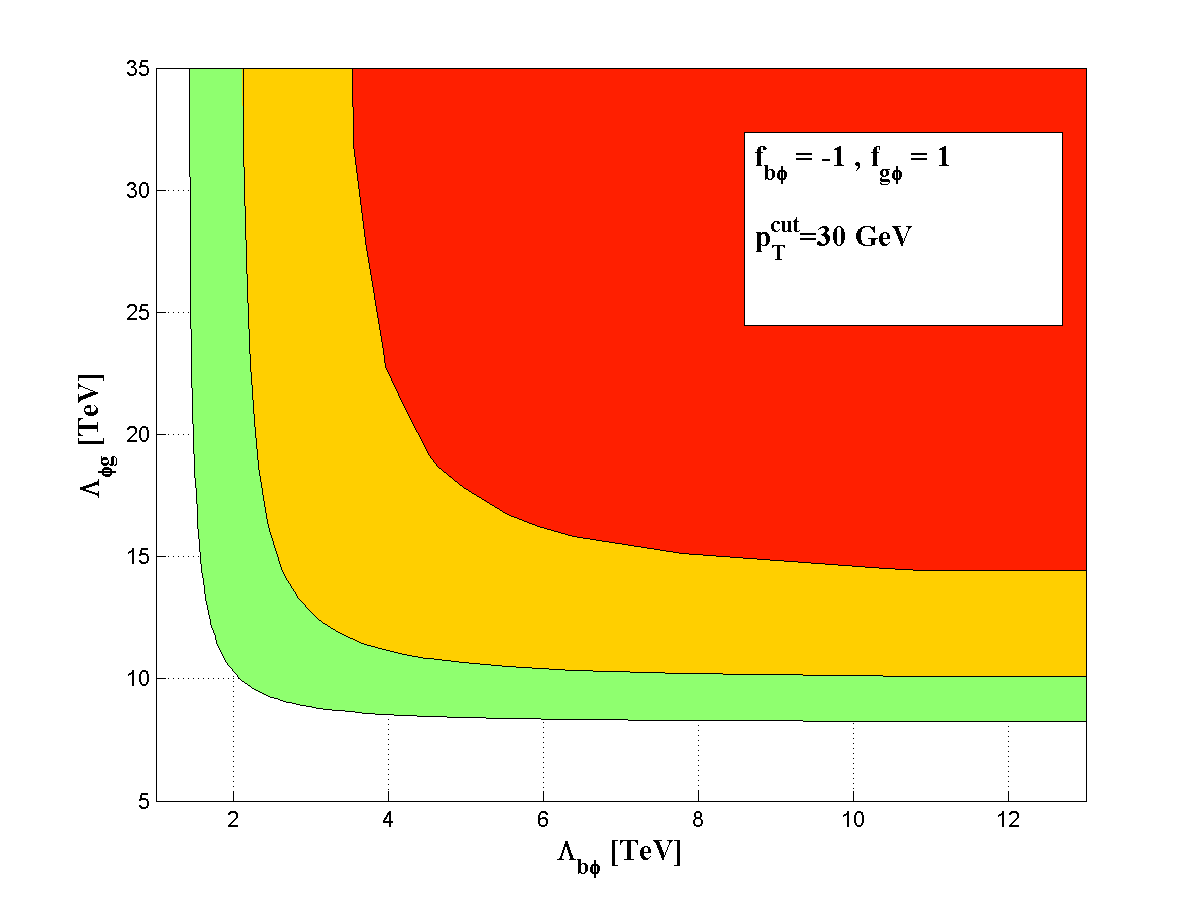}
\includegraphics[scale=0.4]{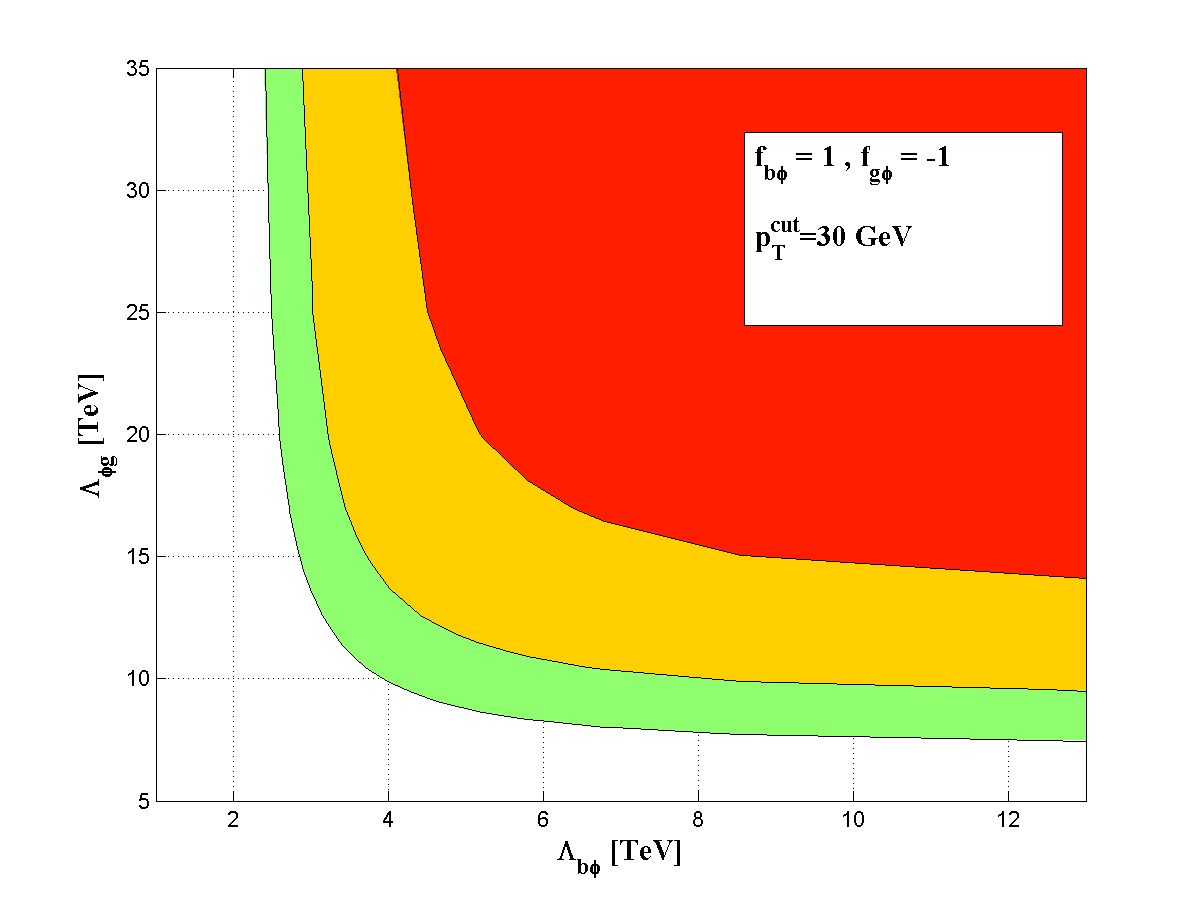}
\includegraphics[scale=0.4]{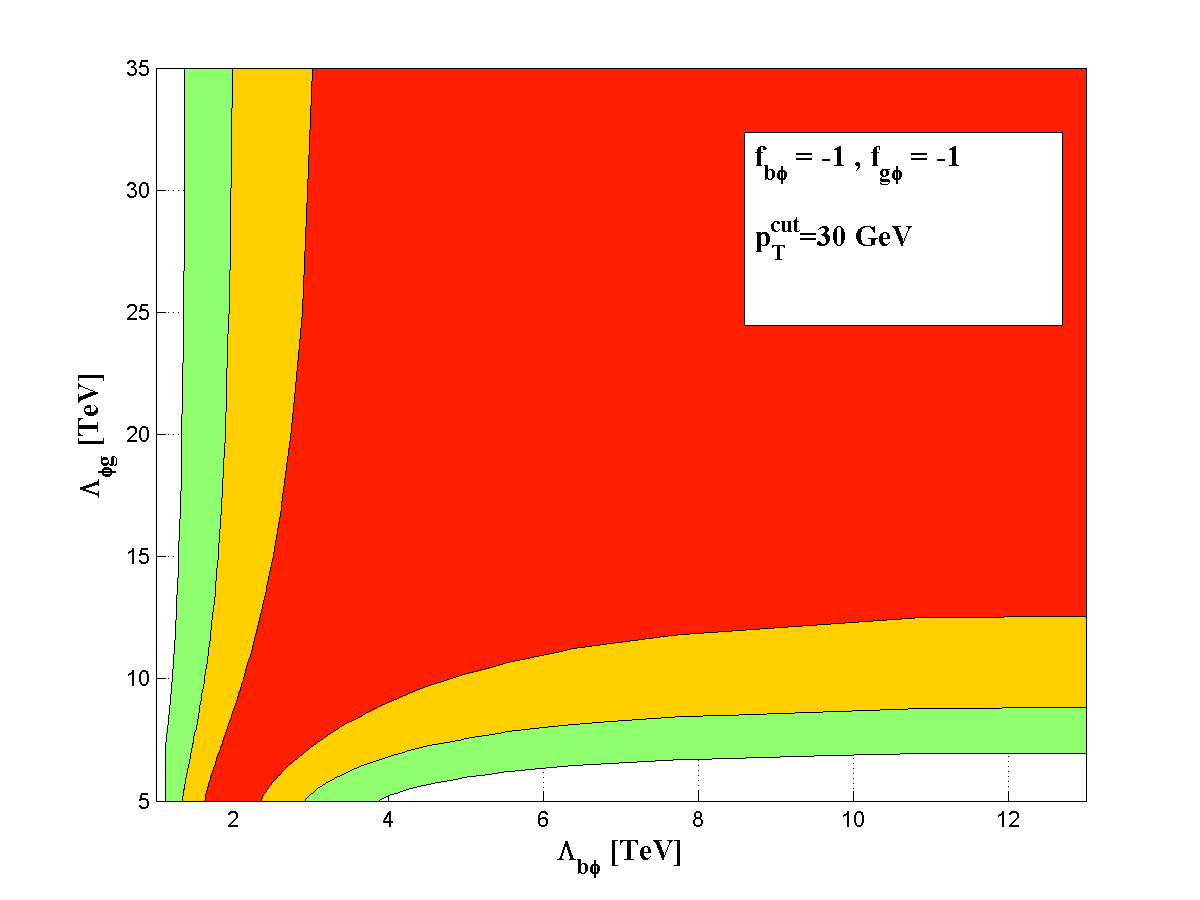}
\end{center}
\caption{The 68\%(red), 95\%(orange) and 99\%(green) CL sensitivity ranges in the
$\Lambda_{b \phi}-\Lambda_{\phi g}$ plane,
corresponding to $\Delta\mu_{hj}^f \equiv \left| \mu_{hj}^f -1 \right| \leq 0.05,~0.1$ and $0.15$, respectively,
with $p_T^{cut}=30$ GeV and
for $f_{b \phi}=1,~f_{\phi g}=1$,
$f_{b \phi}=-1,~f_{\phi g}=1$,
$f_{b \phi}=1,~f_{\phi g}=-1$ and
$f_{b \phi}=-1,~f_{\phi g}=-1$, as indicated in the figures.}
\label{fighjbSMEFT1}
\end{figure*}
\begin{figure*}[htb]
\begin{center}
\includegraphics[scale=0.4]{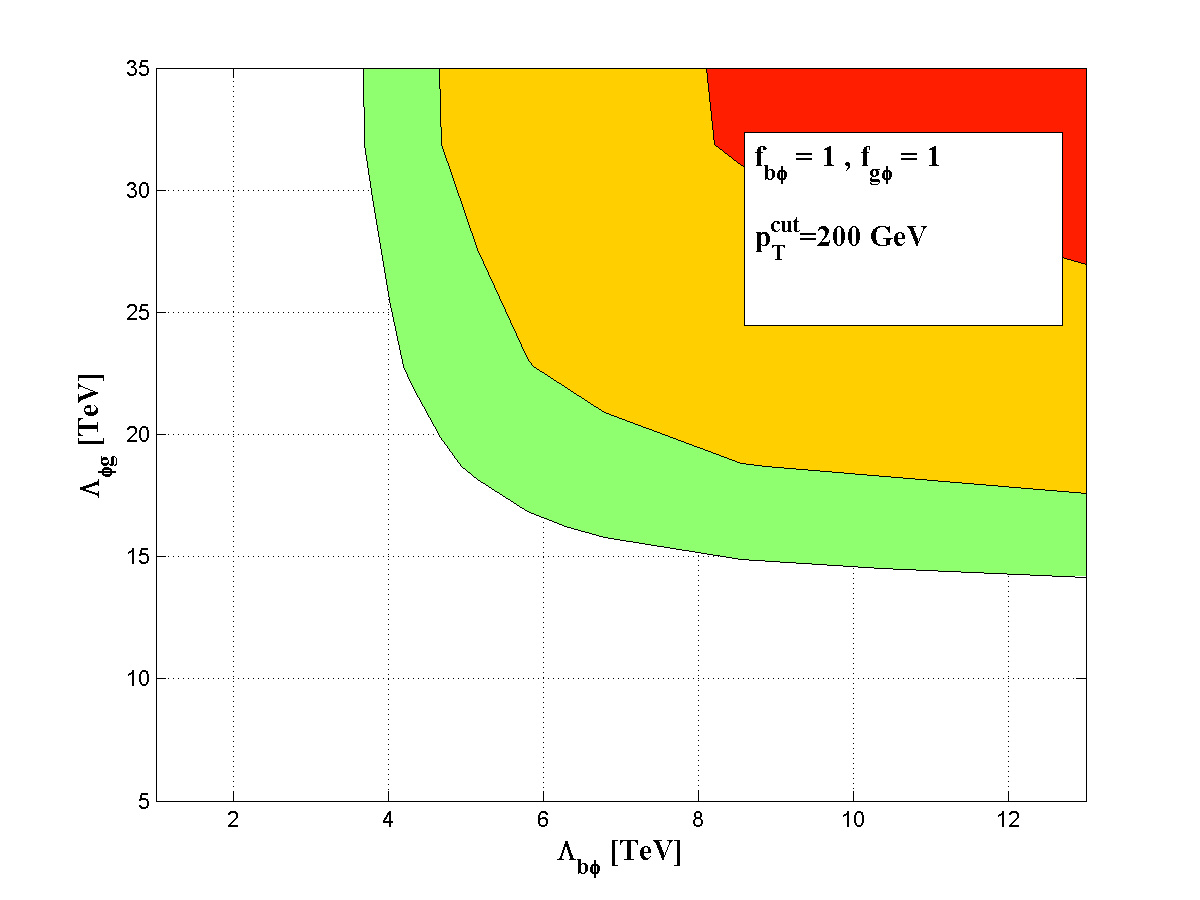}
\includegraphics[scale=0.4]{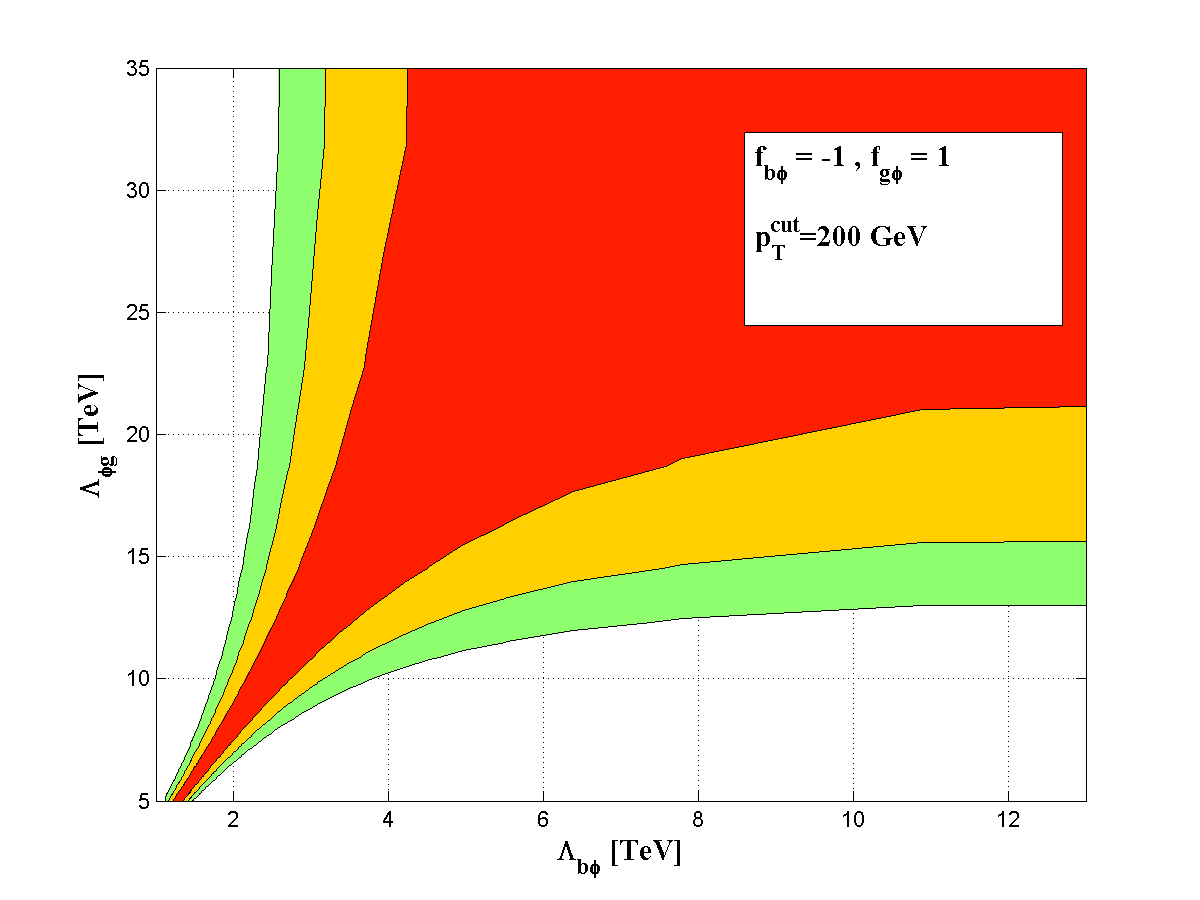}
\includegraphics[scale=0.4]{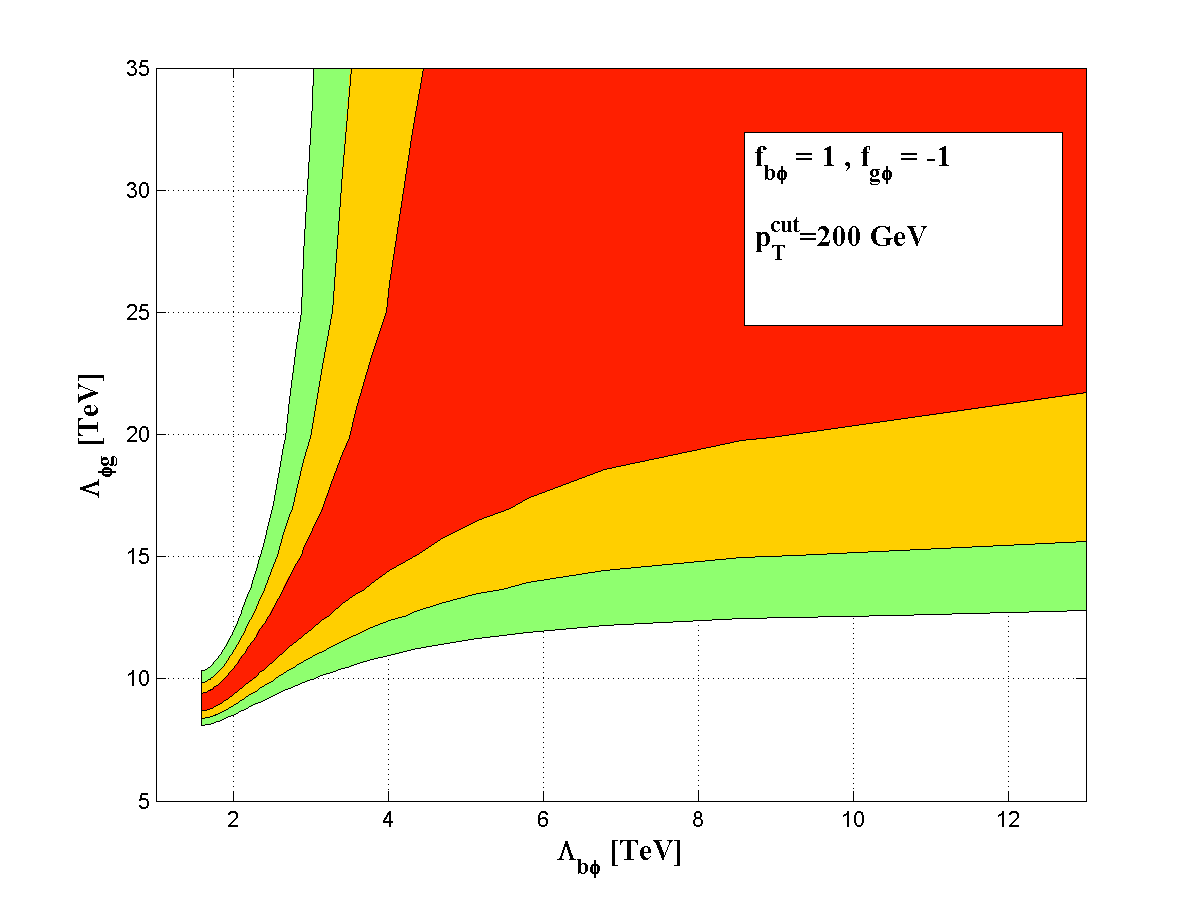}
\includegraphics[scale=0.4]{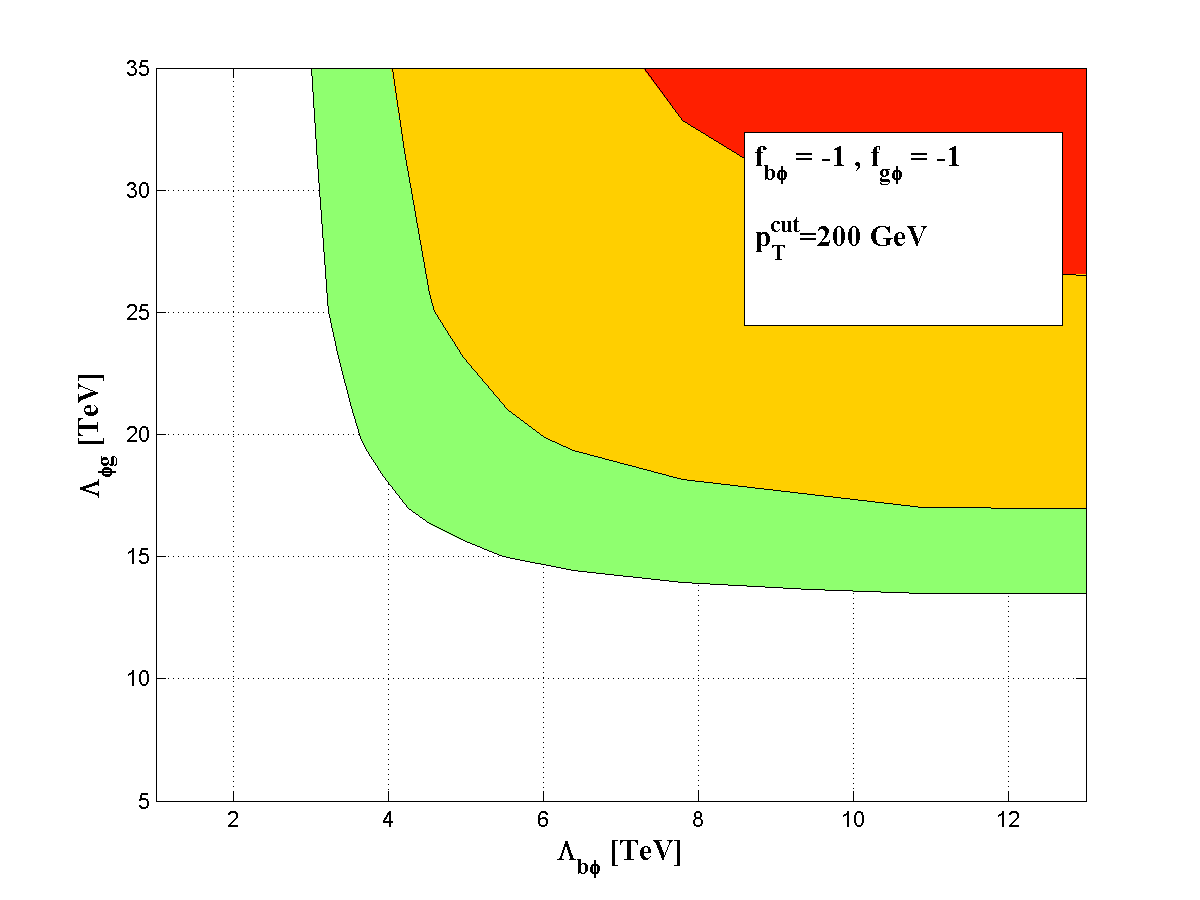}
\end{center}
\caption{Same as Fig.~\ref{fighjbSMEFT1} for $p_T^{cut}=200$ GeV.}
\label{fighjbSMEFT2}
\end{figure*}

As illustrated in Fig.~\ref{CMDMu_pt}, the momentum dependent contribution
from ${\cal O}_{u g}$ drastically changes the $p_T(h)$-dependence
of the cross-section with respect to the SM and also with respect to the case where the
NP is in the form of scaled
couplings (i.e., in the kappa-framework). Indeed, the effect of ${\cal O}_{u g}$
(or any other NP with a similar $p_T(h)$ behaviour) are better isolated in the harder Higgs $p_T$
regime. This can be obtained by using a relatively high $p_T^{cut}$ for the cumulative
cross-section (see below).
\begin{figure}[htb]
\begin{center}
\includegraphics[scale=0.45]{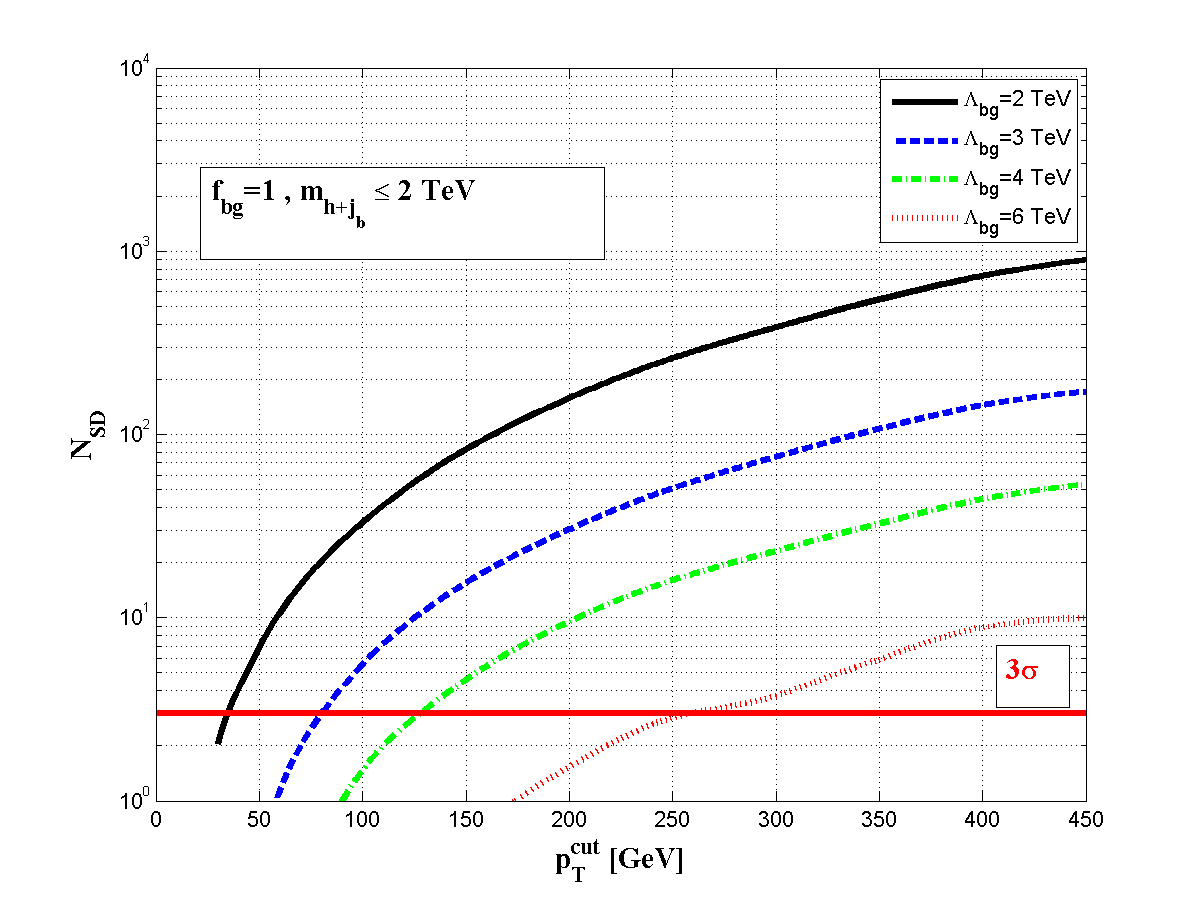}
\end{center}
\caption{The statistical significance of the signal, $N_{SD}=\mu_{hj_b}^f/\delta\mu_{hj_b}^f$,
for $\delta\mu_{hj_b}^f = 0.05(1\sigma)$,
as a function of $p_T^{cut}$,
in the presence of ${\cal O}_{b g}$ (assuming no additional NP in the decay)
for $f_{bg}=1$ and
$\Lambda_{bg}=2,~3,~4$ and 6 TeV and with an invariant $h +j_b$ mass cut
of $m_{h+j_b} \leq 2$ TeV.}
\label{CMDMb_nsd}
\end{figure}

Assuming no additional NP in the decay (the effects of ${\cal O}_{u g}$ in the Higgs decay
is $\propto (m_h/\Lambda_{u g})^4$ and is, therefore, negligible for $\Lambda \sim {\rm few}$ TeV),
the corresponding signal strength is:
\begin{eqnarray}
\mu_{hj}^f ({\cal O}_{u g}) = 1 + \left(\frac{f_{u g}}{\Lambda_{u g}^2}\right)^2
R_{ug}^{hj} ~~,~~ R_{ug}^{hj} \equiv \frac{\sigma_{ug}^{hj}}{\sigma_{SM}^{hj}} ~, \label{muCMDMu}
\end{eqnarray}
so that the NP signal, as defined in Eq.~\ref{deltamu}, is:
\begin{eqnarray}
\Delta\mu_{hj}^f ({\cal O}_{u g}) = \mid \mu_{hj}^f({\cal O}_{u g}) - 1 \mid = \left(\frac{f_{u g}}{\Lambda_{u g}^2}\right)^2
R_{ug}^{hj} ~. \label{deltamu2}
\end{eqnarray}

In Fig.~\ref{CMDMu_dmu} we plot the NP signal, $\Delta\mu_{hj}^f ({\cal O}_{u g})$, as a
function of $\Lambda_{ug}$ with $f_{ug}=1$,
for $p_T^{cut}$ values of 100, 250 and 400 GeV and an invariant mass cut $m_{h+j} \leq 2$ TeV.
As expected (see Fig.~\ref{CMDMu_pt}),
the sensitivity to $\Lambda_{ug}$ is significantly improved
the higher the $p_T^{cut}$ is. In particular, while
$\Delta\mu_{hj}^f/\mu_{hj}^f \gsim 5\%$ for $p_T^{cut}=100$ GeV and $\Lambda_{u g} \lsim 4$ TeV,
for $p_T^{cut}=400$ GeV we obtain
$\Delta\mu_{hj}^f/\mu_{hj}^f \gsim 5\%$ for $\Lambda_{u g} \lsim 8.5$ TeV.

In Fig.~\ref{CMDMu_nsd} we plot the statistical significance of the signal,
$N_{SD}=\mu_{hj}^f/\delta\mu_{hj}^f$, for $\delta\mu_{hj}^f = 0.05(1\sigma)$,
and the expected number of events, again assuming that the Higgs decays via $h \to \gamma \gamma$,
i.e., $N(pp \to h + j \to \gamma \gamma + j)$, as a function of $p_T^{cut}$ and
for $\Lambda_{ug}=2,~4,~6$ and 8 TeV with $f_{ug}=1$ and an invariant mass cut $m_{h+j} \leq 2$ TeV.
$N(pp \to h + j \to \gamma \gamma + j)$ is shown for an integrated luminosity
of 300 fb$^{-1}$ and a signal acceptance of 50\%.
We see, for example, that
if $\Lambda_{ug}=6$ TeV, then a high $p_T^{cut} \sim 350$ GeV is required
in order to obtain a $3 \sigma$ effect, for which
$N(pp \to h + j \to \gamma \gamma + j) \sim {\cal O}(10)$
and ${\cal O}(100)$ is expected at the LHC with ${\cal L}=300$ fb$^{-1}$ and the HL-LHC with ${\cal L}=3000$ fb$^{-1}$, respectively.

Note that the effect of changing the calculation scheme of the SM
cross-section from the point-like $ggh$ interaction to the exact mass dependent 1-loop one
is to change $R_{ug}^{hj} \to r_{ggh} R_{ug}^{hj}$ in Eq.~\ref{muCMDMu}
($r_{ggh}$ is defined in Eq.~\ref{rggh}) and therefore
it also increases the statistical significance $N_{SD}$ by a factor of
$r_{ggh}$ which depends on the $p_T^{cut}$ used (see Fig.~\ref{compsigma}).
Thus, the statistical significance values reported in the upper plot of
Fig.~\ref{CMDMu_nsd} are on the conservative side.

\subsection{The case of Higgs + b-jet production \label{subsec42}}

As mentioned above, the effects of the NP operators ${\cal O}_{b \phi}$ and
${\cal O}_{\phi g}$ in $pp \to h+ j_b$,
can be described using the kappa-framework formalism
of Eq.~\ref{Lkappa}, with the NP factors multiplying the SM $bbh$
Yukawa coupling ($\kappa_b$) and $ggh$ coupling ($\kappa_g$)
as prescribed in Eq.~\ref{SMEFTkappa}.

In Figs.~\ref{fighjbSMEFT1} and \ref{fighjbSMEFT2} we plot
the 68\%, 95\% and 99\% CL sensitivity ranges in the
$\Lambda_{b \phi}-\Lambda_{\phi g}$ plane,
for $(f_{b \phi},f_{\phi g})=(1,1),(1,-1),(-1,1),(-1,-1)$ and
$p_T^{cut}=30$ GeV and 200 GeV,
assuming again that the signal strength had been measured to a $5\%(1\sigma)$ accuracy
with a SM central value, i.e., $\mu_{hj}^f \sim 1 \pm 0.05(1\sigma)$.
As in the kappa-framework analysis of the previous section, we use
the two $p_T^{cut}$ values, $p_T^{cut} = 30$ GeV and $p_T^{cut} = 200$ GeV,
as two representative examples of a high and low statistics
$pp \to h +j_b\to \gamma\gamma +j_b$ signal at the HL-LHC
(see also Fig.~\ref{fighjb1}).
As expected, a better sensitivity to the NP is obtained for
the higher $p_T^{cut} = 200$ GeV, where $\Lambda_{b \phi} \lsim 3$ TeV
and $\Lambda_{\phi g} \lsim {\cal O}(10)$ TeV can be excluded at
$3 \sigma$ if $\mu_{hj_b}^f$ is found to be consistent with the SM
within  15\% ($3 \sigma$).
Here also, similar to the kappa-framework analysis for $pp \to h+j_b$,
the sensitivity ranges
in the $\Lambda_{b \phi}-\Lambda_{\phi g}$ plane for the $p_T^{cut} = 200$ GeV
case mildly depend on whether the SM cross-section is calculated
with the point-like $ggh$ approximation or at 1-loop with a finite top-quark mass.

Finally, we consider the case where the NP in $pp \to h+j_b$ is due
only to the b-quark CMDM-like operator ${\cal O}_{b g}$. The corresponding
tree-level diagrams with the new momentum
dependent CMDM-like
$bbg$ vertex and $bbgh$ contact interaction are
shown in Fig.~\ref{figCMDMu}, where, as opposed to the $pp \to h+j$ case, here
there is an interference (though small - see below) between the CMDM-like diagrams and
the tree-level SM ones (depicted in Fig.~\ref{SMTL}).
In particular, in the presence of ${\cal O}_{b g}$,
the total $pp \to h+j_b$ cross-section can be written as:
\begin{eqnarray}
\sigma^{hj_b} = \sigma_{SM}^{hj_b} +
\frac{f_{b g}}{\Lambda_{b g}^2} \sigma_{bg}^{1,h j_b} +
\left(\frac{f_{b g}}{\Lambda_{b g}^2}\right)^2 \sigma_{bg}^{2, hj_b} ~, \label{sigCMDMb}
\end{eqnarray}
where $\sigma_{SM}^{hj_b}$ is the SM cross-section
(the relevant SM squared amplitude terms are given in Eqs.~\ref{dsigb2},\ref{dsigb3},\ref{dsig3},\ref{dsig4}) and
the NP terms $\sigma_{bg}^{1,2}$ can be obtained from
the following CMDM-like NP squared amplitudes
(summed and averaged over spins and colors):
\begin{eqnarray}
\sum \overline{ \left| {\cal M}_{bg}^{1,b g \to bh} \right|^2} &=&
\frac{8 g_s y_b}{{\cal C}_{qg}}
\left( 4 v C_g \hat t -m_h^2 \right) ~, \label{dsigbg1} \\
\sum \overline{ \left| {\cal M}_{bg}^{2,b g \to bh} \right|^2} &=&
- \frac{8}{{\cal C}_{qg}} \left[\hat s \hat u \left( 1- 4 v C_g
+ 8 v^2 C_g^2 \right) \right. \nonumber \\
&& \left. + y_b^2 v^2 \hat t \right]
~, \label{dsigbg2} \\
\sum \overline{ \left| {\cal M}_{bg}^{1,\bar b g \to \bar bh} \right|^2}  &=&
\sum \overline{ \left| {\cal M}_{bg}^{1,b g \to bh} \right|^2} (\hat u \leftrightarrow \hat t)~,  \label{dsigbg3} \\
\sum \overline{ \left| {\cal M}_{bg}^{2,\bar b g \to \bar bh} \right|^2}  &=&
\sum \overline{ \left| {\cal M}_{bg}^{2,b g \to bh} \right|^2} (\hat u \leftrightarrow \hat t)~,  \label{dsigbg4}
\end{eqnarray}
where again
$\hat s=(p_1+p_2)^2,~\hat t=(p_1+p_3)^2$ and $\hat u=(p_2+p_3)^2$, defined for
$b(-p_1)+ \bar b(-p_2) \to h + g(p_3)$.

We see from Eqs.~\ref{dsigbg1} and \ref{dsigbg3} above that the interference terms
${\cal M}_{bg}^{1,b g \to bh}$ and
${\cal M}_{bg}^{1,\bar b g \to \bar bh}$
(corresponding to
$\sigma_{bg}^{1,hj_b}$ in Eq.~\ref{sigCMDMb}) are
proportional to $y_b \sim {\cal O}(m_b/v)$ and are therefore sub-leading, so that
the dependence of the $pp \to h+j_b$ cross-section on the sign of the CMDM-like
Wilson coefficient, $f_{bg}$, is tenuous.
As a result, $\sigma^{hj_b}$ has a very similar
$p_T$-behaviour as the one depicted in Fig.~\ref{CMDMu_pt} for
the $pp \to h+j$ case.
In particular, here also,
the Higgs $p_T$ spectrum becomes appreciably harder with respect to the SM
and also with respect to the case of the NP operators ${\cal O}_{b \phi}$ and ${\cal O}_{g \phi}$,
due to the momentum-dependent $\sigma_{bg}^{2, hj_b}$ term, which corresponds
to the square of the b-quark CMDM-like diagrams, generated by the
operator ${\cal O}_{b g}$ and depicted in Fig.~\ref{figCMDMu}.

In Fig.~\ref{CMDMb_nsd} we plot the statistical significance of the ${\cal O}_{b g}$
signal
for $\delta\mu_{hj}^f = 0.05(1\sigma)$, as a function of $p_T^{cut}$
for $f_{bg}=1$ and $\Lambda_{bg}=2,~3,~4$ and 6 TeV,
imposing an invariant mass cut
of $m_{h+j_b} \leq 2$ TeV.
The results for $f_{bg}=-1$ are very similar due to the small interference
between the CMDM-like and SM amplitudes (see discussion above).
We see that, as expected,
the sensitivity to the scale of the CMDM-like operator, $\Lambda_{bg}$, is higher
the higher the $p_T^{cut}$ is. We find, for example, that
the effect of ${\cal O}_{bg}$ with a typical scale of $\Lambda_{bg} \sim 4$ TeV
can be probed in $pp \to h+j_b \to \gamma \gamma +j_b$
to the level of $N_{SD} \sim {\cal O}(10\sigma)$ with $p_T^{cut} = 200$ GeV.
The expected number of $pp \to h +j_b \to \gamma \gamma + j_b$
events in this case (i.e., for $\Lambda_{bg} \sim 4$ TeV, $p_T^{cut} = 200$ GeV
and an invariant mass cut of $m_{h+j_b} \leq 2$ TeV),
assuming an integrated luminosity of 3000 fb$^{-1}$,
a signal acceptance of ${\cal A} = 0.5$ and a
b-jet tagging efficiency of 70\%, $\epsilon_b=0.7$, is
$N(pp \to h + j_b \to \gamma\gamma + j_b) \sim 30$ (see also Fig.~\ref{fighjb1}).

As for the sensitivity of the above results to the
calculational scheme: due to the smallness of the interference term
it is similar to that of the u-quark CMDM-like case in $pp \to h+j$.
In particular, the statistical significance $N_{SD}$ shown
in Fig.~\ref{CMDMb_nsd} should also be considered conservative with respect
to the values which would have been obtained using the exact 1-loop induced
SM cross-section, i.e., $N_{SD}$ is naively larger by a factor of
$r_{ggh}$ in the exact 1-loop calculation case.

\section{Summary \label{sec5}}

We have examined the effects of various NP scenarios,
which entail new forms of effective $qqh$ and $qqg$ interactions
in conjunction with beyond the SM Higgs-gluon effective coupling, in exclusive
Higgs + light-jet ($pp \to h+j$) and Higgs + b-jet ($pp \to h+j_b$)
production at the LHC. We have defined the signal strength for
$pp \to h+j (j_b)$ followed by the Higgs decay $h \to ff$, as the ratio
of the corresponding NP and SM rates,
and studied its dependence on the Higgs $p_T$ spectrum.
We specifically focused on $h \to \gamma \gamma$ and assumed
that there is no NP in this decay channel.

We first analyse NP in $pp \to h+j (j_b) \to \gamma \gamma + j(j_b)$ within the kappa-framework, in which
the SM Higgs couplings to the light-quarks ($qqh$) and to the gluons ($ggh$) are assumed to be scaled
by a factor of $\kappa_q$ and $\kappa_g$, respectively. In particular, in our notation
the scale factors $\kappa_q$ for all light-quark's Yukawa couplings ($q=u,d,c,s,b$) are normalized
with respect to the b-quark Yukawa, $\kappa_q = y_q/y_b^{SM}$,
so that in the SM we have e.g., $\kappa_b=1$ and $\kappa_u \sim {\cal O}(10^{-3})$.
This NP setup does not introduce any new Lorentz structure in the underlying hard processes
(i.e., $gg \to gh$, $qg \to q h$, $\bar q g \to \bar q h$, $q \bar q \to g h$ in the case
of $pp \to h+j$ and $bg \to b h$, $\bar b g \to \bar b h$ in the case of $pp \to h+j_b$), thus
retaining the SM $pp \to h+j (j_b)$ kinematics.
In particular, we find that strong bounds can be obtained in the $\kappa_g - \kappa_q$
plane at the LHC, by measuring
a $p_T$-dependent signal strength for
Higgs + jet events at relatively high Higgs $p_T$.
For example, the combination of $\kappa_g < 0.8$ with $\kappa_u > 0.25$
($\kappa_g < 0.8$ with $\kappa_b > 1.5$) can be excluded at more than $7 \sigma$
at the HL-LHC with a luminosity of 3000 fb$^{-1}$, if the signal strength
in the $pp \to h+j(j_b) \to \gamma \gamma +j(j_b)$ channels will be measured
and known to an accuracy of $5\%(1\sigma)$, for high $p_T(h)$ events with
$p_T(h) \geq 400(200)$ GeV.
Recall that in our notation the corresponding SM strengths of these couplings are
$\kappa_b=\kappa_g=1$ and $\kappa_u \sim {\cal O}(10^{-3})$.

We also considered NP effects in $pp \to h+j(j_b)$
in the SMEFT framework, where higher dimensional
effective operators modify the SM $qqh$ Yukawa couplings and the Higgs-gluon $ggh$ interaction
by a scaling factor, similar to the case of the kappa-framework for NP. We thus utilize
an interesting ``mapping" between the SMEFT and kappa-frameworks to derive new bounds on the typical
scale of NP that underlies the SMEFT lagrangian.
We find, for example, that $pp \to h+j(j_b) \to \gamma \gamma +j(j_b)$ events with
high $p_T(h) > 400(200)$ GeV at the HL-LHC,
are sensitive to the new effective operators that modify the $qqh$ (Yukawa) and $ggh$ couplings,
if their typical scale (i.e., with ${\cal O}(1)$
dimensionless Wilson coefficients) is a few TeV and ${\cal O}(10)$ TeV, respectively.

Finally, as a counter example, we study the effects of NP in the form of dimension six
u-quark and b-quark chromo magnetic dipole moment (CMDM)-like effective operators, which induce new derivative
and new contact interactions that significantly distort the $pp \to h+j(j_b)$ SM kinematics and, therefore,
cannot be described in terms of scaled couplings. In particular, in this case,
the high-$p_T$ Higgs spectrum becomes significantly harder with respect to the SM.
We thus show that $pp \to h+j(j_b) \to \gamma \gamma + j(j_b)$ events at the HL-LHC,
with a high Higgs $p_T$ of $p_T(h) \gsim 400(200)$ GeV, can probe the higher dimensional
CMDM-like u-quark and b-quark effective operators, if their typical scale is around
$\Lambda \sim 5$ TeV.

Our main results were obtained using an effective point-like $ggh$ interaction
approximation. To estimate the sensitivity to this approximation, we also compared
samples of our results to the case where the $ggh$ vertex is calculated explicitly at
leading order, which, for Higgs + jet, corresponds to
a 1-loop mass dependent calculation using a finite top-quark mass.

\bigskip
\bigskip
\bigskip
\bigskip
\bigskip
\bigskip

{\bf Acknowledgments:}
The work of AS was supported in part by the US DOE contract \#DE-SC0012704.

\pagebreak

\end{document}